\documentclass[fleqn,usenatbib]{mnras}

\usepackage[T1]{fontenc}
\usepackage{ae,aecompl}

\DeclareRobustCommand{\VAN}[3]{#2}
\let\VANthebibliography\thebibliography
\def\thebibliography{\DeclareRobustCommand{\VAN}[3]{##3}\VANthebibliography}

\usepackage{graphicx}	% Including figure files
\usepackage{amsmath}	% Advanced maths commands
\usepackage{amssymb}	% Extra maths symbols
\usepackage{bm}

\usepackage{pdflscape}
\usepackage{ulem}

\usepackage{newtxtext,newtxmath}
\hypersetup{draft} % use when there are problems with overleaf compilation
\title[CasA NS cooling from all Chandra/ACIS-S data]{
Constraints on neutron star superfluidity from the  cooling neutron star in Cassiopeia~A using all Chandra ACIS-S observations }

\author[P.~S. Shternin et al.]{Peter S. Shternin,$^{1}$\thanks{E-mail: pshternin@gmail.com}
Dmitry D. Ofengeim,$^{2,1}$
Craig O. Heinke,$^{3}$
Wynn C.~G. Ho$^{4}$\\ 
$^{1}$Ioffe Institute, Politekhnicheskaya 26, St. Petersburg, 194021, Russia\\
$^{2}$Racah Institute of Physics, The Hebrew University of Jerusalem, Jerusalem 91904, Israel
\\
$^{3}$Department of Physics, University of Alberta, CCIS 4-181, T6G 2E1, Edmonton, Alberta, Canada\\
$^{4}$Department of Physics and Astronomy, Haverford College, 370 Lancaster Avenue, Haverford, PA, 19041, USA\\
}

\date{Accepted XXX. Received YYY; in original form ZZZ}

\pubyear{2022}

\begin{document}
\label{firstpage}
\pagerange{\pageref{firstpage}--\pageref{lastpage}}
\maketitle

\begin{abstract}
Analysis of Chandra observations of the neutron star (NS) in the centre of the Cassiopeia A supernova remnant taken in the subarray (FAINT) mode of the ACIS detector performed by Posselt and collaborators revealed, after inclusion of the most recent (May 2020) observations, a significant  decrease of the source surface temperature from 2006 to 2020. The obtained cooling rate is consistent with those obtained from analysis of the 2000--2019 data taken in the GRADED mode of the ACIS detector, which is potentially more strongly affected by  instrumental effects. We performed a joint spectral analysis using all ACIS data to constrain the NS  parameters and cooling rate. We constrain the mass of the Cassiopeia~A NS at  $M=1.55\pm0.25~M_\odot$, and its radius at $R=13.5\pm 1.5$~km. The surface temperature cooling rate is found to be $2.2\pm 0.3$ per cent in 10 years if the absorbing hydrogen column density is allowed to vary
and $1.6\pm 0.2$ per cent in 10 years if it is fixed. The observed cooling can be explained by 
enhanced neutrino emission from the superfluid NS interior due to Cooper Pair Formation (CPF) process. Based on analysis of all ACIS data, we constrain the maximal critical temperature of triplet neutron pairing within the NS core at 
$(4-9.5)\times 10^{8}$~K.
In accordance with previous studies, the required effective strength of the CPF neutrino emission is at least a factor of 2 higher than existing microscopic calculations suggest.
\end{abstract}


%%%%%%%%%%%%%%%%%%%%%%%%%%%%%%%%%%%%%%%%%%%%%%%%%%

%%%%%%%%%%%%%%%%% BODY OF PAPER %%%%%%%%%%%%%%%%%%

%%%%%%%%%%%%%%%%%%%%%%%%%%%%%%%%%%%%%%%%%%%%%%%%%%%%%%%%%%%%%%%%%%%%%%%%%%%%%%%%%%%%%%%%%%%%%%%%%%%%
\section{Introduction}\label{S:intro}
%%%%%%%%%%%%%%%%%%%%%%%%%%%%%%%%%%%%%%%%%%%%%%%%%%%%%%%%%%%%%%%%%%%%%%%%%%%%%%%%%%%%%%%%%%%%%%%%%%%%
The point-like X-ray source CXOU~J232327.9+584842  \citep{Tananbaum1999IAUC,Pavlov2000ApJ} in the centre of the Cassiopeia~A supernova remnant is a $\sim 340$~yr old neutron star (hereafter CasA~NS), which is so far the  youngest neutron star known in 
or around 
our Galaxy, except possibly
the candidate neutron star from SN 1987A \citep{Page2020ApJ}. It belongs to a small class of the central compact objects (CCOs) --- weakly-magnetised thermally emitting neutron stars (NSs) found near supernova remnant centres \citep[e.g.,][]{DeLuca2017JPhCS}. It is the first CCO for which a carbon composition was found to be likely for its atmosphere \citep{Ho2009Natur,Ho21}. 

Moreover, \citet{Heinke2010ApJ} found that the CasA~NS exhibits a real-time cooling, with a rate of about $4$ per cent per decade. These findings were subsequently confirmed and refined by additional observations 
with the same instrument and mode 
\citep{Shternin2011MNRAS,Elshamouty2013ApJ,Ho2015PhRvC, Wijngaarden2019MNRAS,Ho21}, lowering 
the temperature decay rate 
over 10 years 
to $2-3$ per cent depending on the model assumptions (see \citealt{Ho21} and \citealt{Shternin2021MNRAS} for details). 
A possible explanation to the observed rapid cooling involves an onset of the triplet-state pairing of neutrons in the NS core in 
the 
near past, which triggered the intensive neutrino emission accompanying the formation of the neutron Cooper pairs \citep{Page2011PhRvL,Shternin2011MNRAS}. Within this interpretation, it is  possible to constrain the physical parameters of the nucleon pairing (aka superfluidity) from the CasA~NS cooling observations. Notice, that this is not the only interpretation of these data, see, e.g., \citet{Shternin2021MNRAS} for a brief review of other proposals. 

Recently \citet{Shternin2021MNRAS}, hereafter Paper I, have developed the model-independent method for extracting the information of the NS superfluidity parameters from the observations of the real-time cooling of a NS. This technique was applied to the CasA~NS cooling data summarised by \citet{Ho21} which contain 14 sets of observations taken over 19 yrs. As a result, the maximal (over the NS core) critical temperature of the triplet neutron pairing, $T_{Cn\mathrm{max}}$, was firmly constrained in the range of $ (5-10)\times 10^8$~K, independent of the equation of state of a NS or superfluidity model, see Paper I for details.

However, all observations analysed in Paper I were taken in the  GRADED mode of the \textit{Chandra} ACIS-S detector, which potentially suffers from  instrumental effects. The main complication comes from the considerable pile-up\footnote{Pile-up is the recording of two adjacent photons during one CCD readout period as a single photon, altering the inferred spectrum; see \url{https:/cxc.harvard.edu/ciao/ahelp/acis\_pileup.html}.}, especially at earlier observation epochs, which needs to be included in the spectral model. There are reasonable doubts that the pile-up effects can be modelled in a complete and unbiased way \citep{Posselt2013ApJ}. 

In contrast, observations in the ACIS-S FAINT, or subarray, mode do not suffer from these problems because of the $\sim 10$ times smaller frame time than in the GRADED mode, dramatically reducing the frequency of pile-up\footnote{Notice, however, that the ACIS detector in any mode suffers from the contaminant problem that can induce additional systematic effects, see \citet{Plucinsky2020SPIE} for details.}.
However, until recently, the  FAINT-mode observations had not provided sufficient temporal coverage to firmly constrain or reject the rapid cooling of the CasA~NS \citep{Posselt2013ApJ,Posselt2018ApJ}.
In May 2020, two additional observations, ObsID 22426 and 23248 in Chandra ACIS FAINT mode (PI B. Posselt), were performed, extending the FAINT mode observations time span to 14 years. Analysis of these observations reported by \citet{PosseltPavlov2022ApJ} resulted in the 10-yr decay rate of about 2 per cent at $\gtrsim 5\sigma$ significance.
Moreover, as we show below, the results of the spectral analysis of the FAINT mode data turn out to be consistent with those for the GRADED mode data, 
suggesting that the whole ACIS-S dataset can be used simultaneously.

In the present study we repeat the analysis of Paper I,  using all available ACIS-S data (taken both in FAINT and GRADED modes), and constrain the parameters of the CasA NS, its temperature evolution, and NS superfluidity models.

The paper is organised as follows. The spectral analysis is described in Sec.~\ref{S:obs}. We first illustrate the similarity between the results obtained in GRADED and FAINT modes fitting the spectral data by models with fixed NS parameters in Sec.~\ref{S:obs:fixMR}. We then perform full multiparametric Bayesian analysis using all spectral data in Sec.~\ref{sec:MCMC}. We briefly recall the method of Paper I and constrain the parameters of the NS superfluidity in Sec.~\ref{Sec:SFanalysis}. We discuss the results in Sec.~\ref{S:Discuss} and conclude in Sec.~\ref{S:conclusions}.

%%%%%%%%%%%%%%%%%%%%%%%%%%%%%%%%%%%%%%%%%%%%%%%%%%%%%%%%%%%%%%%%%%%%%%%%%%%%%%%%%%%%%%%%%%%%%%%%%%%%
\section{Spectral analysis}\label{S:obs}
%%%%%%%%%%%%%%%%%%%%%%%%%%%%%%%%%%%%%%%%%%%%%%%%%%%%%%%%%%%%%%%%%%%%%%%%%%%%%%%%%%%%%%%%%%%%%%%%%%%%

\renewcommand\arraystretch{1.2}
\begin{table*}
\caption{Chandra FAINT mode observations and the results of the simplified spectral fitting described in the text. The quantitiy $t_{\mathrm{exp}}$ is the observation exposure time; for merged observations the sum of the observation times is given. Modified Julian dates for merged observations are exposure-time weighted. Fits with variable $N_{\mathrm{H}}$ are for $M=1.60M_\odot$ and $R=13.7$~km, while for fixed  $N_{\mathrm{H}}$, $M=1.53M_\odot$ and $R=13.5$~km are used. Distance is set at $d=3.33$~kpc. The last column gives the number of spectral energy bins in each spectra. 
Uncertainties correspond to 68 per cent confidence intervals.
}
\label{tab:T_NH_Xspec_spectral}
\begin{tabular}{llcccccccccccc}
\hline
&&&&\multicolumn{4}{c}{$N_\mathrm{H}$ variable} & \multicolumn{3}{c}{$N_{\mathrm{H}}=1.656\times 10^{22}~\mathrm{cm}^{-2}$}\\
 ObsID & Date & MJD & $t_{\mathrm{exp}}$ & $\log_{10} T_s$ & $N_{\mathrm{H}}$ & $\alpha$ & $\chi^2$ &$\log_{10} T_s$ & $\alpha$ & $\chi^2$ & $N_{\mathrm{bins}}$\\
 &&&(ks)& (K) &  ($10^{22}~\mathrm{cm}^{-2}$)&&& (K)
 \\
\hline
6690 		& 2006 Oct 19 			&  54021 	& 62  &	$6.239^{+0.002}_{-0.002} $ & $1.68^{+0.03}_{-0.03}$ & $0.56^{+0.34}_{-0.34}$ & 147 & $6.238^{+0.001}_{-0.001} $ & $0.78^{+0.22}_{-0.32}$ & 148 & 147 \\ 
13783		& 2012 May 5			&  56052 	& 63  &	$6.239^{+0.002}_{-0.002} $ & $1.72^{+0.04}_{-0.04}$ & $0.26^{+0.39}_{-0.26}$ & 130 & $6.236^{+0.001}_{-0.001} $ & $0.67^{+0.33}_{-0.35}$ & 135 & 145 \\ 
16946/17639 & 2015 Apr 28 / May 1 	&  57141.2 	& 111 &	$6.232^{+0.002}_{-0.002} $ & $1.58^{+0.03}_{-0.03}$ & $0.32^{+0.33}_{-0.32}$ & 169 & $6.235^{+0.001}_{-0.001} $ & $0.12^{+0.29}_{-0.12}$ & 175  & 169 \\ 
22426/23248 & 2020 May 11/14		&  58981.1 	& 76  &	$6.224^{+0.002}_{-0.002} $ & $1.54^{+0.04}_{-0.04}$ & $0.44^{+0.53}_{-0.44}$ & 124 & $6.229^{+0.001}_{-0.001} $ & $0.06^{+0.47}_{-0.06}$ & 130  & 125 \\ 
\hline
\end{tabular}
\end{table*}
%--------------------------------------------------

\renewcommand\arraystretch{1.2}
\begin{table*}
\caption{Chandra GRADED mode observations and the results of the simplified spectral fitting described in the text. Column notations and fixed parameters of the fit are similar to those in Table~\ref{tab:T_NH_Xspec_spectral}. The calibration constant $A$ is fixed at $1.081$ for the variable $N_{\mathrm{H}}$ fit, and at $1.075$ for the fit with fixed $N_{\mathrm{H}}$.
Uncertainties correspond to 68 per cent confidence intervals.
}
\label{tab:T_NH_Xspec_spectral_graded}
\begin{tabular}{llcccccccccccc}
\hline
&&&&\multicolumn{4}{c}{$N_\mathrm{H}$ variable} & \multicolumn{3}{c}{$N_{\mathrm{H}}=1.656\times 10^{22}~\mathrm{cm}^{-2}$} \\
 ObsID & Date & MJD & $t_{\mathrm{exp}}$ & $\log_{10} T_s$ & $N_{\mathrm{H}}$ & $\alpha$ &$\chi^2$&$\log_{10} T_s$ & $\alpha$ & $\chi^2$ & $N_{\mathrm{bins}}$\\
 &&&(ks)& (K) &  ($10^{22}~\mathrm{cm}^{-2}$)&&& (K)\\
\hline
114 		 & 2000 Jan 30 		 & 51573.4  & 50 & $6.246^{+0.003}_{-0.003} $ & $1.69^{+0.04}_{-0.04}$ & $0.39^{+0.05}_{-0.05}$ & 137  & $6.244^{+0.002}_{-0.002} $ &  $0.42^{+0.04}_{-0.05}$ & 138 &135 \\
1952 		 & 2002 Feb 6 		 & 	52311.3  &50 & $6.251^{+0.003}_{-0.003} $ & $1.74^{+0.04}_{-0.04}$ & $0.29^{+0.05}_{-0.04}$ & 133  & $6.246^{+0.002}_{-0.002} $ &  $0.34^{+0.05}_{-0.04}$ & 138 & 134 \\
5196 		 & 2004 Feb 8 		 & 	53043.7 &50  & $6.245^{+0.003}_{-0.003} $ & $1.64^{+0.04}_{-0.04}$ & $0.30^{+0.05}_{-0.05}$ & 107  & $6.245^{+0.002}_{-0.002} $ &  $0.31^{+0.04}_{-0.04}$ & 107 & 131\\
9117/9773 	 & 2007 Feb 5/8 	 & 	54439.9 &50  & $6.238^{+0.003}_{-0.003} $ & $1.67^{+0.04}_{-0.04}$ & $0.41^{+0.06}_{-0.06}$ & 131  & $6.237^{+0.002}_{-0.002} $ &  $0.43^{+0.06}_{-0.05}$ & 131 & 125\\
10935/12020  & 2009 Nov 2/3 	 & 	55137.9 &45  & $6.238^{+0.003}_{-0.003} $ & $1.67^{+0.04}_{-0.05}$ & $0.32^{+0.07}_{-0.06}$ & 119  & $6.236^{+0.002}_{-0.002} $ &  $0.35^{+0.06}_{-0.06}$ & 120 & 119\\
10936/13177  & 2010 Oct 31/Nov 2 &  55500.2 &49  & $6.231^{+0.003}_{-0.003} $ & $1.59^{+0.04}_{-0.05}$ & $0.36^{+0.07}_{-0.06}$ & 131  & $6.235^{+0.002}_{-0.002} $ &  $0.33^{+0.06}_{-0.06}$ & 132 & 123 \\
14229		 & 2012 May 15 		 & 56062.4  & 49 & $6.234^{+0.003}_{-0.003} $ & $1.69^{+0.05}_{-0.05}$ & $0.24^{+0.08}_{-0.08}$ & 122  & $6.232^{+0.002}_{-0.002} $ &  $0.28^{+0.07}_{-0.07}$ & 123 & 110\\
14480 		 & 2013 May 20 		 & 56432.6  & 49 & $6.237^{+0.003}_{-0.003} $ & $1.67^{+0.05}_{-0.05}$ & $0.25^{+0.07}_{-0.06}$ & 116  & $6.236^{+0.002}_{-0.002} $ &  $0.28^{+0.06}_{-0.06}$ & 117 & 119 \\
14481		 & 2014 May 12 		 & 56789.1  &49  & $6.237^{+0.003}_{-0.003} $ & $1.73^{+0.05}_{-0.05}$ & $0.14^{+0.07}_{-0.06}$ & 113  & $6.233^{+0.002}_{-0.002} $ &  $0.20^{+0.06}_{-0.06}$ & 116 & 113\\
14482 		 & 2015 Apr 30 		 & 57142.5  &49  & $6.228^{+0.003}_{-0.003} $ & $1.58^{+0.05}_{-0.05}$ & $0.24^{+0.08}_{-0.07}$ & 119  & $6.232^{+0.002}_{-0.002} $ &  $0.21^{+0.07}_{-0.06}$ & 121 & 114\\
19903/18344  & 2016 Oct 20/21 	 & 	57681.2 & 51 & $6.228^{+0.003}_{-0.003} $ & $1.56^{+0.05}_{-0.05}$ & $0.21^{+0.07}_{-0.07}$ &  98  & $6.232^{+0.002}_{-0.002} $ &  $0.17^{+0.07}_{-0.07}$ & 100 & 111\\
19604 		 & 2017 May 16 		 & 57889.7  & 50 & $6.234^{+0.003}_{-0.003} $ & $1.66^{+0.05}_{-0.05}$ & $0.13^{+0.07}_{-0.07}$ & 108  & $6.234^{+0.002}_{-0.002} $ &  $0.15^{+0.07}_{-0.06}$ & 108 & 110\\
19605 		 & 2018 May 15 		 & 58253.7  & 49 & $6.229^{+0.003}_{-0.003} $ & $1.55^{+0.05}_{-0.05}$ & $0.15^{+0.08}_{-0.07}$ &  91  & $6.234^{+0.002}_{-0.002} $ &  $0.11^{+0.07}_{-0.07}$ &  95 & 107\\
19606 		 & 2019 May 13 		 & 58616.5  & 49 & $6.230^{+0.003}_{-0.003} $ & $1.63^{+0.06}_{-0.05}$ & $0.18^{+0.08}_{-0.08}$ &  78  & $6.231^{+0.002}_{-0.002} $ &  $0.18^{+0.08}_{-0.08}$ &  78 & 102\\
\hline
\end{tabular}
\end{table*}
%--------------------------------------------------

We use all \textit{Chandra} ACIS-S 
observations of CasA~NS, both in FAINT and GRADED modes. 
In some cases the observations were taken only a few days apart. Following previous works \citep{Ho21,PosseltPavlov2022ApJ}, we merged such observations into a single observation epoch.
Respective ObsIDs, dates and exposure times for FAINT and GRADED mode observations are given in Table~\ref{tab:T_NH_Xspec_spectral} and Table~\ref{tab:T_NH_Xspec_spectral_graded}, respectively.
The FAINT mode data contain 6 observations at 4 observation epochs 
spanning 14 years from 2006 Oct 19 to 2020 May 14 \citep{PosseltPavlov2022ApJ}, while the GRADED mode data contain 18 observations at 14 observation epochs from 2000 Jan 30 to 2019 May 13.
Thus in total we have 18 observation epochs over 20 years.

To get a uniform dataset, we reprocessed all spectra (including the GRADED mode ones analysed in Paper I) with  \textsc{ciao} 4.14 using  \texttt{CALDB} 4.9.8, 
and then binned to ensure a minimum of 25 counts per energy bin. Accordingly, we use $\chi^2$ statistics as the likelihood for our data.\footnote{We checked that the use of the $C$-statistic \citep{Cash1979ApJ}, with data binned by a minimum of 1 count per energy bin, gives similar results, see Appendix~\ref{app:spectral} for details}. The fits were performed using the X-ray spectral fitting package \textsc{xspec} v 12.11.1 \citep{XSPEC1996}. The $0.5-7.0$~keV spectral interval was used for fitting. The spectral model is the same as in Paper I  (see also \citealt{Heinke2010ApJ,Elshamouty2013ApJ, Wijngaarden2019MNRAS, Ho21}). 
It contains a thermal component modelled by the non-magnetized carbon atmosphere model (\textsc{nsx} in \textsc{xspec}, \citealt{Ho2009Natur}) which has these parameters:  (non-redshifted) surface temperature $T_{s}$, NS mass $M$, radius $R$, the distance $d$ to the star and normalisation which is set to one suggesting that the emission is coming from the entire stellar surface.
This model is supported by the non-detection of pulsations (pulsed fraction upper limit is less than 10 per cent for periods $>10$~ms) from CasA NS \citep{Murray2002ApJ,Ransom2002ASPC,Pavlov2009ApJ,Halpern2010ApJ}. 
Neutron star mass, radius and distance clearly are the same for all observations. The thermal spectrum emerging from the NS atmosphere is distorted by the  interstellar absorption which is accounted for by the  \textsc{tbabs} model \citep{Wilms2000ApJ} and parametrized by the effective hydrogen column density $N_\mathrm{H}$. Correspondingly,  the {\it wilm} abundance set for the photoelectric absorption model is used \citep{Wilms2000ApJ}.  We also include the dust scattering model \textsc{spexpcut} \citep{Predehl2003AN} as the dust scattering is not accounted for in the \textsc{tbabs} model. In principle, inclusion of the \textsc{spexpcut} model only leads to the renormalization of the effective hydrogen column density $N_\mathrm{H}$ \citep{Posselt2018ApJ} so it can be omitted. It is included for consistency with the previous studies. 
Finally we accounted for possible pile-up employing the \textsc{pileup} model \cite{Davis2001ApJ}, in which the grade migration parameter $\alpha$ was allowed to vary independently for each epoch \citep[see][for details]{Shternin2021MNRAS,Ho21}.
For completeness, the \textsc{pileup} component is included both for the GRADED and FAINT mode data, since it was found that  even a small pileup probability can lead to a subtle, but noticeable effects \citep[e.g.,][]{Bogdanov2016ApJ}, see Appendix~\ref{app:spectral} for more details.

The surface temperature $T_s$ is allowed to vary between observations. In principle, it is possible that  the hydrogen column density $N_{\mathrm{H}}$ can vary between observations as well. Therefore we traditionally investigate here both the varying and fixed $N_{\mathrm{H}}$ possibilities. The surface temperature correlates with other spectral parameters (i.e., $M$, $R$, $d$) and  when the latter are free to vary,  the uncertainties on the individual temperature measurements are too large for the temperature evolution to be seen. However, the temperature evolution is actually significant and can be readily observed when $M$, $R$ and $d$ are fixed at certain values. This approach is often used in the literature \citep[e.g.,][and references therein]{Ho21,PosseltPavlov2022ApJ} and we follow it for illustrative purposes in  Section~\ref{S:obs:fixMR}. However such an approach is simplified, in the sense that it does not allow accounting for the correlations of the temperature decline with other parameters of the model. Therefore, following Paper I, in Section~\ref{sec:MCMC} we implement the temperature evolution law at the level of the spectral fitting and perform a joint analysis of all spectra within the Bayesian framework.

\subsection{Fit with fixed NS parameters}\label{S:obs:fixMR}
For illustration, 
we first fix $d=3.33$~kpc 
(\citealt*{Alarie2014MNRAS}; \citealt{Reed1995ApJ}),
$R=13.5$~km and $M=1.53\ M_\odot$ ($R=13.7$~km and $M=1.60\ M_\odot$) for the models where $N_{\mathrm{H}}$ is allowed to vary ($N_{\mathrm{H}}$ is fixed) between the  observations. 
The adopted values for $M$, $R$ and $d$ correspond to the best fits (maximal a posteriori estimates) for the more detailed multiparametric model described in Section ~\ref{sec:MCMC}.

%%%%%%%%%%%%%%%%%%%%%%%%%%%%%%%%%%%%%%%%%%%%%%%%%%%%%%%%%%%%%%%%%%%%%%%%%%%%%%%%%%%%%%%%%%%%%%%%%%%%
\begin{figure*}
    \centering
    \includegraphics[width=0.45\textwidth]{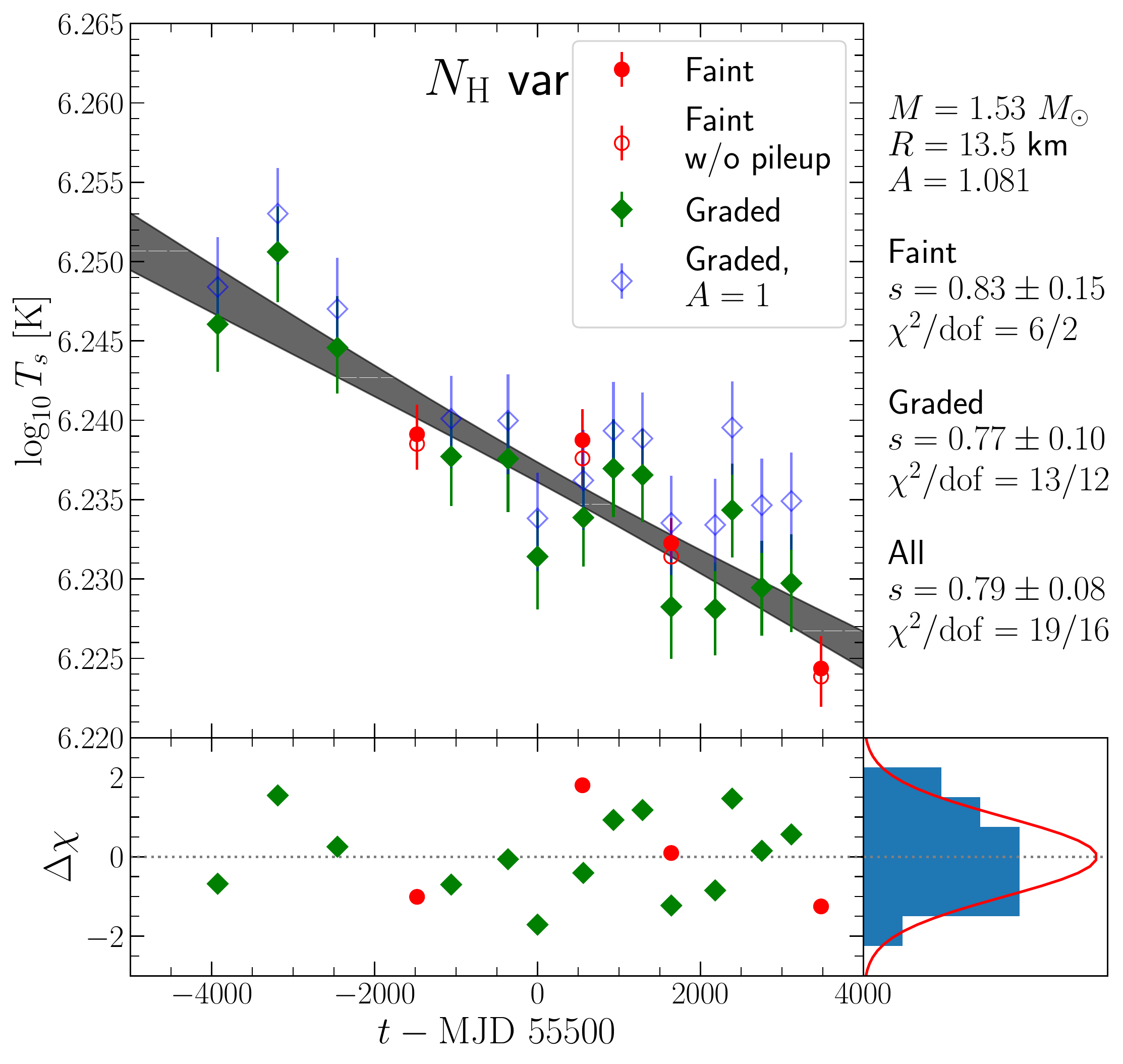}
    \includegraphics[width=0.45\textwidth]{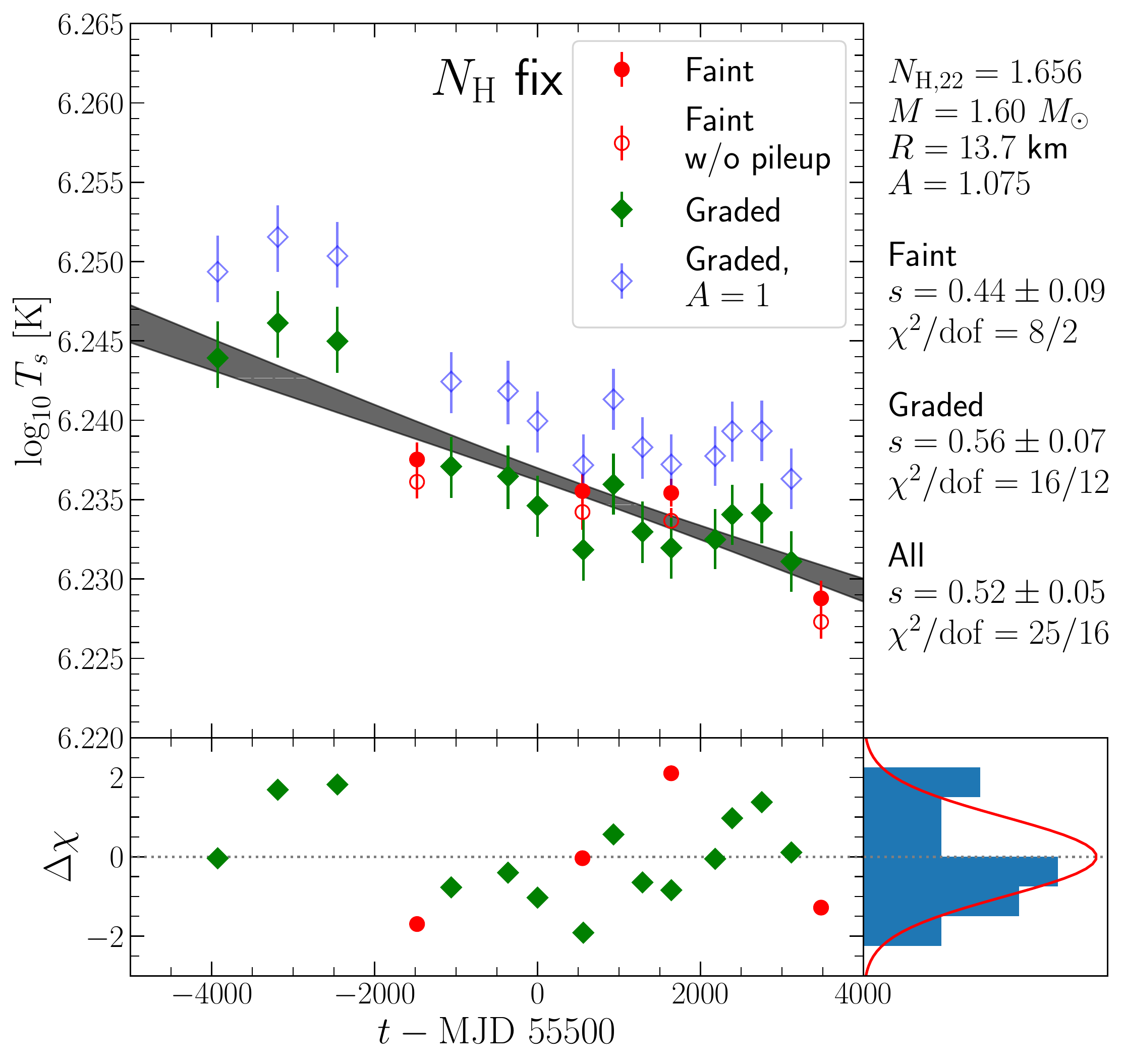}
    \caption{Surface temperature evolution for modes with fixed NS parameters and
    variable $N_{\mathrm{H}}$ (left) or fixed $N_{\mathrm{H}}$ (right). The employed values of the NS parameters are given in the top right corner in each panel. There, $N_{\mathrm{H},22}\equiv N_{\mathrm{H}}/(10^{22}~\mathrm{cm}^{-2})$. Horizontal axes show time in modified Julian days with respect to MJD 55500. Filled dots, open dots, filled diamonds and open diamonds correspond to the FAINT mode data, FAINT mode data without pile-up, GRADED mode data and GRADED mode data fitted without introducing the calibration factor $A$, respectively. Filled strips show 68 per cent prediction intervals for the regression model obtained by fitting equation~(\ref{eq:T-t_linear}) to all data (filled dots and diamonds). 
    Regression slope $s$ estimates and best-fit $\chi^2$ values are indicated at each panel for regression using all data and the FAINT and GRADED data points alone.
    Lower panels show the standardised residuals around the best-fit regression curve for all data. To guide the eye, the distribution of residuals is compared to the standard normal distribution shown by the red solid curve. }\label{fig:slopes}
\end{figure*}
%%%%%%%%%%%%%%%%%%%%%%%%%%%%%%%%%%%%%%%%%%%%%%%%%%%%%%%%%%%%%%%%%%%%%%%%%%%%%%%%%%%%%%%%%%%%%%%%%%%%

%%%%%%%%%%%%%%%%%%%%%%%%%%%%%%%%%%%%%%%%%%%%%%%%%%%%%%%%%%%%%%%%%%%%%%%%%%%%%%%%%%%%%%%%%%%%%%%%%%%%
\begin{figure}
    \centering
    \includegraphics[width=\columnwidth]{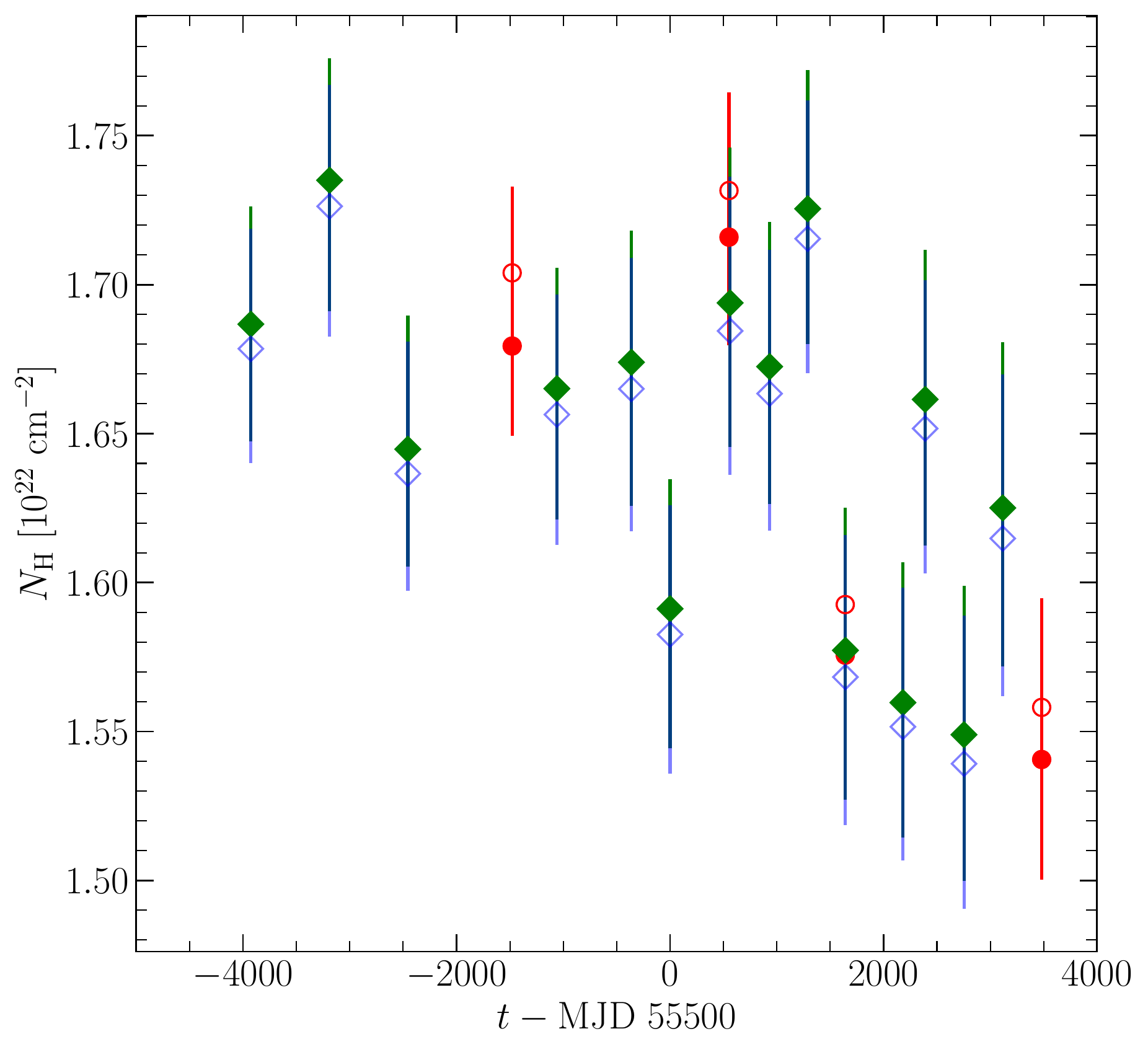}
\caption{Effective hydrogen column densities for different observations obtained from the simplified spectral fits.  Horizontal axis shows time in modified Julian days with respect to MJD 55500. As in Fig.~\ref{fig:slopes}, filled dots, open dots, filled diamonds and open diamonds correspond to the FAINT mode data, FAINT mode data without pile-up, GRADED mode data and GRADED mode data fitted without introducing the calibration factor $A$, respectively.}\label{fig:NH}
\end{figure}
%%%%%%%%%%%%%%%%%%%%%%%%%%%%%%%%%%%%%%%%%%%%%%%%%%%%%%%%%%%%%%%%%%%%%%%%%%%%%%%%%%%%%%%%%%%%%%%%%%%%

 The fit results are presented in Tables~\ref{tab:T_NH_Xspec_spectral}, 
 \ref{tab:T_NH_Xspec_spectral_graded} and \ref{tab:T_NH_Xspec_spectral_graded_nocal} and illustrated in Figs.~\ref{fig:slopes} and \ref{fig:NH}. 
 Temperature evolution for models with variable $N_{\mathrm{H}}$ and fixed $N_{\mathrm{H}}$ are shown in the left and right panels of  Fig.~\ref{fig:slopes}, respectively, while the $N_{\mathrm{H}}$ evolution for the $N_{\mathrm{H}}$-variable models is shown in Fig.~\ref{fig:NH}. FAINT mode results are shown with red filled dots in Figs.~\ref{fig:slopes} and \ref{fig:NH}, while the GRADED mode results are shown with blue open diamonds. The results of the FAINT mode data fitting without pileup component are shown with red open dots.
Temperature decline is quantified by  fitting  $T_s(t)$ data with the linear regression law (in log scale)
\begin{equation}\label{eq:T-t_linear}
\log T_s(t)=\log T_{s0} - s \log{t/t_0},
\end{equation}
where $t$ is the NS age, $t_0=330$~yr corresponds to  MJD=55500 (Oct. 31, 2010), $T_{s0}\equiv T_s(t_0)$ and $s$ is the cooling slope. The fit results are indicated in Fig.~\ref{fig:slopes}. Clearly, the GRADED and FAINT mode data show compatible decline rates, and the temperature decline is somewhat larger in the case when $N_{\mathrm{H}}$ is allowed to vary as compared to the $N_{\mathrm{H}}$-fixed fits, in accordance with previous studies. The individual  $N_{\mathrm{H}}$ values inferred for the two modes 
are also similar, as shown in Fig.~\ref{fig:NH}. This similarity suggests that a joint fit for all ACIS-S spectral data can and should be performed. On the other hand, as seen in Fig.~\ref{fig:slopes}, inferred temperatures for the GRADED mode are systematically larger than those for the FAINT mode \citep[see also][]{Heinke2010ApJ}. A possible reason for this discrepancy could be, for instance, imperfect modelling of the pileup 
or other unknown, uncalibrated effects. Notice that inclusion of the pileup model component for the FAINT mode data reduces the difference, but only slightly, i.e. compare red filled and open dots in Fig.~\ref{fig:slopes} (see Appendix~\ref{app:spectral} for a more detailed comparison). 
In order to account for this discrepancy in the  simplest way, we introduce an additional calibration constant $A$ into a spectral model for the GRADED mode data.
Clearly, this constant can be determined only in the joint fit where both GRADED and FAINT data are linked by the evolution law (\ref{eq:T-t_linear}), and this is done in the next Sec.~\ref{sec:MCMC}. 

In the present Section we treat the calibration constant in the same way as the other joint parameters, and fix it at the best-fit values $A=1.081$ ($A=1.075$) for $N_{\mathrm{H}}$-variable (-fixed) models, and refit the GRADED data. The resulting temperatures and $N_{\mathrm{H}}$ are shown, respectively, in Figs.~\ref{fig:slopes} and \ref{fig:NH} with green filled diamonds, and given in Table~\ref{tab:T_NH_Xspec_spectral_graded}. For brevity, a similar Table~\ref{tab:T_NH_Xspec_spectral_graded_nocal} but for $A=1$ (corresponding to the open diamonds in Fig.~\ref{fig:slopes}) is placed in the Appendix~\ref{app:spectral}. 
Introduction of the calibration factor shifts the GRADED mode temperatures towards the FAINT mode ones. With the filled regions in Fig.~\ref{fig:slopes} we show the 68 per cent prediction intervals of the regression curve calculated from all data, while the lower panels in Fig.~\ref{fig:slopes} show the standardised residuals ($\Delta\chi=(data-model)/error$).
 The regression for the variable $N_{\mathrm{H}}$ is formally slightly preferable over the  
 $N_{\mathrm{H}}$-fixed fit, although the direct comparison of the $\chi^2$ values for the temperature regression curves is somewhat misleading, since one should add the $\chi^2$ values obtained from the spectral fitting and the regression models to obtain  the total log-likelihood for our data. 
 
The inferred cooling slopes based on all FAINT$+$GRADED data are $s=0.79\pm0.08$ if $N_\mathrm{H}$ is allowed to vary and $s=0.52\pm0.05$ if $N_\mathrm{H}$ is fixed, see Fig.~\ref{fig:slopes}. This translates to the relative temperature decline of $2.40\pm 0.24$ per cent and $1.57\pm 0.15$ per cent in 10 years, respectively, 
consistent with the FAINT mode results of  \citet{PosseltPavlov2022ApJ} and the GRADED mode results of Paper I. 
Notice, that, if necessary (for instance, for direct comparison with theoretical cooling curves), the data points given in Tables~\ref{tab:T_NH_Xspec_spectral}, \ref{tab:T_NH_Xspec_spectral_graded} and Fig.~\ref{fig:slopes} can be rescaled to different values of $R$.%, in the first approximation. 
This can be done, in the first approximation, by using $T_s^4R^2=\mathrm{const}$ law.

According to Fig.~\ref{fig:NH}, neither introduction of the calibration factor $A$ nor inclusion of the pileup component for the FAINT mode data changes $N_{\mathrm{H}}$ considerably. Overall, on the over hand, Fig.~\ref{fig:NH} may indicate some $N_{\mathrm{H}}$ evolution
due to an astrophysical origin or 
yet-uncalibrated instrumental effects related to the ACIS-S  contamination \citep{Plucinsky2020SPIE}.

\subsection{Fit within the Bayesian framework}\label{sec:MCMC}
This simplified analysis does not allow us to fully explore the parameter space and infer the constraints on the parameters of the NS core superfluidity. Following Paper I, we now set up the Bayesian framework and fit all spectra simultaneously employing the relation given by equation~(\ref{eq:T-t_linear}) in the spectral model. This allows us to infer the cooling slope and its correlations with other spectral parameters without relying on the individual data points. Additionally, this makes it possible to perform a joint FAINT+GRADED fit as described above. For fitting we use the affine-invariant Markov Chain Monte Carlo (MCMC) sampler \textsc{emcee} \citep{Foreman-Mackey2013} which is connected to \textsc{xspec} via the Python wrapper \textsc{pyxspec}. 
The priors on the spectral parameters for our Monte Carlo runs were similar to Paper I, except that the distance prior is taken to be  Gaussian with the mean $3.33$~kpc and the standard deviation $0.1$~kpc \citep{Alarie2014MNRAS} instead of a broader prior based on the earlier results of \citet{Reed1995ApJ}. For other parameters, we employed uniform priors in the ranges $-5<s<5$, $5.89 <\log_{10} T_{s0}/(1~\mathrm{K})< 6.6 $, $0.5 M_\odot < M< 3.0 M_\odot$, and  $1~\mathrm{km}<R<30~\mathrm{km}$. We also do not allow acausal models (with $M/M_\odot>0.24\, R/(1\,\text{km})$, e.g., \citealt*{Lattimer2016PhysRep})
and  parameter sets with surface gravity outside the range available for the \textsc{nsx} model\footnote{\url{https://www.slac.stanford.edu/~wynnho/nsx_models.dat}.}. 
For the fixed-$N_{\mathrm{H}}$ model we employ a broad uniform prior for a common column density $N_{\mathrm{H}0}$, $10^{21}~\mathrm{cm}^{-2}<N_{\mathrm{H}0}<3\times10^{22}~\mathrm{cm}^{-2}$. 
For the variable-$N_{\mathrm{H}}$ model we 
use the hierarchical prior set. That is, we
assume that the column densities $N_{\mathrm{H}i}$, $i=1\dots N_{\mathrm{obs}}$, 
have the normal prior distributions with same mean $N_{\mathrm{H}0}$ and variance $\sigma_{N_{\mathrm{H}}}^2$, where $N_{\mathrm{H}0}$ and $\sigma_{N_{\mathrm{H}}}^2$
are 
additional
 model
parameters (so-called hyperparameters, see, e.g., \citealt{GelmanBook}).
For the latter,
we assume a noninformative prior distribution  $\sigma_{N_{\mathrm{H}}}^2>0$ on the variance, while a uniform prior distribution on $N_{\mathrm{H}0}$ is the same as for the model with fixed $N_{\mathrm{H}}$.\footnote{Such hierarchical approach allows us to estimate the level of variability in  $N_{\mathrm{H}}$ over the time, at the same time censoring the possible outliers.} 
Individual  grade migration parameters $\alpha_i$ in the pileup model had uniform priors $0<\alpha_i<1$ and the weak uniform prior was set for the calibration constant $A$ in case of the joint FAINT+GRADED fit.

%%%%%%%%%%%%%%%%%%%%%%%%%%%%%%%%%%%%%%%%%%%%%%%%%%%%%%%%%%%%%%%%%%%%%%%%%%%%%%%%%%%%%%%%%%%%%%%%%%%%%%%%
\renewcommand\arraystretch{1.2}
\begin{table*}
    \centering
    \caption{Results of the spectral fit. Uncertainties correspond to the 68 per cent highest posterior density credible intervals.}
    \label{tab:spectral_params}
    \begin{tabular}{llcccccccccc}
        \hline
       Mode &  $N_{\mathrm{H}}$ &$\log_{10} T_{s0}$ & $s$ & $M$ & $R$ & $d$ & $N_{\mathrm{H}0}$ & $\sigma_{N_{\mathrm{H}}}$& $A$ &$\chi^2/(\mathrm{d.o.f.})$ \\
              & &(K)&&$(M_\odot)$& (km) &(kpc)&($10^{22}$~cm$^{-2}$)& ($10^{20}$~cm$^{-2}$)&&\\
        \hline
        All & Var & $6.23^{+0.02}_{-0.02}$ & $0.66^{+0.09}_{-0.07}$ & $1.53^{+ 0.14}_{-0.14}$ & $13.5^{+1.2}_{-1.3}$ & $3.35^{+0.09}_{-0.11}$ & $1.642^{+0.035}_{-0.034}$ &  $3.7^{+1.4}_{-1.1}$ &$1.081^{+0.018}_{-0.015}$& 2171/(2211)\\
        All & Fix & $6.24^{+0.03}_{-0.02}$ & $0.53^{+0.07}_{-0.05}$ & $1.60^{+0.17}_{-0.12}$ & $13.7^{+1.1}_{-1.8}$ & $3.33^{+0.11}_{-0.09}$ & $1.656^{+0.044}_{-0.026}$  & -- & $1.075^{+0.013}_{-0.010}$  & 2220/(2228)\\
        FAINT & Var&
      $6.26^{+0.09}_{-0.04}$ & $0.81^{+0.15}_{-0.14}$ & $1.87^{+0.16}_{-0.24}$ & $10.7^{+2.5}_{-2.4}$ & $3.33^{+0.11}_{-0.09}$ & $1.720^{+0.120}_{-0.150}$ &  $2.8^{+50.0}_{-2.8}$ &--&$566/(577)$ \\ 
        FAINT & Fix& $6.40^{+0.02}_{-0.10}$ & $0.51^{+0.09}_{-0.09}$ & $1.93^{+0.15}_{-0.14}$ & $8.4^{+3.0}_{-0.6}$ & $3.34^{+0.09}_{-0.11}$ & $1.764^{+0.039}_{-0.050}$ & -- & -- & $586/(580)$ \\
        GRADED & Var &$6.21^{+0.05}_{-0.02}$ & $0.62^{+0.12}_{-0.08}$ & $1.53^{+0.11}_{-0.22}$ & $14.8^{+1.6}_{-3.2}$ & $3.34^{+0.09}_{-0.11}$ & $1.617^{+0.046}_{-0.038}$ & $3.4^{+1.6}_{-1.4}$ & --& 1597/(1620)\\
        GRADED & Fix &  $6.21^{+0.05}_{-0.02}$ & $0.56^{+0.08}_{-0.08}$ & $1.53^{+0.15}_{-0.17}$ & $14.7^{+3.0}_{-2.3}$ & $3.34^{+0.09}_{-0.10}$ & $1.624^{+0.046}_{-0.033}$ & -- &--& 1624/(1633)\\
        \hline
    \end{tabular}
\end{table*}
%%%%%%%%%%%%%%%%%%%%%%%%%%%%%%%%%%%%%%%%%%%%%%%%%%%%%%%%%%%%%%%%%%%%%%%%%%%%%%%%%%%%%%%%%%%%%%%%%%%%%%%%%%%

%%%%%%%%%%%%%%%%%%%%%%%%%%%%%%%%%%%%%%%%%%%%%%%%%%%%%%%%%%%%%%%%%%%%%%%%%%%%%%%%%%%%%%%%%%%%%%%%%%%%
\begin{figure*}
    \centering
    \includegraphics[width=0.45\textwidth]{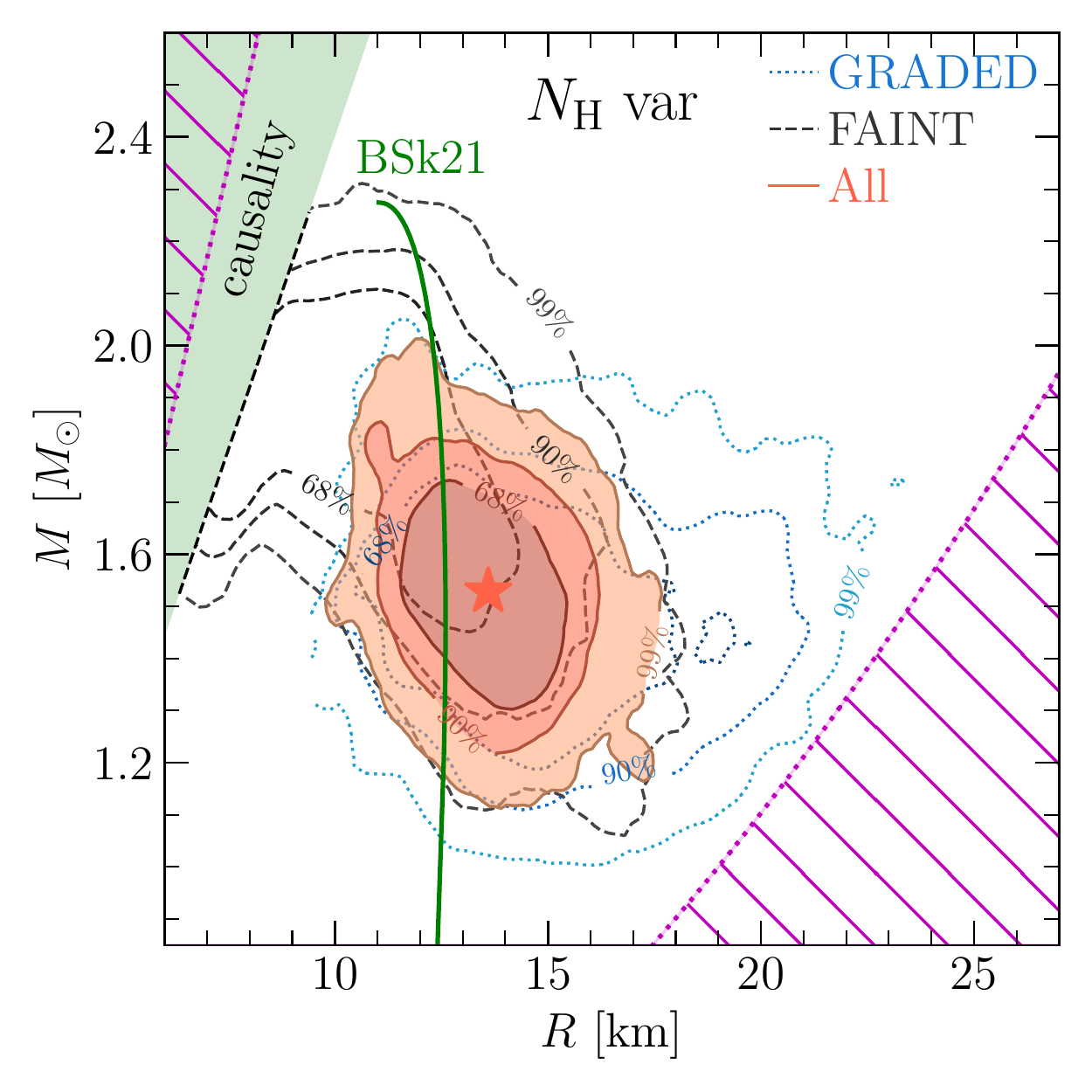}
    \includegraphics[width=0.45\textwidth]{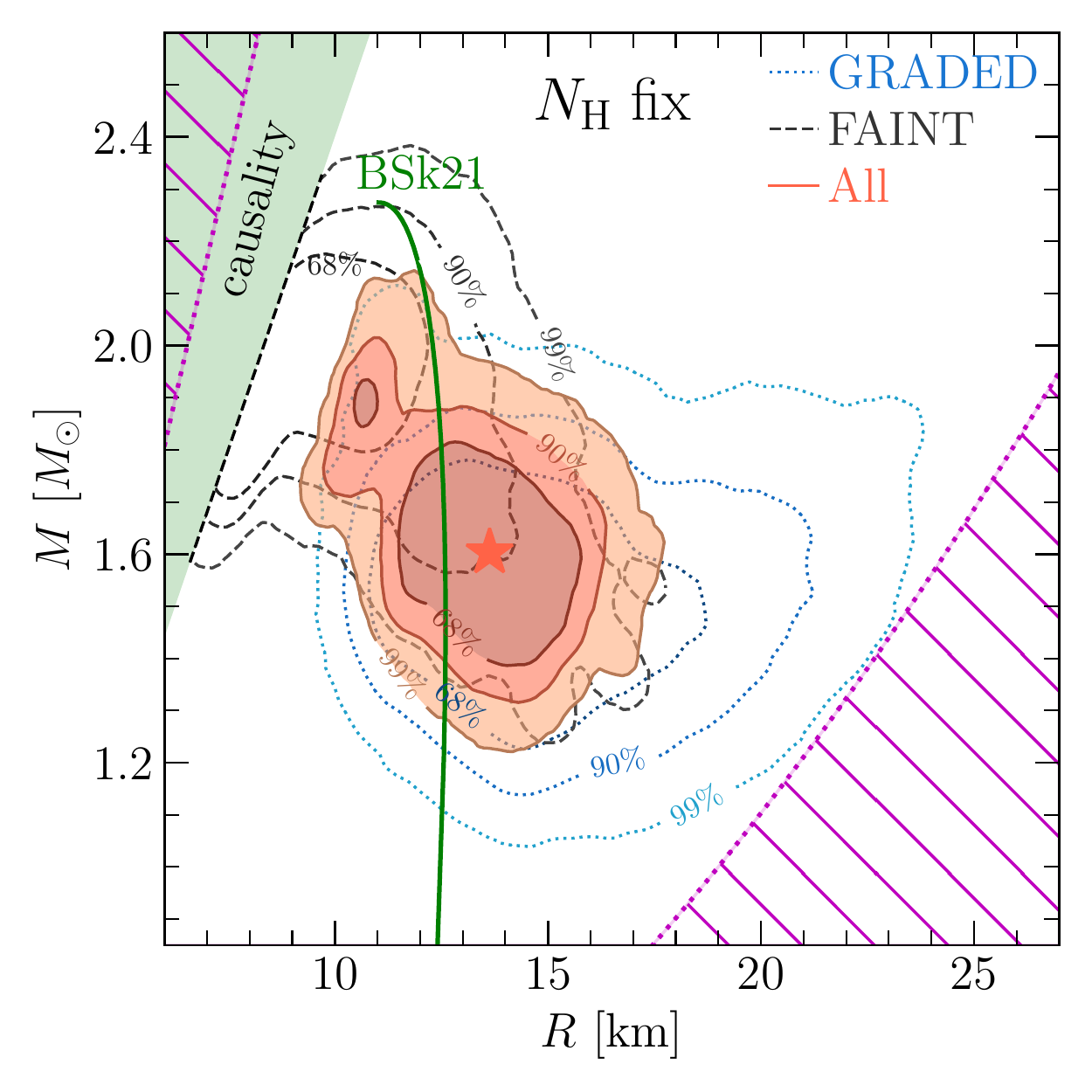}
        \caption{Mass-radius credible contours obtained from spectral fits with variable $N_{\mathrm{H}}$ (left) and fixed $N_{\mathrm{H}}$ (right). Dashed and dotted contours correspond, respectively,  to the FAINT mode and GRADED mode data analysed alone. Filled solid contours correspond to the joint fit. Contours are labelled with their 68, 90  and 99 per cent credibility levels.  Red stars show the point estimates from Table~\ref{tab:spectral_params} used for the simplified fit in Sec.~\ref{S:obs:fixMR}.
        The thick solid line indicates the $M-R$ relation for the BSk21 EOS. Green filled regions show the causality restrictions, while hatched areas show regions outside the available \textsc{nsx} model parameterization of surface gravity.
    }\label{fig:MR}
\end{figure*}
%%%%%%%%%%%%%%%%%%%%%%%%%%%%%%%%%%%%%%%%%%%%%%%%%%%%%%%%%%%%%%%%%%%%%%%%%%%%%%%%%%%%%%%%%%%%%%%%%%%%

The inferences on the model parameters from MCMC runs are summarised in Table~\ref{tab:spectral_params} and also in Table~\ref{tab:spectral_nuisance}. The point estimates in Table~\ref{tab:spectral_params}  correspond to modes of 1D marginalised posteriors. They are slightly different but consistent with the linear regression analysis. The uncertainties given in Table~\ref{tab:spectral_params} correspond to the 68 per cent highest posterior density credible intervals. For completeness we present here the results from the joint fit (mode designation `All') as well as the results for the FAINT and GRADED data alone. The GRADED mode results are consistent with the results of Paper I, slight differences are due to the different distance prior used and the updated CALDB. 
Overall, the FAINT and GRADED mode results are different, but consistent within uncertainties (Table~\ref{tab:spectral_params}). As a consequence, the joint fit results in similar spectral parameter inferences, but with smaller uncertainties for most of the parameters. The calibration constant $A$ introduced in the joint fit for the GRADED mode is well-constrained at a reasonably small value $A\approx 1.1$ (Table~\ref{tab:spectral_params}). 
The overall goodness of the fit is illustrated by the $\chi^2$ values\footnote{Here the $\chi^2$ value is calculated with respect to the mean posterior prediction values for the data points, see Appendix~\ref{app:spectral} for details.} in Table~\ref{tab:spectral_params}, and the quality of the fitting to individual spectra is illustrates in Appendix~\ref{app:spectral}.

Let us examine first the inferred masses and radii. 
The mass-radius credible contours (68 per cent, 90 per cent and 99 per cent credibility levels) are shown in  Fig.~\ref{fig:MR} for the variable $N_{\mathrm{H}}$ model (left panel) and the fixed $N_{\mathrm{H}}$ model (right panel). The contours are compared for GRADED (dotted lines), FAINT (dashed lines) and All (filled contours) modes. To guide the eye, we also show in Fig.~\ref{fig:MR} the theoretical $M-R$ relation for the particular equation of state (EOS) of the neutron star matter, namely the BSk21 model \citep{Potekhin2013A&A}. We also show the prior restrictions on $M$ and $R$ due to causality and availability of \textsc{nsx} model parameterization by green filled regions and hatched regions, respectively. The results for variable and fixed  $N_{\mathrm{H}}$ are consistent within each mode, and directly plotting them on the same figure would make it unreadable. 
The FAINT mode data constraints on $M$ and $R$ are consistent with those for the GRADED mode, although the latter constraints show a considerably larger extent towards higher values of $R$. This can be attributed to complications induced by the pileup model components required for the GRADED mode data analysis, see Paper I. The FAINT mode results are slightly affected by the causality restrictions. Would these restrictions be lifted, slightly more compact solutions will be possible according to atmospheric model. The $M-R$ contours for the joint fit basically follow the intersection of GRADED and FAINT modes contours. As seen in Fig.~\ref{fig:MR} and Table~\ref{tab:spectral_params}, the joint fit allows us to constrain the CasA~NS radius to about 10 per cent accuracy.

%%%%%%%%%%%%%%%%%%%%%%%%%%%%%%%%%%%%%%%%%%%%%%%%%%%%%%%%%%%%%%%%%%%%%%%%%%%%%%%%%%%%%%%%%%%%%%%%%%%%
\begin{figure*}
    \centering
    \includegraphics[width=0.8\textwidth]{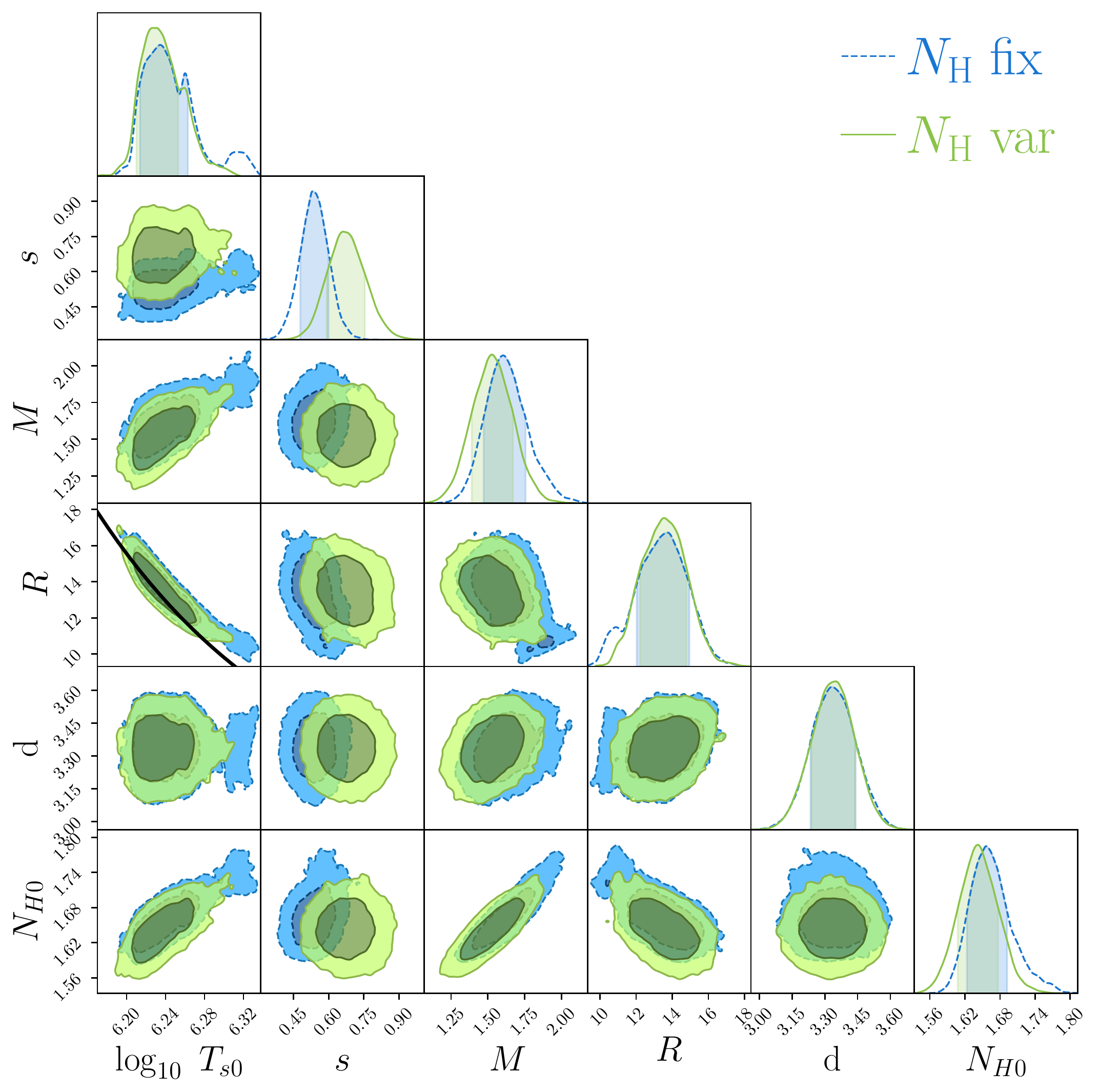}
    \caption{ 1D and 2D posterior distributions for the spectral parameters for the joint fit using data in all modes. Shaded areas on 1D distributions correspond to 68 per cent credible intervals, while contours on 2D distributions correspond to 68 per cent and 90 per cent levels. Solid and dashed contours correspond to variable $N_{\mathrm{H}}$ and fixed $N_{\mathrm{H}}$ models respectively. The thick black line shows the $T_{s0}^4R^2=\mathrm{const}$ relation, see text for details. }\label{fig:tri_spectral}
\end{figure*}
%%%%%%%%%%%%%%%%%%%%%%%%%%%%%%%%%%%%%%%%%%%%%%%%%%%%%%%%%%%%%%%%%%%%%%%%%%%%%%%%%%%%%%%%%%%%%%%%%%%%

The 2D and 1D marginalised posterior densities for spectral parameters of the joint fit are shown in Fig.~\ref{fig:tri_spectral}. 
One observes that variable-$N_\mathrm{H}$ and fixed-$N_\mathrm{H}$ models give similar results for all spectral parameters except for the cooling slope $s$. The latter is larger (but also consistent) for the variable $N_\mathrm{H}$ model than for the fixed $N_\mathrm{H}$ model. According to Table~\ref{tab:spectral_params}, this is mainly due to the contribution of the FAINT mode data.
The cooling slope values for the joint fit in Table~\ref{tab:spectral_params} correspond to the surface temperature decline of $2.2\pm 0.3$ per cent in 10 years for variable-$N_{\mathrm{H}}$ model and to $1.6\pm 0.2$ for fixed-$N_{\mathrm{H}}$ model.
Notice that the spectral slope $s$ does not correlate significantly with other spectral parameters.
The strongest correlation in Fig.~\ref{fig:tri_spectral} is observed between $T_{s0}$ and $R$. Clearly, the main reason for this correlation is that the total flux is actually proportional to $T_{s0}^4R^2$ combination. This is illustrated by the thick black line plotted in the $T_{s0}-R$ panel in Fig.~\ref{fig:tri_spectral}, which shows the relation $T_{s0}^4R^2=\mathrm{const}$, where the value of const is fixed by the point estimates for the variable-$N_\mathrm{H}$ model\footnote{We are grateful to L.~B. Leinson for the suggestion to explicitly emphasise this dependence.}. This supports the rough rescaling recipe given in the end of Sec.~\ref{S:obs:fixMR} for individual temperature data points.

We will further use the results of the joint fit as the base model for the analysis of the superfluid parameters. The comparison between the  marginalised posterior densities for spectral parameters for FAINT, GRADED and joint fits is given in Fig.~\ref{fig:tri_spectral_comp} in Appendix~\ref{app:spectral}. Inferences on the individual grade migration parameters $\alpha_i$ and column densities $N_{\mathrm{H},i}$ are presented in Table~\ref{tab:spectral_nuisance}.

%%%%%%%%%%%%%%%%%%%%%%%%%%%%%%%%%%%%%%%%%%%%%%%%%%%%%%%%%%%%
\section{Constraining superfluid parameters}\label{Sec:SFanalysis}
%%%%%%%%%%%%%%%%%%%%%%%%%%%%%%%%%%%%%%%%%%%%%%%%%%%%%%%%%%%%
The procedure for inferring constraints on the NS superfluidity parameters from the CasA NS data is described in detail in Paper I.  Here we only present a short summary.

According to the standard NS cooling theory (e.g., \citealt{Nomoto1981ApJ}, \citealt{YakovlevPethick2004ARA}, \citealt{Page2009ApJ}), the CasA NS is in the neutrino cooling stage and has an isothermal interior. The latter means that the condition $\widetilde{T}=\mathrm{const}$, where $\widetilde{T}$ is the local temperature redshifted for a distant observer \citep{Thorne1966}, holds almost everywhere inside the star except for the thin outer heat blanketing envelope. 
Since CasA~NS is assumed to be a weakly magnetized and slowly rotating star, 
the NS cooling is described by a simple heat balance equation
\begin{equation}\label{eq:ell_def}
\frac{\mathrm{d} \widetilde{T}}{\mathrm{d} t}  = - \frac{L^\infty_\nu(\widetilde{T})}{C(\widetilde{T})}\equiv - \ell(\widetilde{T}),
\end{equation}
where $L^\infty_\nu(\widetilde{T})$ is the (redshifted) integrated neutrino luminosity, $C(\widetilde{T})$ is the integrated heat capacity, and their ratio defines the so-called neutrino cooling function $\ell(\widetilde{T})$. 

The relation $T_s(\widetilde{T})$ between the surface temperature and the internal temperature of the star depends on the properties of the heat blanket envelope, noticeably on its composition and magnetic field,  \citep*[see, e.g,][for a review]{Beznogov2021PhR}. Following  Paper I, here we adopt the $T_s(\widetilde{T})$ relations given by \citet*{Potekhin1997A&A} and \citet*{Beznogov2016MNRAS} for an insignificant amount of light elements (carbon) above the iron envelope, see Paper I and \citet{Shternin2015MNRAS} for discussion. In this case, the heat-blanketing relation for relevant temperatures can be approximated by a power law $T_s\propto \widetilde{T}^\beta$ with $\beta\approx 0.53$.
The slope of the cooling curve $T_s(t)$ is then [see equation~(\ref{eq:T-t_linear})]
\begin{equation}\label{eq:slope def}
s=-\frac{\mathrm{d} \ln T_s}{\mathrm{d} \ln t}\approx -\beta \frac{\mathrm{d} \ln \widetilde{T}}{\mathrm{d} \ln t}=\frac{\beta t}{\widetilde{T}}\ell(\widetilde{T}).
\end{equation}

Before the onset of the triplet neutron pairing in the core, the NS cools due to standard slow cooling processes (e.g., \citealt{YakovlevPethick2004ARA}; \citealt{Yakovlev2001physrep}; \citealt*{Potekhin2015SSRv}).  Let us denote the neutrino cooling function at this initial stage by $\ell_0(\widetilde{T})$.
The slow cooling processes mainly include neutron-neutron bremsstrahlung, and, in the regions of the core where the proton pairing is absent, the modified Urca and nucleon-proton bremsstrahlung.  The neutrino cooling function for these 
processes obeys $\ell(\widetilde{T})=\ell_0(\widetilde{T})\propto \widetilde{T}^n$ with $n=7$. Then $s=\beta/(n-1)\approx 0.09$. Therefore the CasA NS cannot be on the slow cooling stage today \citep{Heinke2010ApJ}.

Eventually, the NS cools down to the point where $\widetilde{T}=\widetilde{T}_{Cn\mathrm{max}}$.  Recall that $\widetilde{T}_{Cn\mathrm{max}}$ is the maximal redshifted critical temperature of the triplet neutron pairing throughout the core. Then the formation of the Cooper pairs (CPF) results in the additional process of neutrino emission which quickly becomes the dominant cooling agent (the neutron-involving electroweak bremsstrahlung processes become suppressed by superfluid effects, e.g., \citealt{Yakovlev2001physrep}; \citealt{Schmitt2018}). According to the results of Paper I, the neutrino cooling function for the CPF process can be written as
\begin{equation}\label{eq:lCPddo}
\ell_\mathrm{CPF}(\widetilde{T})= 
q 
\frac{\Lambda_\mathrm{CPF}}{\Sigma_{n\ell}} \widetilde{T}^6 F(\tau),
\end{equation}
where $\tau=\widetilde{T}/\widetilde{T}_{Cn \mathrm{max}}$ and $\Lambda_\mathrm{CPF}$ and $\Sigma_{n\ell}$ are the quantities related to the neutrino luminosity and heat capacity, respectively. They do not depend on the profile of the neutron triplet critical temperature $\widetilde{T}_{Cn}(\rho)$, where $\rho$ is the density inside the star, and contain the main dependence of $\ell_\mathrm{CPF}(\widetilde{T})$ on the  model of the star  (i.e. on EOS and $M$). In Paper I it was shown that $\Lambda_\mathrm{CPF}$ and $\Sigma_{n\ell}$ weakly depend on the EOS and can be reliably approximated by model-independent expressions which 
contain only $M$ and $R$. Similar expressions for other neutrino cooling processes were constructed earlier by \citet{Ofengeim2017PhRvD}. 

The factor $\Lambda_\mathrm{CPF}$ in equation~(\ref{eq:lCPddo}) is calculated based on the expressions for the CPF neutrino emissivity \citep{Yakovlev2001physrep} which do not include a response of the Cooper pair condensate. The microscopic calculations by \citet{Leinson2010PhRvC} show that the correct analysis reduces the CPF neutrino emissivity by a factor of 0.19 in comparison with the expressions given by \citet{Yakovlev2001physrep}. This factor is, however, calculated in the limit of non-relativistic neutrons, while the relativistic corrections can alter it in an unknown direction. Therefore, here we describe the modification due to the condensate response, as well as other possible corrections due to collective effects to the basic formula, via the phenomenological factor $q$ (see Paper I for more detailed discussion), and will try to constrain it from observations.

Finally, the dimensionless function $F(\tau)$ in equation~(\ref{eq:lCPddo}) 
depends on the shape of $\widetilde{T}_{Cn}(\rho)$ (and also on the stellar model), but not on its amplitude. $\widetilde{T}_{Cn\mathrm{max}}$.
This function is not universal, in the sense that it also depends on the EOS, $M$ and $R$, however, it has a few appealing properties investigated in detail in Paper I which make possible a largely model-independent analysis. The function $F(\tau)$ has a bell-like shape, with a maximum at about $\tau\sim 0.2-0.4$. The maximal value $F_{\mathrm{max}}=\max F(\tau)$ is model-dependent, however the results of Paper I show that virtually always one constrains $F_{\mathrm{max}}<F_m$, where $F_m\approx2$ is the supremum of $F_{\mathrm{max}}$ over models. Moreover, as shown by \citet{Shternin2015MNRAS}, an early part of $F(\tau)$, while $\tau\gtrsim 0.6$, is described by a universal expression
\begin{equation}\label{eq:FSY}
F(\tau)\approx F_{\mathrm{SY}}(\tau)\equiv 117.6 \mu_{\mathrm{max}} \tau(1-\tau)^2,    
\end{equation}
which is parametrized by $\mu_{\mathrm{max}}=\max [\tau^6 F(\tau)]$; this maximum is reached at $\tau=\tau_\mu\approx 0.8$. A universal form of equation~(\ref{eq:FSY}) allows to construct self-similar NS cooling solutions \citep[see details in][]{Shternin2015MNRAS}. According to equation~(\ref{eq:lCPddo}), $\mu_{\mathrm{max}}$ characterises a maximal dimensionless CPF neutrino cooling function. Like $F_{\mathrm{max}}$,
the parameter $\mu_{\mathrm{max}}$ is also constrained from above via $\mu_{\mathrm{max}}<0.18$. These two constraints combine in the following restriction on any possible function $F(\tau)$:
\begin{equation}\label{eq:Fconst}
    \left\{
    \begin{array}{ll}
    F(\tau)<F_{\mathrm{max}}<F_m\approx 2, & \\
 \textbf{}    F(\tau)<F_{\mathrm{SYmax}}(\tau)=21.2\ \tau (1-\tau)^2, & \tau>0.6.
    \end{array}
    \right. 
\end{equation}

\begin{figure*}
    \centering
        \includegraphics[width=0.47\textwidth]{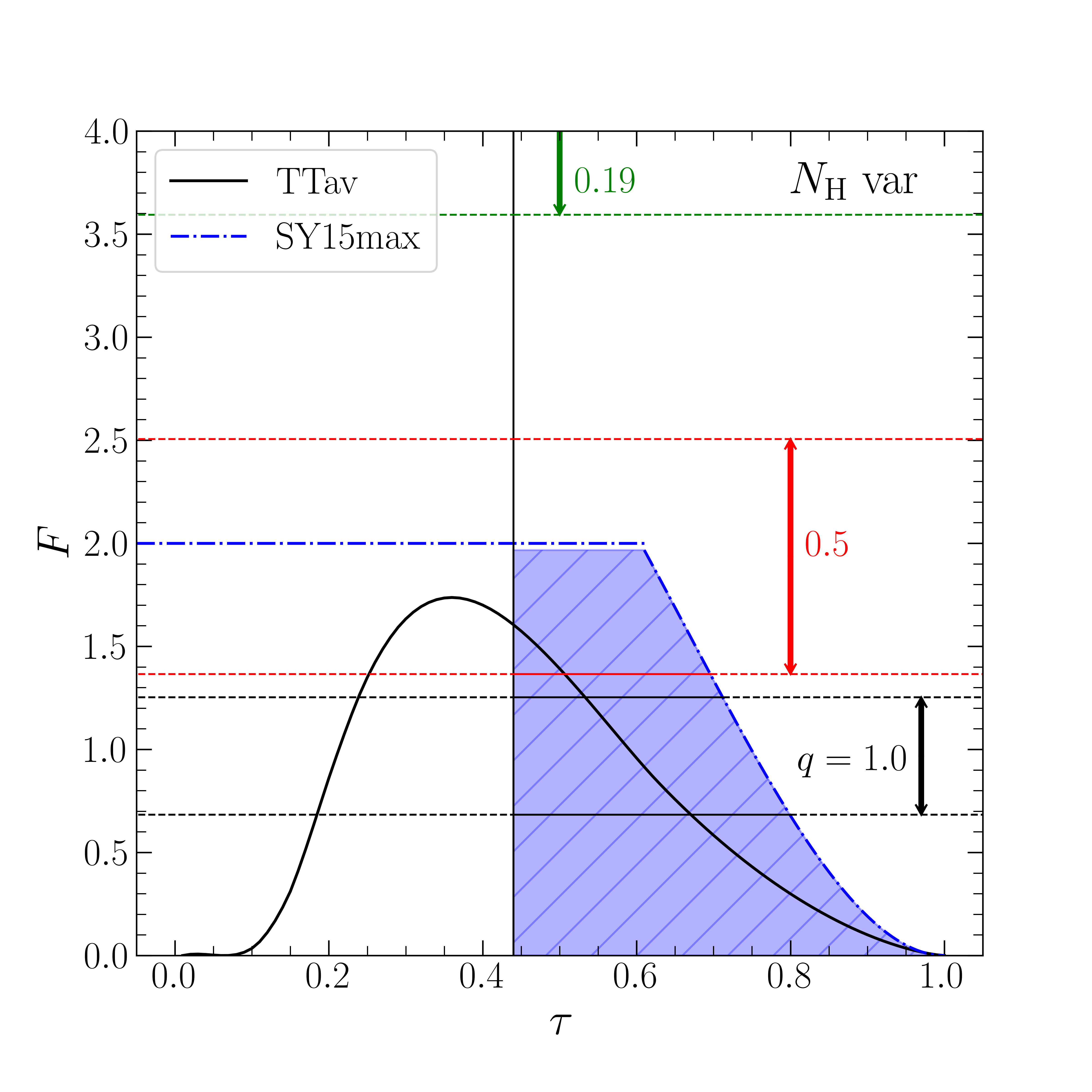}
    \includegraphics[width=0.47\textwidth]{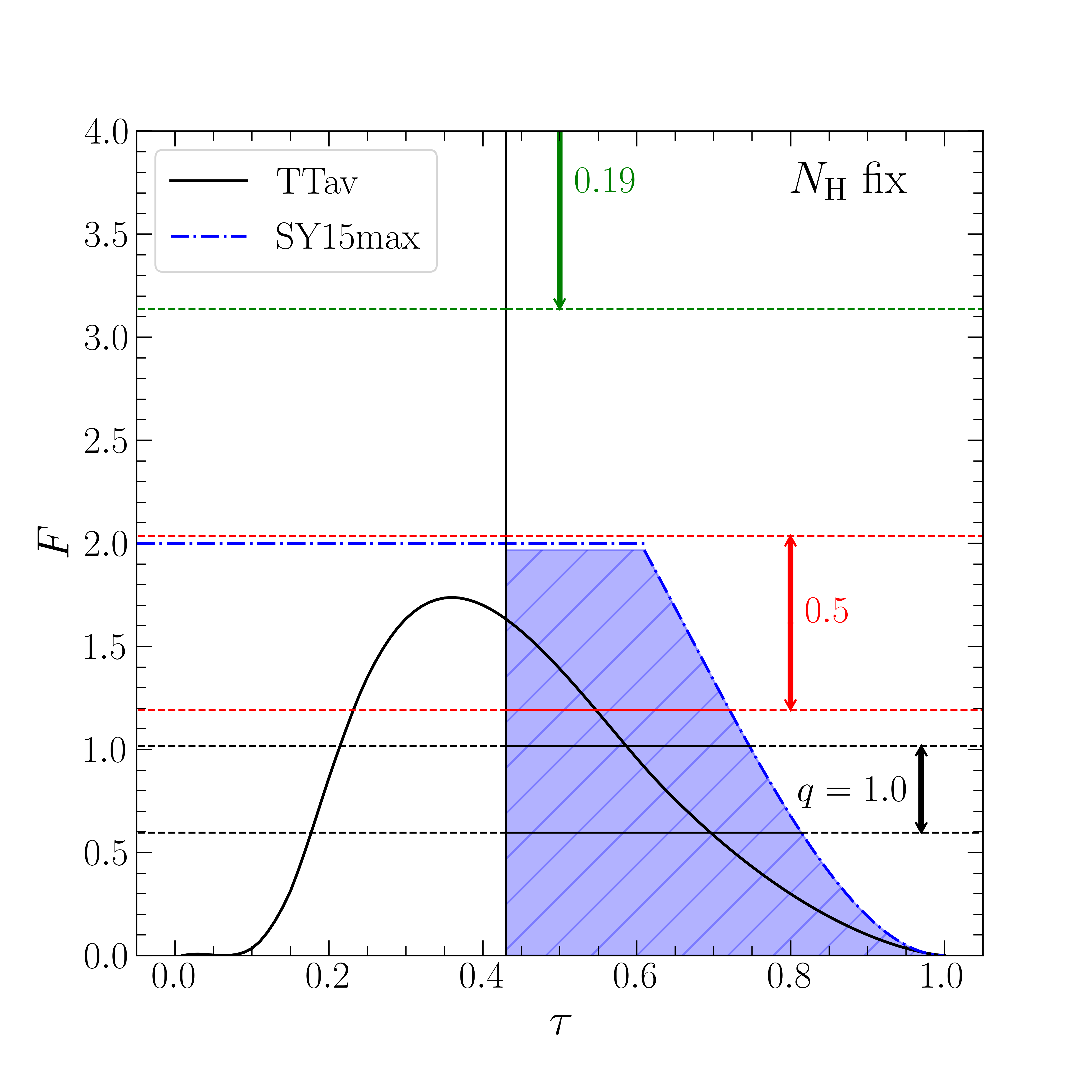}
        \caption{Various constraints on $F$ and $\tau$ based on the joint fit. The left and right panels correspond to the models with variable and fixed $N_\mathrm{H}$, respectively. Dash-dotted lines show the constraints given in equation~(\ref{eq:Fconst}). Black solid curves show the example of the 
        TTav superfluidity model profile for $M=1.5 M_\odot$ and BSk21 EOS. The black, red and green arrows between the coloured horizontal lines indicate the 90 per cent credible intervals for $F_d$ if $q=1$, 0.5, and 0.19, respectively, as indicated near the arrows. Vertical lines indicate the lowest $\tau_d$ (90 per cent credibility) inferred from the data. Accordingly, the hatched boxes show the possible values of $(F_d,\tau_d)$ which are allowed before the inference on $F_d$ is accounted for. }\label{fig:boxes}
\end{figure*}

Fig.~\ref{fig:boxes} illustrates these properties. There, in both panels,  the dot-dashed lines on the $F-\tau$ plane  show the constraints (\ref{eq:Fconst}). Any dimensionless emissivity profile $F(\tau)$ should reside below the dash-dotted lines. Indeed, the solid curves show the function $F(\tau)$ calculated for a specific superfluidity model [specific $\widetilde{T}_{Cn}(\rho)$ shape], provided  by \citet{Takatsuka2004PThPh} and denoted as TTav by \citet{Ho2015PhRvC}, and a $M=1.5\ M_\odot$ NS having the BSk21 EOS. The other content of figure~\ref{fig:boxes} is explained later below.

Using equations~(\ref{eq:slope def}) and (\ref{eq:lCPddo})
we can express the cooling slope as
\begin{equation}\label{eq:slope_Ftau}
    s(\tau)=q\beta t \frac{\Lambda_\mathrm{CPF}}{\Sigma_{n\ell}}\widetilde{T}^5  F(\tau)=q\beta t \frac{\Lambda_\mathrm{CPF}}{\Sigma_{n\ell}}\widetilde{T}_{Cn\mathrm{max}}^5 \tau^5 F(\tau).
\end{equation}
The function $s(\tau)$ has a bell-like shape with a maximum around $\tau\approx \tau_\mu$ \citep{Shternin2015MNRAS}. The height of this bell is regulated by the contrast between the initial slow cooling governed by $\ell_0(\widetilde{\tau})$] and strength of the CPF emission. This contrast is conveniently described by the parameter
\begin{equation}\label{eq:delta_def}
\delta = \frac{\max \ell_\mathrm{CPF}}{\ell_C} = \frac{q\Lambda_\mathrm{CPF} \widetilde{T}_{Cn\mathrm{max}}^6}{\Sigma_{n\ell} \ell_C} \mu_\text{max},
\end{equation}
where $\ell_C=\ell_0(\widetilde{T}_{Cn\mathrm{max}})$ is the neutrino cooling function at the superfluidity onset. A larger $\delta$ leads to a taller and wider $s(\tau)$ peak. Actually, at $\tau\gtrsim 0.6$, $\delta$ is the only parameter which regulates self-similar cooling 
of the CPF-mediated NSs \citep{Shternin2015MNRAS}.
At $\tau\lesssim 0.6$ the cooling curve shapes start to depend on the exact shape of $F(\tau)$.

The described properties allow one to constrain the parameters of the NS superfluidity. Let us denote the values for the current (detection) epoch with subscript $d$. Then equation~(\ref{eq:slope_Ftau}) taken at the present epoch ($t=t_d=330~\mathrm{yr}$, $\widetilde{T}=\widetilde{T}_d$, $s=s_d$) gives the present-day value of the combination 
\begin{equation}\label{eq:Fd}
G_d\equiv q F_d=\frac{s_d\,
\Sigma_{n\ell}
}{\beta t_d \Lambda_\mathrm{CPF} \widetilde{T}_d^5},
\end{equation}
where $F_d\equiv F(\tau_d)$. This means that the data alone can constrain $F_d$ only up to the unknown factor $q$. Upper and lower boundaries of the $90$ per cent highest-posterior-density credible intervals for $G_d$ are given in Table~\ref{tab:TaudFdBox}. 
If we knew $q$, we could constrain the possible values of $F_d$ as illustrated in Fig.~\ref{fig:boxes}
(left and right panels in Fig.~\ref{fig:boxes} correspond to variable-$N_{\mathrm{H}}$ and fixed-$N_{\mathrm{H}}$ models respectively). There, based on the joint spectral fit,  we show $90$ per cent credible intervals on $F_d$, given $q=1,\,0.5$ and $0.19$, with black, red and green horizontal lines, respectively. 
Only the parts of the intervals that satisfy equation~(\ref{eq:Fconst}) are allowed. Clearly, lowering $q$ makes a range of the possible superfluid models (possible $F(\tau)$ profiles) shallower. At some critical value of $q$ the corresponding interval becomes incompatible with the constraint $F_d<F_m$ given in the first line of equation~(\ref{eq:Fconst}). 

Further constraints on $q$ or $F_d$ require separation of these variables,  since only their combination $G_d$ (equation~\ref{eq:Fd}) is constrained by the data. We give more discussion on this in Sec.~\ref{sec:q_analysis}. 
The simplest constraint is the following. Let $G_{d\alpha}$ be a one-side lower $\alpha$-quantile for $G_d$. Then, using the first line in equation~(\ref{eq:Fconst}) and equation~(\ref{eq:Fd}), one can reject $q<q_\alpha\equiv G_{d\alpha}/F_m$ at least at level $1-\alpha$.
In other words, since $F_d$ is at most $F_m$, any possible lower value of $F_d$ reduces the probability for $q<q_\alpha$, i.e. $\mathrm{Pr}(q<q_\alpha)<\alpha$. According to Table~\ref{tab:TaudFdBox}, for the joint fit we get $\mathrm{Pr}(q<0.34)<0.95$
($\mathrm{Pr}(q<0.30)<0.95$)
for the  model with variable (fixed) $N_{\mathrm{H}}$.
We also include the significance limits understood in this way for the theoretical value $q=0.19$ in Table~\ref{tab:TaudFdBox}.

The second line in equation~(\ref{eq:Fconst}) can be used to constrain the present-day dimensionless temperature $\tau_d=\widetilde{T}_d/\widetilde{T}_{Cn\mathrm{max}}$ from above. Indeed, the point $(\tau_d,\, F_d)$ cannot reside above and to the right of the blue dash-dotted lines in Fig.~\ref{fig:boxes}. A specific upper limit on $\tau_d$, and hence on the $\widetilde{T}_{Cn\mathrm{max}}$, depends thus on $F_d$. However, the second equality in equation~(\ref{eq:slope_Ftau}) actually sets a lower limit on $q^{1/5} \widetilde{T}_{Cn\mathrm{max}}$, 
since according to theory $\max [\tau^5 F(\tau)]<0.23$ (Paper I); this limit is reached at some $\tau=\tau_5\approx 0.75$. Given the posterior distribution of $s$, $M$ and $R$, the lower limit on $q^{1/5} \widetilde{T}_{Cn\mathrm{max}}$ can be inferred from equation~(\ref{eq:slope_Ftau}). The corresponding boundaries of 90 per cent one-side credible intervals for $q^{1/5} \widetilde{T}_{Cn\mathrm{max}}$ are given in Table~\ref{tab:sf_res}. 
 
 Finally, we can constrain $\widetilde{T}_{Cn\mathrm{max}}$ from above, or $\tau_d$ from below. The idea is as follows. Lowering $\tau_d$ for a given $s_d$ requires amplification of the $s(\tau)$ peak (provided $\tau_d<\tau_\mu$, which is always the case), in other words it requires an increase in $\delta$.  Recall that it is assumed that the initial neutrino cooling function $\ell_0(\widetilde{T})\propto \widetilde{T}^{\, 7}$, therefore, according to equation~(\ref{eq:delta_def}), $\delta\propto \widetilde{T}_{Cn\mathrm{max}}^{\, -1}\propto \tau_d$ for a given $\widetilde{T}_d$, so that lowering $\tau_d$ while keeping the same $\ell_0(\widetilde{T})$ actually lowers $\delta$. Therefore, in order to increase $\delta$ one needs to suppress $\ell_0(\widetilde{T})$. 
 From a physical point of view, in our model there is a lowest possible $\ell_0(\widetilde{T})$ given mainly by  neutron-neutron bremsstrahlung neutrino emission (and lepton bremsstrahlung as less important processes). This situation is realised when the strong proton pairing completely suppresses the modified Urca processes.
 The presence of the lower limit on $\ell_0(\widetilde{T})$ thus constrains $\tau_d$ from below and $\widetilde{T}_{Cn\mathrm{max}}$ from above, see Paper I for the technical details. The resulting 90 per cent one-side credible boundaries on the largest $\widetilde{T}_{Cn\mathrm{max}}$ according to the data are given in Table~\ref{tab:TaudFdBox}, and shown for the joint fit in Fig.~\ref{fig:boxes} by the vertical lines there. Notice that this limit does not depend on $q$ since it is governed by the actual present-day cooling rate.
 
 These arguments, in principle, can be inverted to constrain the amount of the proton-paired matter in the core. Indeed, moderate proton pairing (i.e., protons are strongly paired in some part of the core) leads to the moderate $\ell_0(\widetilde{T})$, increasing $\tau_d$. Since, clearly, it should be $\tau_d<\widetilde{T}_d/\widetilde{T}_{Cn\mathrm{max}}^{\mathrm{low}}$, the upper limit on $\ell_0(\widetilde{T})$ can be set. 
 
In Fig.~\ref{fig:boxes} we showed the result based on the joint fit for all ACIS-S data. The similar figures plotted using either FAINT or GRADED mode alone are given in Appendix~\ref{app:SF} for completeness.

%%%%%%%%%%%%%%%%%%%%%%%%%%%%%%%%%%%%%%%%%%%%%%%%%%%%%%%%%%%%%%%%%%%%%%%%%%%%%%%%%%%%%%%%%%%%
\begin{table}
    \centering
    \setlength{\tabcolsep}{5pt}
    \caption{$G_d^\mathrm{low}$ and $G_d^\mathrm{up}$ are the lower and upper boundaries, respectively, of the 90 per cent credible regions for $G_d$.
        Upper and lower limits on $\widetilde{T}_{Cn\mathrm{max}}$ correspond to the 90 per cent one-sided credible intervals for a corresponding quantity (see text for details). The last column gives the upper boundary for the $q<0.19$ hypothesis significance, i.e. $\mathrm{Pr}(q<0.19)<\alpha$, see text for details. }
    \label{tab:TaudFdBox}
    \begin{tabular}{llccccccc}
        \hline
		Mode & $N_{\mathrm{H}}$ & $G_d^\mathrm{low}$ & $G_d^\mathrm{up}$  & $q^{1/5} \widetilde{T}_{Cn\mathrm{max}}^\mathrm{low}$ &  $\widetilde{T}_{Cn\mathrm{max}}^\mathrm{up}$ &$\alpha$  \\
		 & &  & &  ($10^8$~K) &  ($10^8$~K) &  $q<0.19$ 		\\
		\hline
		ALL & Var & 0.68 & 1.25 & 3.3 & 6.2 & $8\times 10^{-6}$ \\
		ALL & Fix & 0.60 & 1.02 & 3.0 & 6.2 & $2\times 10^{-4}$\\ 
		FAINT & Var &  0.26 &  1.56 &    2.0 &  5.9 & $0.14$\\
		FAINT & Fix & 0.15  &  0.92 &  1.7 &  5.6 & 0.43\\ 
		GRADED & Var & 0.41 & 1.06 & 3.3 & 6.6 & 0.029 \\
		GRADED & Fix & 0.36 & 0.96 & 3.2 & 6.6 & 0.064\\
\hline
    \end{tabular}
\end{table}
%%%%%%%%%%%%%%%%%%%%%%%%%%%%%%%%%%%%%%%%%%%%%%%%%%%%%%%%%%%%%%%%%%%%%%%%%%%%%%%%%%%%%%%%

%%%%%%%%%%%%%%%%%%%%%%%%%%%%%%%%%%%%%%%%%%%%%%%%%%%%%%%%%%%%%%%%%%%%%%%%%%%%%%%%%%%%%%%%%%%%%%%%%%%%
%%%%%%%%%%%%%%%%%%%%%%%%%%%%%%%%%%%%%%%%%%%%%%%%%%%%%%%%%%%%%%%%%%%%%%%%%%%%%%%%%%%%%%%%%%%%%%%%%%%%

\subsection{Constraints on $q$}\label{sec:q_analysis}
In this Section we take the point of view that the independent parameters in our model are $G_d$, which represents an actual measured cooling rate of the star, and $F_d$, for which we have theoretical constraints. Then $q=G_d/F_d$ can be viewed as the derived parameter, which functionally depends on $q$ and $F_d$.%
\footnote{Notice, that in Paper I we considered a slightly different setup. 
Namely,
we estimated $F_d$ given $q$. This represents a different perspective on the same statistical data.} 
We can now write the conditional probability of finding $(F_d,\,G_d)$ given the data $D$ as $p(F_d,G_d|D)=\pi(F_d) p(G_d|D)$, where $\pi(F_d)$ is the prior probability
{density}
of $F_d$. Then the marginalized probability for observing $q$ given the data is
\begin{equation}\label{eq:pq_gen}
    p(q|D)=\int\int \mathrm{d}F_d \mathrm{d} G_d\ \delta(q-G_d/F_d)\  \pi(F_d) p(G_d|D).
\end{equation}

Assuming $\pi(F_d)$ 
has
a support on
the interval $[0,F_m]$
and that $G_d>0$,
\begin{equation}\label{eq:pqD}
    p(q|D)=\frac{1}{q^2} \int\limits_{0}^{F_m q}\mathrm{d} G_d\,  G_d\ \pi(G_d/q)\ p(G_d|D).
\end{equation}
The probability for $q<q_0$ is then obtained by integrating equation~(\ref{eq:pqD}). Changing the integration order and performing integration over $q$ explicitly, we get
\begin{equation}\label{eq:Pr_q_q0}
    \mathrm{Pr}(q<q_0|D)= \int\limits_{0}^{F_m q_0}\mathrm{d} G_d\,  p(G_d|D) \left[ 1 - \Pi(G_d/q_0) \right],
\end{equation}
where 
\begin{equation}
    \Pi(F)=\int\limits_0^{F} \mathrm{d} F'\, \pi(F').
\end{equation}
This expression shows that  $\mathrm{Pr}(q<q_0|D)$ is indeed lower than the limit discussed above (which is given by the first integral). Importantly, in this model the posterior distribution of $q$ is proper and allows for calculating its credible intervals without additional assumptions.

For the final analysis we include both  $(F_d,\tau_d)$ as  parameters of the model with uniform priors in the part of the plane set by equation~(\ref{eq:Fconst}).
Because of equation~(\ref{eq:Fconst}), an upper limit for $F_d$ at $\tau>0.6$ depends on $\tau$, which results in a slight modification of equations~(\ref{eq:pq_gen})--(\ref{eq:Pr_q_q0}) by replacing 
$F_m\to \min[F_m,\,21.2\tau_d(1-\tau_d)^2]$ 
and introducing additional integration over $\tau_d$. We do not write these explicit expressions for simplicity. The posterior distribution is then updated using restrictions on $\ell_0(\widetilde{T})$ as described above.
The resulting marginalised posterior distributions of the parameters of interest are shown in  Fig.~\ref{fig:tri_SF} for the joint fit and in the similar Fig.~\ref{fig:tri_SF_app} in Appendix~\ref{app:SF} for FAINT and GRADED modes. The corresponding credible intervals are given in Table~\ref{tab:sf_res}. The parameters shown are as follows. We show $F_d$, which is weakly constrained, and its posterior distribution is close to the 
prior one\footnote{Hence we do not give the intervals for $F_d$ in Table~\ref{tab:sf_res} as they are basically meaningless.}, and $\tau_d$, for which we infer the lower limit as described above. We also show the relative CPF emission strength $\log_{10}\delta$ which regulates the self-similar cooling solutions at the CPF stage \citep{Shternin2015MNRAS}. The neutrino emission at the initial cooling stage is parametrized  in Table~\ref{tab:sf_res} and Fig.~\ref{fig:tri_SF} relative to the so-called standard neutrino candles \citep{Yakovelv2011MNRAS} via the parameter $\log_{10} f_\ell$. Basically, $f_\ell=\ell_0(\widetilde{T})/\ell_{\mathrm{SC}}(\widetilde{T})$, where $\ell_{\mathrm{SC}}(\widetilde{T})$ is the standard neutrino candle cooling function of the star with the same $M$
 and $R$. The temperature dependencies of $\ell_0(\widetilde{T})$ and $\ell_{\mathrm{SC}}(\widetilde{T})$ are assumed to be identical (i.e. both are realizations of the slow cooling, $n=7$), hence $f_\ell$ does not depend on temperature. The next parameter is the maximal redshifted neutron critical temperature $\widetilde{T}_{C\mathrm{n\max}}$, one of the main parameters of interest.\footnote{In Fig.~\ref{fig:tri_SF} it is shown as $\widetilde{T}_C$ for readability of the axis captions.} We also show $M$ and $R$ in Fig.~\ref{fig:tri_SF} to indicate the dependence on the model of the star.
 Finally we show, in logarithmic scale, the relative CPF emission strength $\log_{10}q$.

%%%%%%%%%%%%%%%%%%%%%%%%%%%%%%%%%%%%%%%%%%%%%%%%%%%%%%%%%%%%%%%%%%%%%%%%%%%%%%%%%%%%%%%%%%%%%%%%%%%%
\begin{figure*}
    \centering
    \includegraphics[width=0.7\textwidth]{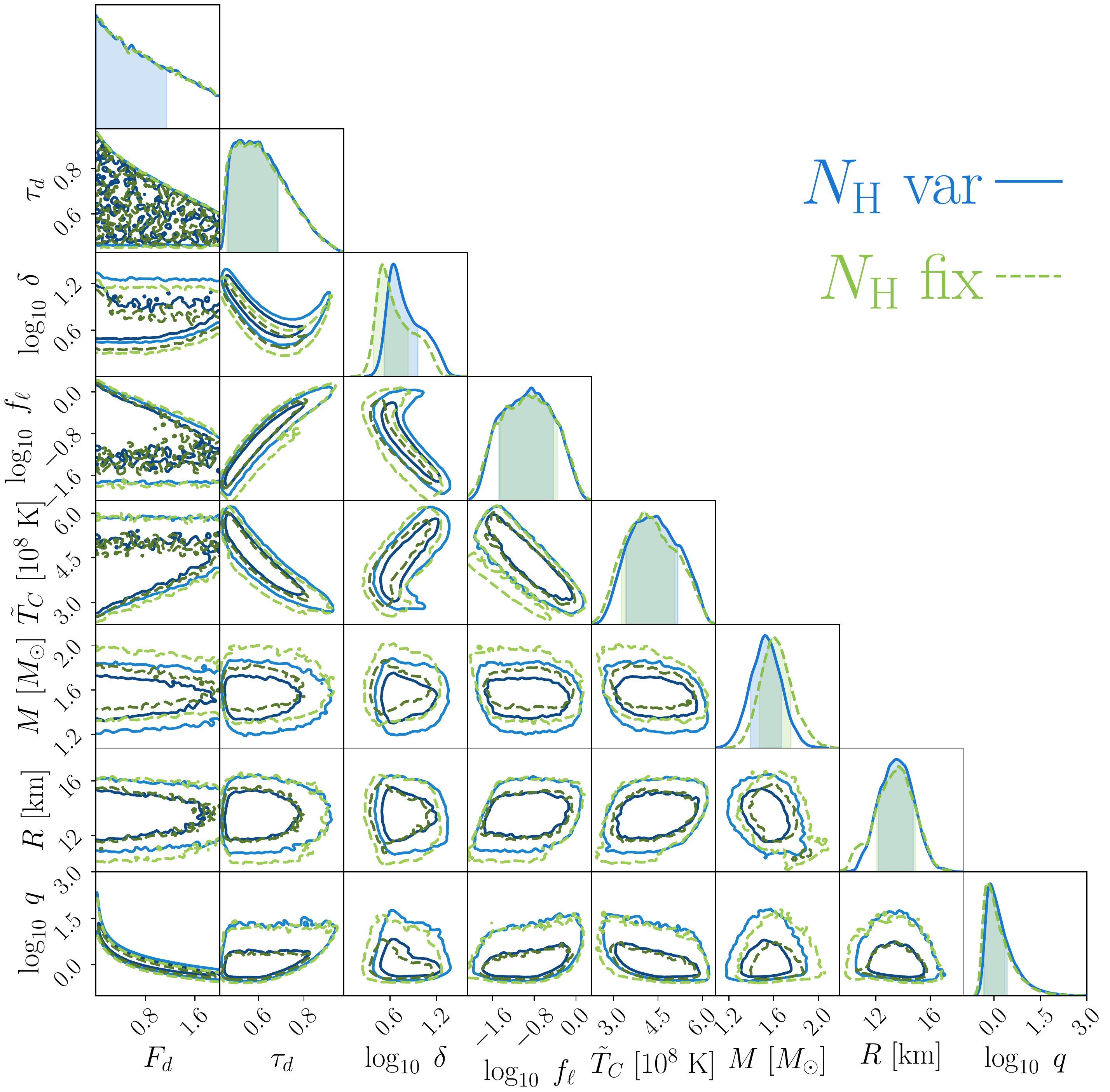}
    \caption{1D and 2D posterior distributions for the parameters related to the CPF cooling model. Solid and dashed lines correspond to the models with variable and fixed $N_{\mathrm{H}}$, respectively. Shaded areas on 1D distributions correspond to 68 per cent credible intervals, while contours on 2D distributions correspond to 68 per cent and 90 per cent levels. }\label{fig:tri_SF}
\end{figure*}
%%%%%%%%%%%%%%%%%%%%%%%%%%%%%%%%%%%%%%%%%%%%%%%%%%%%%%%%%%%%%%%%%%%%%%%%%%%%%%%%%%%%%%%%%%%%%%%%%%%%

\begin{table*}
    \centering
    \caption{Parameters related to the superfluidity cooling models as described in the text. All uncertainties correspond to the 68 per cent highest posterior density credible intervals. 
    }
    \label{tab:sf_res}
    \begin{tabular}{llcccccc}
       \hline
		Mode &$N_{\mathrm{H}}$&  $\tau_d$ &  $\log_{10}\delta$ & $\log_{10}\ f_{\ell 0}$ & $\widetilde{T}_{Cn\mathrm{max}}$ &$\log_{10}q$\\
		 & & &  & &   ($10^8$~K)  	&	\\
		\hline
		ALL & Var &   $0.54^{+0.14}_{-0.08}$ & $0.64^{+0.31}_{-0.12}$ & $-0.79^{+0.33}_{-0.70}$ & $4.2^{+0.9}_{-0.8}$ &  $-0.14^{+0.56}_{-0.19}$\\  
		ALL & Fix &  $0.60^{+0.08}_{-0.14}$ & $0.51^{+0.32}_{-0.13}$ & $-0.82^{+0.42}_{-0.66}$ & $4.1^{+1.0}_{-0.8}$  & $-0.22^{+0.55}_{-0.19}$ \\
        FAINT &Var &  $0.60^{+0.12}_{-0.11}$ & $0.78^{+0.27}_{-0.15}$ & $-1.47^{+0.70}_{-0.43}$ & $3.4^{+1.1}_{-0.7}$ &  $-0.02^{+0.56}_{-0.27}$ \\
        FAINT & Fix & $0.63^{+0.11}_{-0.12}$ & $0.49^{+0.24}_{-0.18}$ & $-1.72^{+0.66}_{-0.36}$ & $2.9^{+1.0}_{-0.5}$ &  $-0.25^{+0.55}_{-0.33}$ \\
        GRADED & Var&   $0.58^{+0.12}_{-0.09}$ & $0.62^{+0.28}_{-0.14}$ & $-0.83^{+0.52}_{-0.51}$ & $4.4^{+1.1}_{-0.7}$ &  $-0.20^{+0.55}_{-0.25}$ \\
        GRADED & Fix&  $0.59^{+0.10}_{-0.12}$ & $0.55^{+0.27}_{-0.16}$ & $-0.77^{+0.50}_{-0.56}$ & $4.4^{+1.0}_{-0.8}$ &  $-0.25^{+0.55}_{-0.25}$ \\
        \hline
     \end{tabular}
\end{table*}

%%%%%%%%%%%%%%%%%%%%%%%%%%%%%%%%%%%%%%%%%%%%%%%%%%%%%%%%%%%%%%%%%%%%%%%%%%%%%%%%%%%%%%%%%%%%%%%%%%%%
\section{Discussion}\label{S:Discuss}
%%%%%%%%%%%%%%%%%%%%%%%%%%%%%%%%%%%%%%%%%%%%%%%%%%%%%%%%%%%%%%%%%%%%%%%%%%%%%%%%%%%%%%%%%%%%%%%%%%%%

The spectral analysis of the FAINT mode data leads to similar results with the GRADED mode observations.
The main difference is that the FAINT mode data analysis leads to larger masses and smaller radii than the GRADED mode one, however the inferred ranges overlap in a region of currently commonly adopted value of $R\sim 11-14$~km for intermediate-mass NSs.
Interestingly, in contrast to the prior expectations before the May 2020 observation was taken, 
the model with variable $N_{\mathrm{H}}$  for the FAINT mode data results even in slightly faster, but consistent, cooling compared to the results obtained from the GRADED mode data. However there are only four observation epochs for the FAINT mode data and the next observations in this mode, if performed, 
may 
change this. 

This similarity between the inferred spectral parameters suggests that both modes now give consistent results, and that the main systematic effects which contaminate the GRADED mode data are to a large extent 
accounted for 
by the calibration and pileup models, allowing us to perform a joint analysis of the CasA~NS cooling data. The joint spectral modelling assumes that the CasA NS follows a regular temperature decrease track and that the systematic difference between the FAINT and GRADED mode data is a result of the incomplete pileup modelling, or incomplete absolute calibration of the ACIS-S detector in different instrumental modes. We find that this systematic difference can be well accounted for by multiplying the model for the GRADED mode data by the calibration factor $A\approx 1.1$. 

The joint fit reduces the uncertainties of the parameters and allows to constrain the models of superfluidity applying the method developed in Paper I. The inferred range for the maximal redshifted superfluid critical temperature is lower but consistent for the FAINT mode data than those for the GRADED mode data (Table~\ref{tab:sf_res}). 
The joint fit results in  the intermediate range
$\widetilde{T}_{Cn\mathrm{max}}=(4.3\pm 1.0)\times 10^{8}$~K (fixed and variable $N_{\mathrm{H}}$ models united). 
This is because of the correlation between $R$ and $\widetilde{T}_{Cn\mathrm{max}}$ apparent in Fig.~\ref{fig:tri_SF}. The FAINT mode data tends to lower $R$ and, as a consequence, to somewhat lower  $\widetilde{T}_{Cn\mathrm{max}}$. The actual (non-redshifted) maximal critical temperature is a factor of $1.2-1.8$ higher than $\widetilde{T}_{Cn\mathrm{max}}$ depending on the position of the critical temperature peak within the core, so that $T_{Cn\mathrm{max}}=(4-9.5)\times 10^{8}$~K. This range is compatible with the bulk of the theoretical estimates of the neutron triplet pairing critical temperatures \citep{Sedrakian2019EPJA,Ho2015PhRvC} as well as other measures  \citep*[e.g.,][]{Kantor2020PhRvL} thus indicating the CPF  paradigm of the CasA~NS cooling is quite plausible. 

On the other hand, existing microscopic calculations \citep{Leinson2010PhRvC} suggest that the power of the CPF neutrino cooling is 
a factor of $0.19$ weaker than the (incorrect) benchmark result given in \citet{Yakovlev2001physrep}. Our present results show that, like in the case of the GRADED mode data analysed in Paper I, $q=0.19$ is hardly compatible with the data. The limiting probability of obtaining $q<0.19$ for the joint fit is extremely low, see Table~\ref{tab:TaudFdBox}. Such a low value of $q$ can be marginally consistent with observations if we (for any reason)  disregard the GRADED mode data, and base an analysis solely on the FAINT mode data, see Fig.~\ref{fig:Faintboxes} and Table~\ref{tab:TaudFdBox}. This is because of the presence of the high temperature and low radius tail in the spectral posteriors for the FAINT data. Given that temperature enters the denominator in equation~(\ref{eq:Fd}) to a high power, this lowers $G_d$ and hence weakens the constraints on $q$. On the other hand, this is based really on the tail of the distribution. According to Table~\ref{tab:sf_res}, the 68 per cent highest posterior density credible regions for $\log_{10} q$ for all considered models lie above $0.3$. 

This discrepancy is more pronounced for the models with varying $N_\mathrm{H}$, see Table~\ref{tab:TaudFdBox}. This means that either the existing theoretical calculations of the $q$ factor are incomplete and the improvements at this side, e.g. going beyond the non-relativistic limit, can produce increased $q$ (Paper I), or that the pure CPF model is not complete. 
For instance, recently \citet{Leinson2022MNRAS} proposed the explanation of the observed fast cooling by additional enhancement of the neutrino luminosity by the existence of the direct Urca process in the tiny inner core of the NS star. However, this explanation requires very tight constraints on the NS mass -- i.e. in this scenario one is extremely lucky to catch the young NS with a mass precisely just above the direct Urca threshold. 
Finally, it is possible that the CPF neutrino emission explanation is wrong and other explanations 
(\citealt*{Yang2011ApJ};  \citealt{Blaschke2012PhRvC}; \citealt{Negreiros2013PhLB}; \citealt{Noda2013ApJ}; \citealt{Sedrakian2013AA}; \citealt{Bonanno2014AA}; \citealt{Leinson2014JCAP}; \citealt{Hamaguchi2018PhRvD}, which all have their own pros and cons)
may be correct.

%%%%%%%%%%%%%%%%%%%%%%%%%%%%%%%%%%%%%%%%%%%%%%%%%%%%%%%%%%%%%%%%%%%%%%%%%%%%%%%%%%%%%%%%%%%%%%%%%%%%
\section{Conclusions}\label{S:conclusions}
%%%%%%%%%%%%%%%%%%%%%%%%%%%%%%%%%%%%%%%%%%%%%%%%%%%%%%%%%%%%%%%%%%%%%%%%%%%%%%%%%%%%%%%%%%%%%%%%%%%%
We performed a joint analysis of all \textit{Chandra} ACIS-S observations of CasA~NS taken in the GRADED as well as in the FAINT modes.
We confirm the recent findings of \citet{PosseltPavlov2022ApJ} that the FAINT mode data shows a significant decrease of the X-ray flux, which can be interpreted as the NS cooling in real time. The cooling rate is similar to that  obtained from the GRADED mode data \citep[][and references therein]{Ho21,Shternin2021MNRAS} indicating that the analysis of both modes has probably reached consistency. Specifically, based on the joint ACIS data analysis we find the surface temperature decline of $2.12\pm 0.3$ per cent in 10 years for variable $N_{\mathrm{H}}$ model and to $1.6\pm 0.2$ for fixed $N_{\mathrm{H}}$ model. However, both modes are observation modes of the same ACIS detector, and one cannot be completely sure that the apparent cooling is not caused by 
ACIS sensitivity degradation that has not yet been completely calibrated \citep{Plucinsky2020SPIE}.  
An observational campaign using other instruments \citep[e.g. with HRC; see ][]{Elshamouty2013ApJ} would help to check this (a new HRC-S observation is planned for 2023).

The joint modelling of the FAINT and GRADED mode spectra allowed us to relatively well constrain the CasA~NS mass at $M=1.55\pm0.25~M_\odot$ and radius at $R=13.5\pm 1.5$~km.\footnote{Here we united the results from $N_{\mathrm{H}}$ fixed and variable models given in Table~\ref{tab:spectral_params}. }

We applied  the model-independent analysis developed in Paper I to all ACIS-S data including those obtained both in FAINT and GRADED modes
and inferred the NS superfluidity parameters. 
We constrain the maximal critical temperature of the triplet neutron pairing within the core at $T_{Cn\mathrm{max}}=(4-9.5)\times 10^{8}$~K. However the required effective strength of the CPF neutrino emission $q=0.5 -2.6$ $(0.4 - 2.1)$ for $N_{\mathrm{H}}$ variable (fixed), 68 per cent credibility, is at least a factor of 2 higher than $q=0.19$ suggested by the existing
microscopic calculations \citep{Leinson2010PhRvC}. Further theoretical and observational  studies are required to finally resolve the CasA~NS puzzle.

\section*{Acknowledgements}

This work is supported by the Russian Science Foundation, grant 19-12-00133. The authors are indebted to Serge Balashev, Bettina Posselt and Dima Yakovlev for numerous discussions.
WCGH appreciates the use of computer facilities at the Kavli Institute for Particle Astrophysics and Cosmology. COH is supported by the Natural Sciences and Engineering Research Council of Canada (NSERC) via Discovery Grant RGPIN-2016-04602. 

\section*{Data Availability}

The data underlying this article will be shared on reasonable request to the corresponding author.

%%%%%%%%%%%%%%%%%%%%%%%%%%%%%%%%%%%%%%%%%%%%%%%%%%



\bibliographystyle{mnras}
\bibliography{casa} % if your bibtex file is called example.bib

% \begin{thebibliography}{99}
% \bibitem[\protect\citeauthoryear{Author}{2012}]{Author2012}
% Author A.~N., 2013, Journal of Improbable Astronomy, 1, 1
% \bibitem[\protect\citeauthoryear{Others}{2013}]{Others2013}
% Others S., 2012, Journal of Interesting Stuff, 17, 198
% \end{thebibliography}

%%%%%%%%%%%%%%%%%%%%%%%%%%%%%%%%%%%%%%%%%%%%%%%%%%

%%%%%%%%%%%%%%%%% APPENDICES %%%%%%%%%%%%%%%%%%%%%

\clearpage
%\columnbreak
\appendix
\section{Additional details of the spectral analysis.}\label{app:spectral}

\renewcommand\arraystretch{1.2}
\begin{table*}
\caption{Same as Table~\ref{tab:T_NH_Xspec_spectral_graded} for Chandra GRADED mode observations, but without the calibration constant, i.e. $A=1$. Uncertainties correspond to the 68 per cent confidence intervals. 
}
\label{tab:T_NH_Xspec_spectral_graded_nocal}
\begin{tabular}{llcccccccccc}
\hline
&&&&\multicolumn{4}{c}{$N_\mathrm{H}$ variable} & \multicolumn{3}{c}{$N_{\mathrm{H},22}=1.656$}\\
 ObsID & Date & MJD & $t_{\mathrm{exp}}$ & $\log_{10} T_s$ & $N_{\mathrm{H}}$ & $\alpha$ & $\chi^2$ &$\log_{10} T_s$ & $\alpha$  & $\chi^2$  & $N_{\mathrm{bins}}$\\
&&&(ks)& (K) &  ($10^{22}~\mathrm{cm}^{-2}$)&&& (K)\\
\hline
114 		 & 2000 Jan 30 		 & 51573.4  & 50 & $6.248^{+0.003}_{-0.003} $ & $1.68^{+0.04}_{-0.04}$ & $0.38^{+0.05}_{-0.05}$ & 137  & $6.249^{+0.002}_{-0.002} $ &  $0.41^{+0.04}_{-0.05}$  & 138 &135 \\
1952 		 & 2002 Feb 6 		 & 	52311.3  &50 & $6.253^{+0.003}_{-0.003} $ & $1.73^{+0.04}_{-0.04}$ & $0.28^{+0.05}_{-0.04}$ & 133  & $6.252^{+0.002}_{-0.002} $ &  $0.33^{+0.05}_{-0.04}$  & 138 & 134 \\
5196 		 & 2004 Feb 8 		 & 	53043.7 &50  & $6.247^{+0.003}_{-0.003} $ & $1.64^{+0.04}_{-0.04}$ & $0.29^{+0.05}_{-0.05}$ & 107  & $6.250^{+0.002}_{-0.002} $ &  $0.30^{+0.04}_{-0.04}$  & 107 & 131\\
9117/9773 	 & 2007 Feb 5/8 	 & 	54439.9 &50  & $6.240^{+0.003}_{-0.003} $ & $1.66^{+0.04}_{-0.04}$ & $0.41^{+0.06}_{-0.06}$ & 131  & $6.242^{+0.002}_{-0.002} $ &  $0.42^{+0.06}_{-0.05}$  & 131 & 125\\
10935/12020  & 2009 Nov 2/3 	 & 	55137.9 &45  & $6.240^{+0.003}_{-0.003} $ & $1.67^{+0.04}_{-0.05}$ & $0.32^{+0.07}_{-0.06}$ & 119  & $6.242^{+0.002}_{-0.002} $ &  $0.34^{+0.06}_{-0.06}$  & 120 & 119\\
10936/13177  & 2010 Oct 31/Nov 2 &  55500.2 &49  & $6.234^{+0.003}_{-0.003} $ & $1.58^{+0.04}_{-0.05}$ & $0.35^{+0.07}_{-0.06}$ & 131  & $6.240^{+0.002}_{-0.002} $ &  $0.32^{+0.06}_{-0.05}$  & 132 & 123 \\
14229		 & 2012 May 15 		 & 56062.4  & 49 & $6.236^{+0.003}_{-0.003} $ & $1.68^{+0.05}_{-0.05}$ & $0.23^{+0.08}_{-0.08}$ & 122  & $6.237^{+0.002}_{-0.002} $ &  $0.27^{+0.07}_{-0.07}$  & 123 & 110\\
14480 		 & 2013 May 20 		 & 56432.6  & 49 & $6.239^{+0.003}_{-0.003} $ & $1.66^{+0.05}_{-0.05}$ & $0.25^{+0.07}_{-0.06}$ & 116  & $6.241^{+0.002}_{-0.002} $ &  $0.26^{+0.06}_{-0.06}$  & 117 & 119 \\
14481		 & 2014 May 12 		 & 56789.1  &49  & $6.239^{+0.003}_{-0.003} $ & $1.72^{+0.05}_{-0.05}$ & $0.13^{+0.07}_{-0.06}$ & 113  & $6.238^{+0.002}_{-0.002} $ &  $0.18^{+0.06}_{-0.06}$  & 116 & 113\\
14482 		 & 2015 Apr 30 		 & 57142.5  &49  & $6.234^{+0.003}_{-0.003} $ & $1.57^{+0.05}_{-0.05}$ & $0.23^{+0.08}_{-0.07}$ & 119  & $6.237^{+0.002}_{-0.002} $ &  $0.20^{+0.07}_{-0.06}$  & 121 & 114\\
19903/18344  & 2016 Oct 20/21 	 & 	57681.2 & 51 & $6.233^{+0.003}_{-0.003} $ & $1.55^{+0.05}_{-0.04}$ & $0.20^{+0.07}_{-0.07}$ &  98  & $6.238^{+0.002}_{-0.002} $ &  $0.16^{+0.07}_{-0.06}$  & 100 & 111\\
19604 		 & 2017 May 16 		 & 57889.7  & 50 & $6.240^{+0.003}_{-0.003} $ & $1.65^{+0.05}_{-0.05}$ & $0.12^{+0.07}_{-0.07}$ & 108  & $6.239^{+0.002}_{-0.002} $ &  $0.13^{+0.07}_{-0.06}$  & 108 & 110\\
19605 		 & 2018 May 15 		 & 58253.7  & 49 & $6.235^{+0.003}_{-0.003} $ & $1.54^{+0.05}_{-0.05}$ & $0.14^{+0.08}_{-0.07}$ &  91  & $6.239^{+0.002}_{-0.002} $ &  $0.10^{+0.07}_{-0.07}$  &  95 & 107\\
19606 		 & 2019 May 13 		 & 58616.5  & 49 & $6.235^{+0.003}_{-0.003} $ & $1.61^{+0.06}_{-0.05}$ & $0.17^{+0.08}_{-0.08}$ &  78  & $6.236^{+0.002}_{-0.002} $ &  $0.16^{+0.08}_{-0.08}$  &  78 & 102\\
\hline
\end{tabular}
\end{table*}
%--------------------------------------------------
Here we present additional figures and tables illustrating the results of the spectral fits described in Sec.~\ref{S:obs}.

In Sec.~\ref{S:obs:fixMR} we tabulated the results for the GRADED mode data in Table~\ref{tab:T_NH_Xspec_spectral_graded} where the calibration constant $A$ was set to the best-fit value. The unmodified fit, for which $A=1$, gives slightly shifted temperatures but a similar temperature decline. Therefore we put the corresponding Table~\ref{tab:T_NH_Xspec_spectral_graded_nocal} here, in the appendix.

\begin{figure*}
    \centering
    \includegraphics[width=0.47\textwidth]{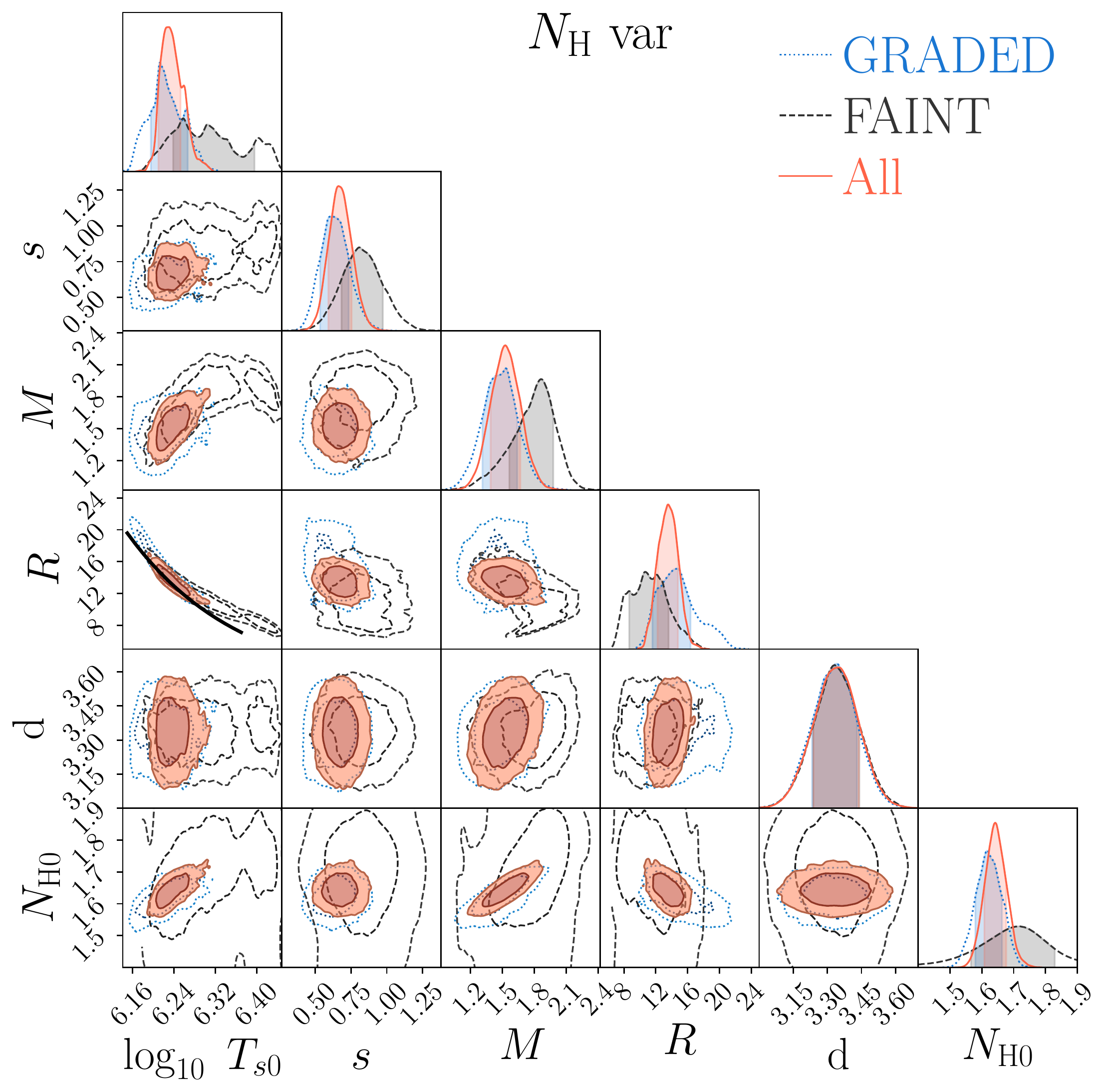}
    \includegraphics[width=0.47\textwidth]{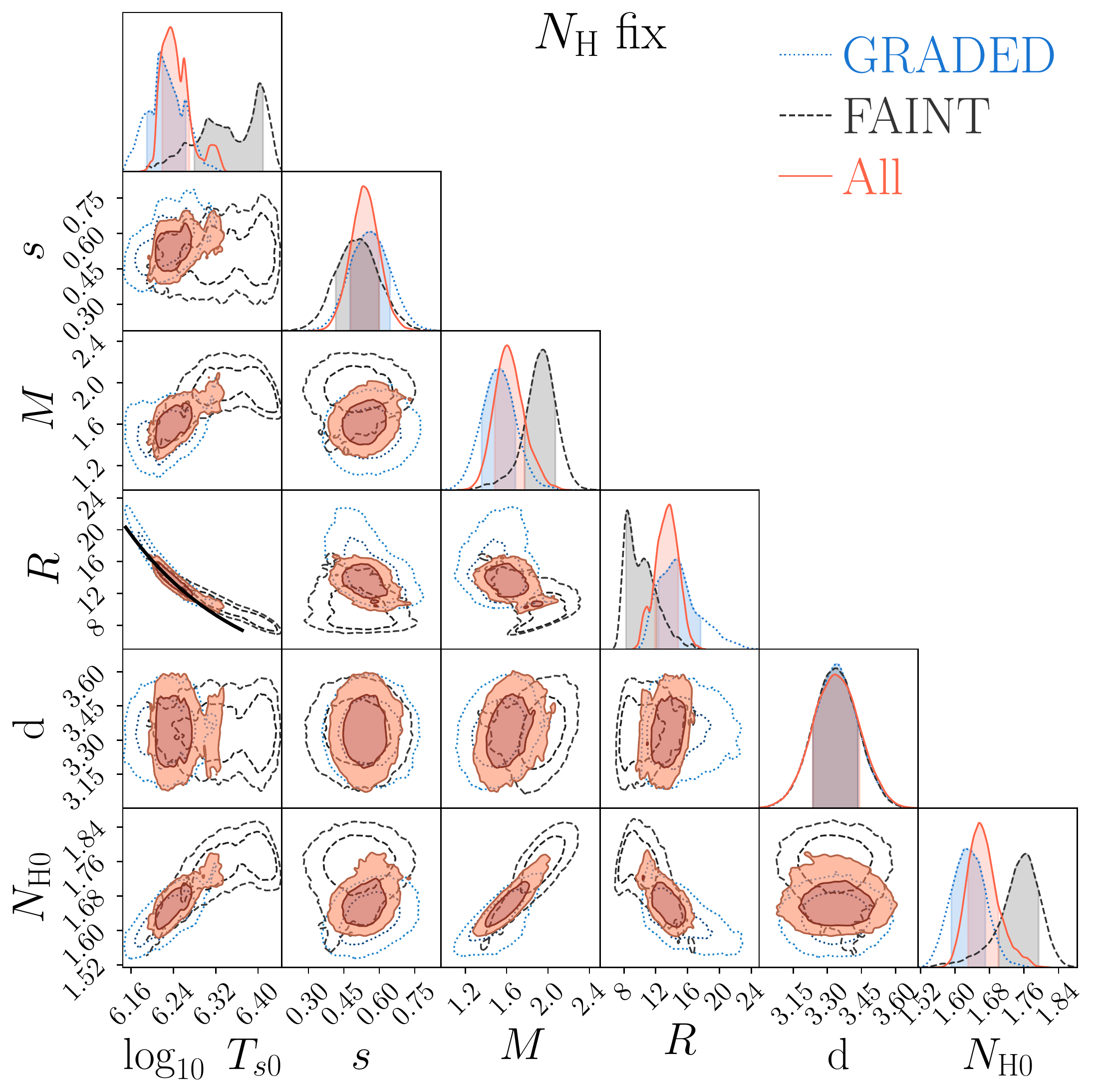}
    \caption{Comparison of the 1D and 2D posterior distributions obtained in different modes. The left and  right panels correspond to models with variable and fixed $N_{\mathrm{H}}$, respectively. Dotted and dashed contours correspond to GRADED and FAINT modes fitted alone, respectively, while filled solid  contours correspond to the joint fit using all data. The thick black lines show the $T_{s0}^4R^2=\mathrm{const}$ relation for the same parameters as in Fig.~\ref{fig:tri_spectral}, i.e. the ones for the joint fits.}
    \label{fig:tri_spectral_comp}
\end{figure*}

\renewcommand\arraystretch{1.2}
\begin{table*}
\caption{Posterior inferences on $N_{\mathrm{H}}$ and $\alpha$ for individual observations from the MCMC fits. Uncertainties correspond to 68 per cent highest posterior credible intervals. In some cases, 68 per cent upper or lower limits on the grade migration parameter ($0<\alpha<1$) are given.
}
\label{tab:spectral_nuisance}
\begin{tabular}{lllcccccc}
\hline
&&Fit mode: &\multicolumn{3}{c}{Combined}& \multicolumn{3}{c}{Single}\\
&&&\multicolumn{2}{c}{$N_\mathrm{H}$ var} & $N_\mathrm{H}$ fix &\multicolumn{2}{c}{$N_\mathrm{H}$ var} & $N_\mathrm{H}$ fix\\
Dataset & ObsId & MJD & $N_{\mathrm{H},22}$ & $\alpha$ & $\alpha$ & $N_{\mathrm{H},22}$ & $\alpha$ & $\alpha$\\
\hline
         GRADED & 114 		 & 51573.4    &  $1.668^{+0.042}_{-0.033}$ & $0.38^{+0.04}_{-0.04}$ & $0.41^{+0.04}_{-0.04}$ & $1.639^{+0.047}_{-0.036}$ & $0.40^{+0.04}_{-0.06}$ & $0.40^{+0.06}_{-0.05}$ \\ 
		 GRADED & 1952 		 & 52311.3    &  $1.660^{+0.038}_{-0.037}$ &$0.36^{+0.04}_{-0.04}$ & $0.38^{+0.04}_{-0.04}$  & $1.637^{+0.044}_{-0.042}$ & $0.36^{+0.05}_{-0.05}$ & $0.37^{+0.05}_{-0.05}$ \\ 
		 GRADED & 5196 		 & 53043.7    &  $1.623^{+0.045}_{-0.031}$ &$0.33^{+0.04}_{-0.05}$ & $0.35^{+0.05}_{-0.04}$  & $1.608^{+0.045}_{-0.047}$ & $0.32^{+0.05}_{-0.05}$ & $0.35^{+0.05}_{-0.06}$ \\ 
		 GRADED & 9117/9773 	 & 54439.9    &  $1.669^{+0.037}_{-0.041}$ &$0.39^{+0.05}_{-0.05}$ & $0.41^{+0.04}_{-0.05}$  & $1.640^{+0.045}_{-0.043}$ & $0.38^{+0.07}_{-0.04}$ & $0.39^{+0.06}_{-0.06}$ \\ 
		 GRADED & 10935/12020 & 55137.9    &  $1.658^{+0.043}_{-0.037}$ &$0.32^{+0.06}_{-0.05}$ & $0.33^{+0.05}_{-0.05}$  & $1.627^{+0.055}_{-0.035}$ & $0.32^{+0.06}_{-0.06}$ & $0.32^{+0.07}_{-0.05}$ \\ 
		 GRADED & 10936/13177 & 55500.2    &  $1.648^{+0.038}_{-0.042}$ &$0.29^{+0.05}_{-0.05}$ & $0.29^{+0.05}_{-0.05}$  & $1.621^{+0.051}_{-0.041}$ & $0.28^{+0.06}_{-0.06}$ & $0.29^{+0.06}_{-0.06}$ \\ 
		 GRADED & 14229		 & 56062.4    &  $1.676^{+0.048}_{-0.035}$ &$0.22^{+0.06}_{-0.07}$ & $0.21^{+0.07}_{-0.06}$  & $1.659^{+0.040}_{-0.049}$ & $0.23^{+0.06}_{-0.09}$ & $0.21^{+0.08}_{-0.07}$ \\ 
		 GRADED & 14480 		 & 56432.6    &  $1.645^{+0.036}_{-0.044}$ &$0.29^{+0.06}_{-0.06}$ & $0.28^{+0.07}_{-0.04}$  & $1.617^{+0.046}_{-0.046}$ 		 & $0.29^{+0.07}_{-0.06}$ & $0.28^{+0.07}_{-0.06}$ \\ 
		 GRADED & 14481		 & 56789.1    &  $1.666^{+0.044}_{-0.036}$ &$0.18^{+0.06}_{-0.06}$ & $0.17^{+0.06}_{-0.05}$  & $1.643^{+0.046}_{-0.044}$ & $0.19^{+0.05}_{-0.08}$ & $0.18^{+0.06}_{-0.07}$ \\ 
		 GRADED & 14482 		 & 57142.5    &  $1.635^{+0.043}_{-0.039}$ &$0.18^{+0.06}_{-0.06}$ & $0.19^{+0.06}_{-0.06}$  & $1.620^{+0.046}_{-0.049}$ & $0.19^{+0.07}_{-0.07}$          & $0.18^{+0.07}_{-0.06}$ \\ 
		 GRADED & 19903/18344 & 	57681.2   &  $1.617^{+0.043}_{-0.043}$ 		  &$0.17^{+0.06}_{-0.06}$ & $0.17^{+0.06}_{-0.07}$  & $1.602^{+0.045}_{-0.058}$ & $0.18^{+0.06}_{-0.07}$ & $0.15^{+0.09}_{-0.05}$ \\ 
		 GRADED & 19604 		 & 57889.7    &  $1.627^{+0.043}_{-0.040}$ &$0.18^{+0.07}_{-0.06}$ & $0.18^{+0.06}_{-0.07}$  & $1.616^{+0.039}_{-0.060}$ & $0.17^{+0.08}_{-0.07}$ & $0.18^{+0.07}_{-0.08}$ \\ 
		 GRADED & 19605 		 & 58253.7    &  $1.587^{+0.052}_{-0.042}$ &$0.14^{+0.08}_{-0.06}$ & $0.15^{+0.07}_{-0.07}$  & $1.581^{+0.051}_{-0.062}$ & $0.14^{+0.08}_{-0.07}$ & $0.13^{+0.09}_{-0.07}$ \\ 
		 GRADED & 19606 		 & 58616.5    &  $1.629^{+0.044}_{-0.045}$ &$0.12^{+0.07}_{-0.08}$ & $0.18^{+0.07}_{-0.08}$  & $1.610^{+0.051}_{-0.053}$ & $0.17^{+0.10}_{-0.06}$ & $0.19^{+0.08}_{-0.09}$ \\ 
		 FAINT & 6690 		 &   54021    &  $1.680^{+0.037}_{-0.036}$ &$0.53^{+0.23}_{-0.35}$ & $0.51^{+0.32}_{-0.25}$ 		  & $1.787^{+0.055}_{-0.067}$ & $0.37^{+0.39}_{-0.20}$ & $>0.53$ \\ 
		 FAINT & 13783		 &   56052    &  $1.650^{+0.043}_{-0.032}$ &$0.69^{+0.28}_{-0.29}$ & $0.62^{+0.36}_{-0.22}$ 		  & $1.756^{+0.048}_{-0.076}$ & $>0.46$ & $>0.56$ \\ 
		 FAINT & 16946/17639 &   57141.2  &  $1.590^{+0.036}_{-0.041}$ &$0.25^{+0.28}_{-0.24}$ & $0.31^{+0.24}_{-0.28}$ 		  & $1.671^{+0.056}_{-0.071}$ & $0.38^{+0.28}_{-0.30}$ & $0.55^{+0.24}_{-0.35}$ \\ 
		 FAINT & 22426/23248 &   58981.1  &  $1.612^{+0.044}_{-0.042}$ &$0.00^{+0.44}_{-0.00}$ & $0.008^{+0.371}_{-0.005} $  & $1.689^{+0.055}_{-0.089}$ & $<0.54$ & $<0.43$ \\
\hline
\end{tabular}
\end{table*}

Comparison of the posterior distributions for different modes corresponding to  Table~\ref{tab:spectral_params} is presented in Fig.~\ref{fig:tri_spectral_comp}. It is clear that the parameter inferences for different modes are broadly consistent. Notice a wide posterior for the $N_{\mathrm{H}0}$ hyperparameter for the FAINT mode in the left panel in Fig.~\ref{fig:tri_spectral_comp}, and corresponding larger uncertainties for this parameter in Table~\ref{tab:spectral_params}, in comparison to two other modes. This is because the FAINT mode dataset contains  only four epochs, so the hyperparameters $N_{\mathrm{H}0}$ and $\sigma_{N_{\mathrm{H}}}$ of the hyperprior Gaussian distribution are not well-constrained. Of course, the individual $N_{\mathrm{H},i}$ values for each of the four observations in FAINT mode are well-constrained by the spectral data. The latter applies to the GRADED mode $N_{\mathrm{H},i}$ values as well.
Individual values of $N_{\mathrm{H},i}$ and $\alpha_i$ based on the MCMC fits are summarised in Table~\ref{tab:spectral_nuisance}. The ``Combined'' column block corresponds to the results from the joint spectral fit, while the ``Single'' column block corresponds to the results from the GRADED of FAINT dataset fitted alone. In the latter case the graded migration parameters $\alpha_i$ are not strongly constrained by the fit due to lower pileup fraction.

\begin{figure*}
    \centering
    \includegraphics[width=0.45\textwidth]{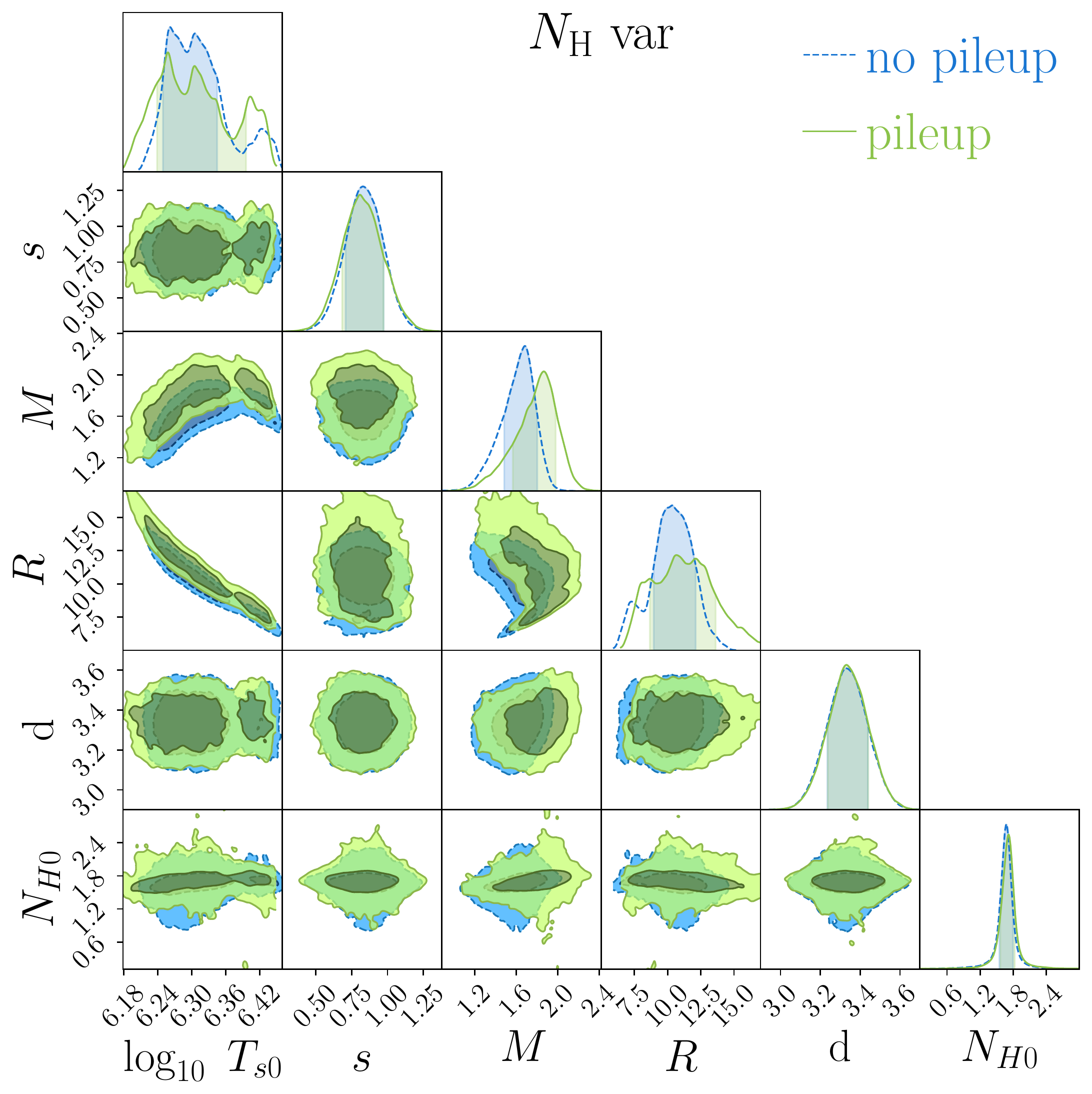}
    \includegraphics[width=0.45\textwidth]{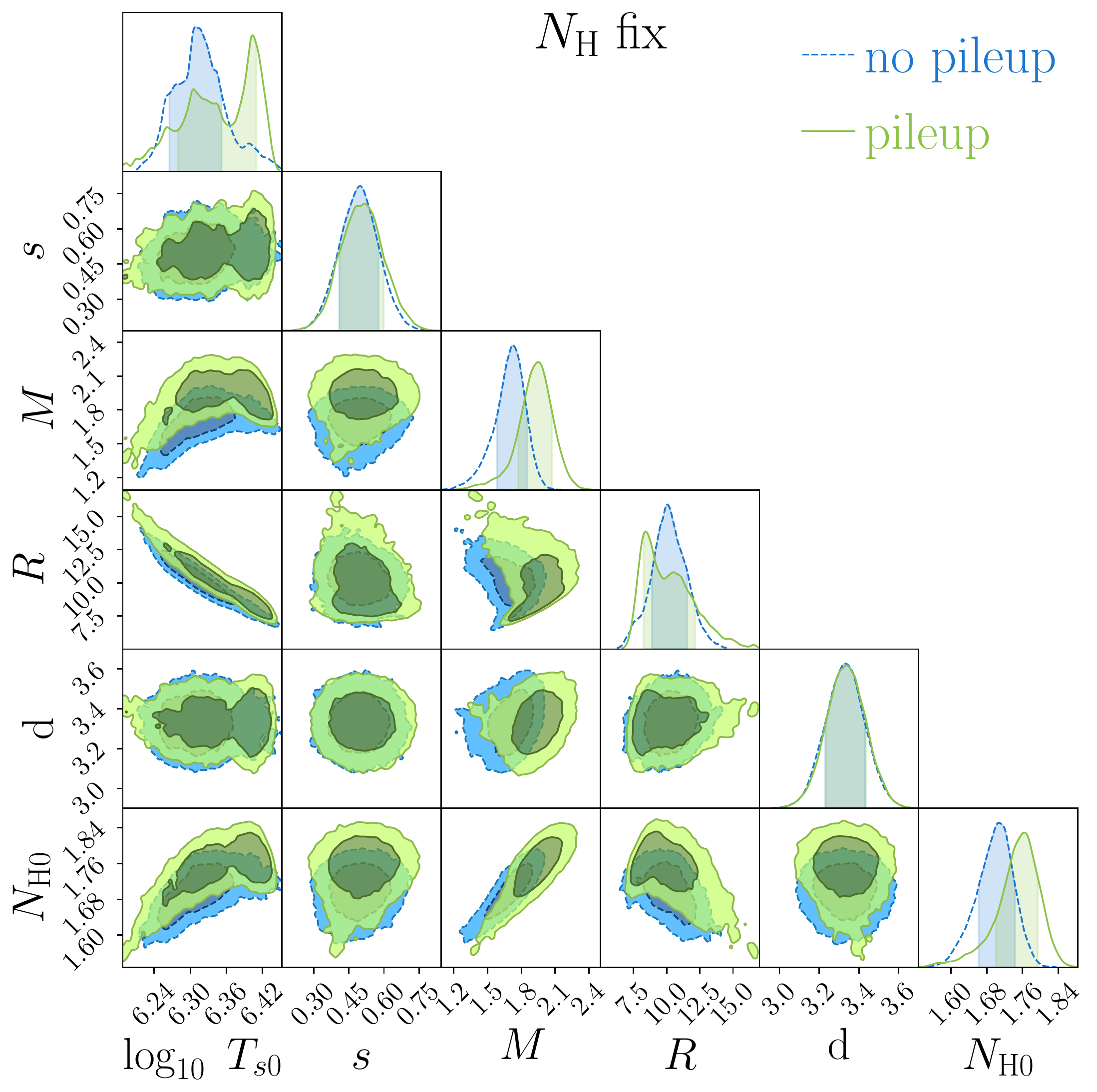}
    \caption{Comparison of the 1D and 2D posterior distributions obtained in the FAINT mode with and without accounting for pileup. The left and  right panels correspond to models with variable and fixed $N_{\mathrm{H}}$, respectively.}
    \label{fig:tri_faint_pileup_comp}
\end{figure*}
\begin{figure*}
    \centering
    \includegraphics[width=0.45\textwidth]{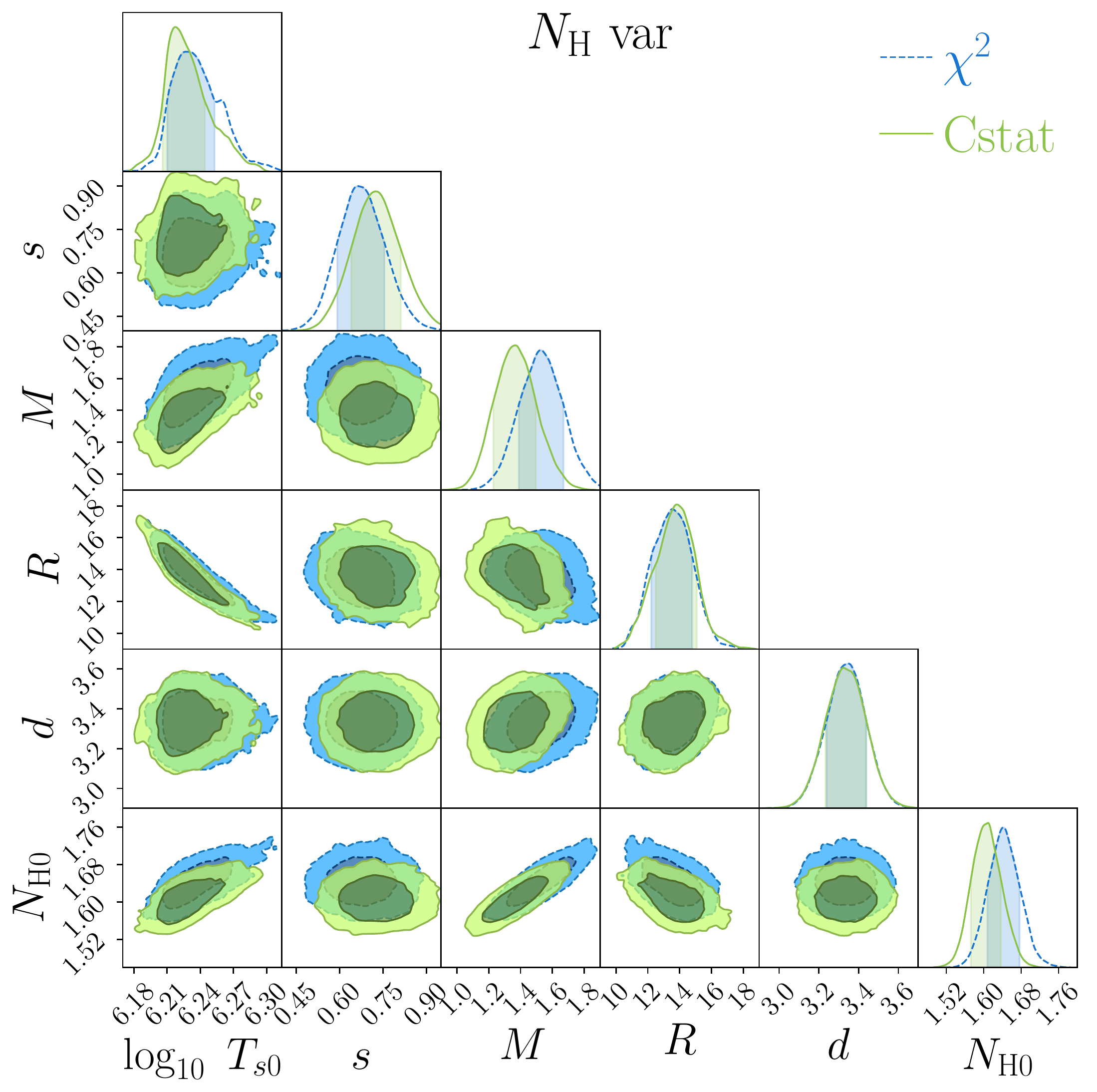}
    \includegraphics[width=0.45\textwidth]{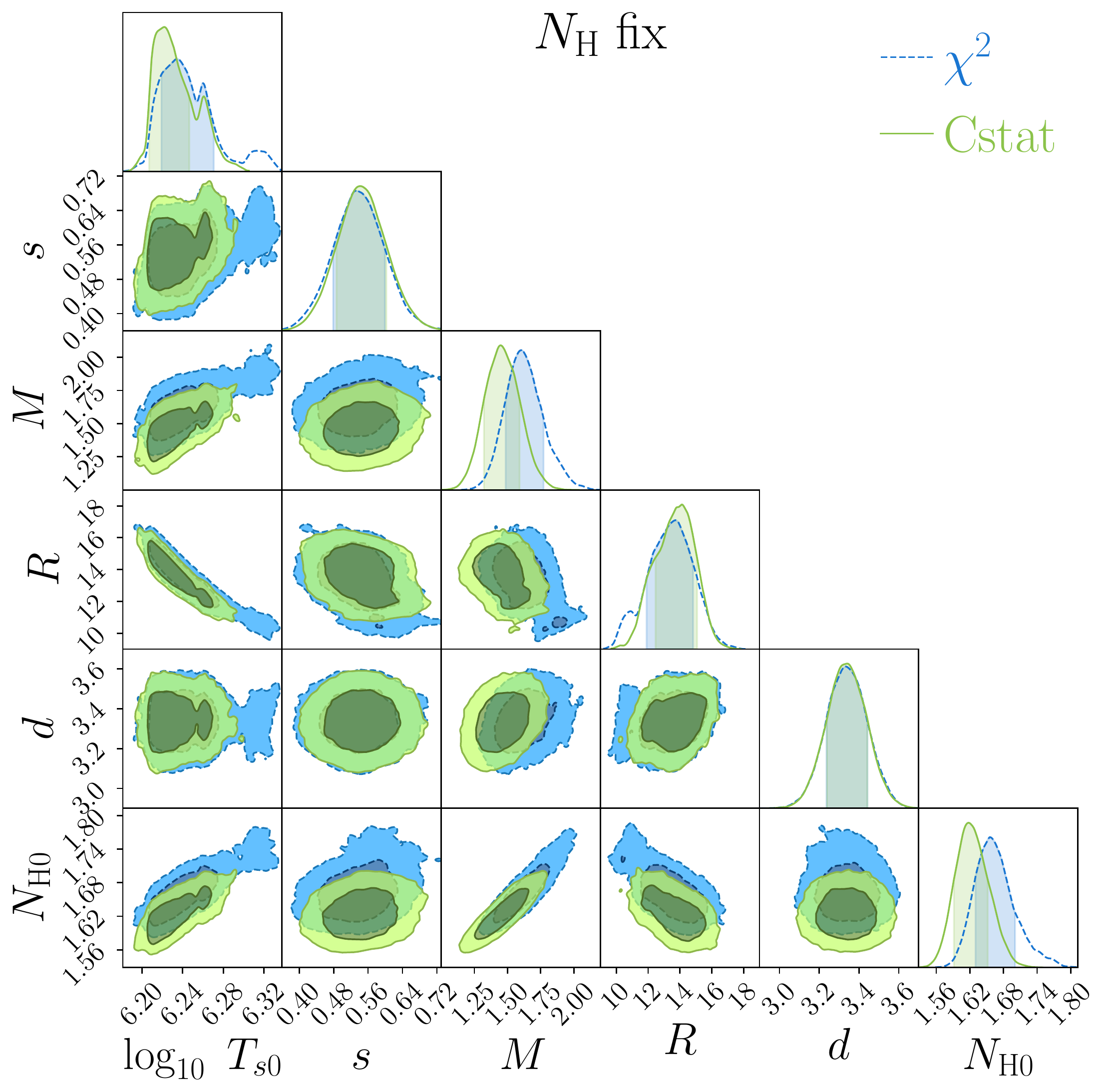}
    \caption{Comparison of the 1D and 2D posterior distributions for the joint fit under the $\chi^2$ and Cstat statistics.The left and  right panels correspond to models with variable and fixed $N_{\mathrm{H}}$, respectively.}
    \label{fig:tri_chi_cstat_comp}
\end{figure*}

%\new{
Inspecting Fig.~\ref{fig:tri_spectral_comp} one can notice broader and more `wiggled' posterior distribution for the temperature in the FAINT mode in comparison to GRADED mode and joint fit results. This is especially clear for the fixed-$N_{\mathrm{H}}$ models (Fig.~\ref{fig:tri_spectral_comp}, right panel).  The wiggled structure is mainly related to the details of the \textsc{XSPEC} implementation of the the \textsc{nsx} model (discretization and interpolation scheme). In addition, inclusion of the pileup component to the model for the FAINT mode data shifts and broadens the posterior distributions for some of the parameters slightly. This is illustrated in Fig.~\ref{fig:tri_faint_pileup_comp}. One observes that the inclusion of the pileup mainly affects NS mass inference, shifting $M$ towards higher values, less affecting  thermal evolution (see also Section~\ref{S:obs:fixMR}). These results support the conclusion that even a modest amount of pileup can affect $M-R$ inferences based on  NS atmospheric models and should be taken seriously \citep[e.g.,][]{Bogdanov2016ApJ}.
%}

We also checked how the choice of the fit statistics (or likelihood) affects our results. We illustrate this for the joint fit using the Poisson likelihood \cite{Cash1979ApJ} (statistics Cstat in \textsc{Xspec}) for the spectra binned to ensure at least 1 count per energy bin. The results are given in Table~\ref{tab:spectral_params_cstat} and compared with those obtained with $\chi^2$ statistics in Fig.~\ref{fig:tri_chi_cstat_comp}. The results are generally compatible with C-statistics ones giving about $1\sigma$ smaller masses. However, in case of the pileup where only one-photon events are counted and attempt is taken to account for false one-photon events, it is not clear if the Poisson distribution is better approximation for the true likelihood than the normal (Gaussian) approximation (for large enough binning). We can therefore consider the differences for $M$ inference in Tables~\ref{tab:spectral_params_cstat} and \ref{tab:spectral_params} as an estimate for a systematic error.

\begin{figure*}
\includegraphics[width=0.45\textwidth]{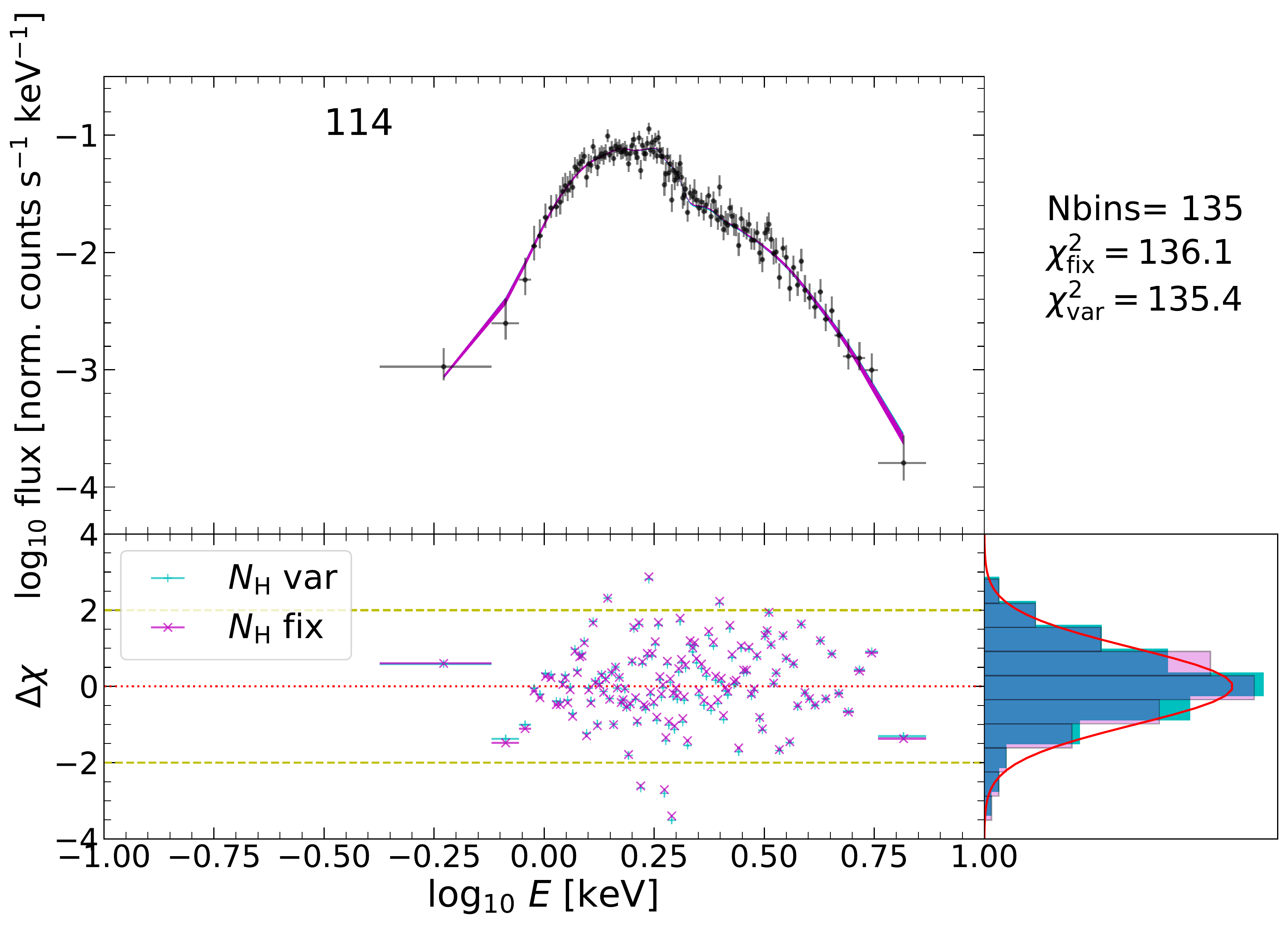}
\includegraphics[width=0.45\textwidth]{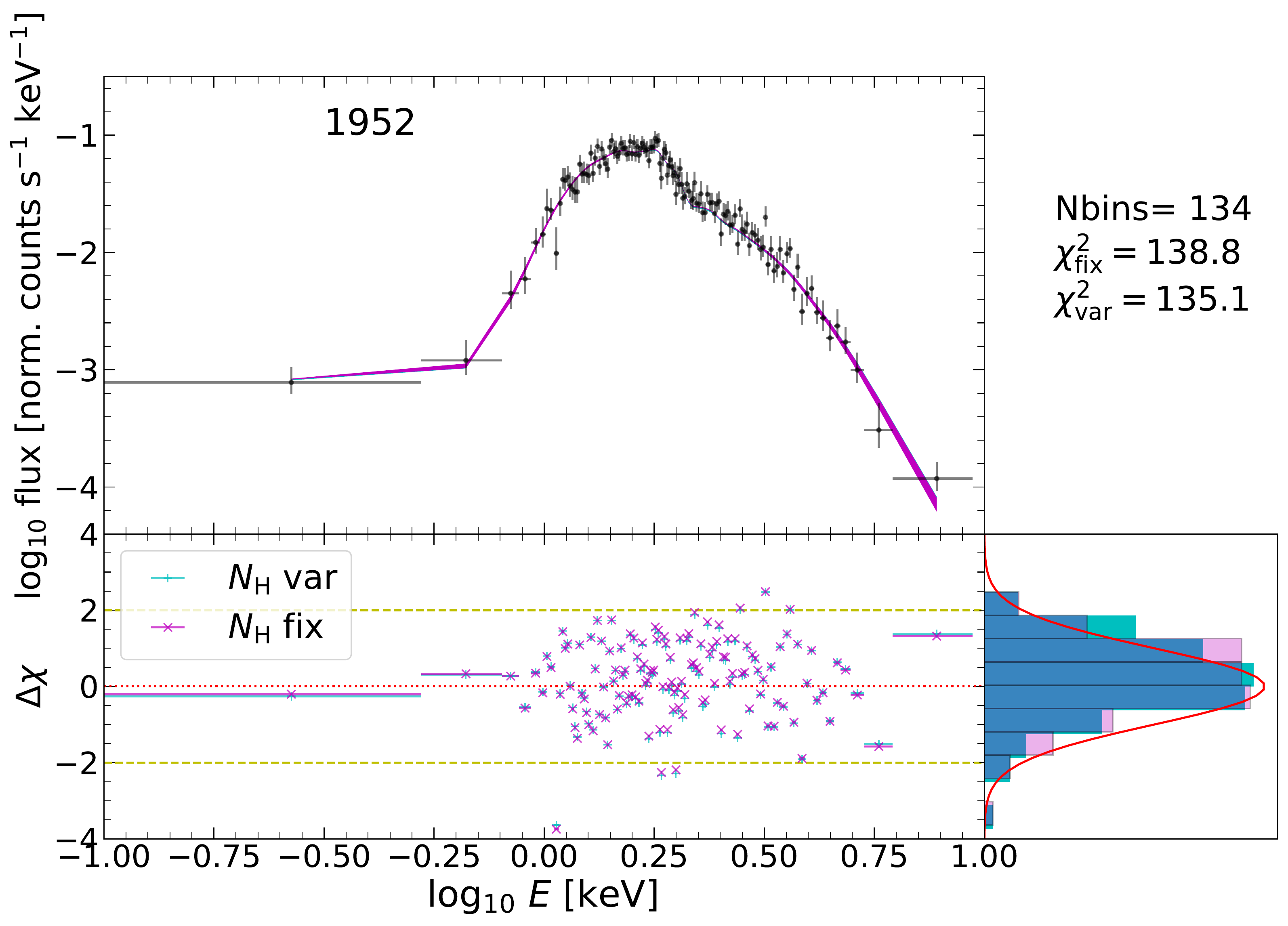}
\includegraphics[width=0.45\textwidth]{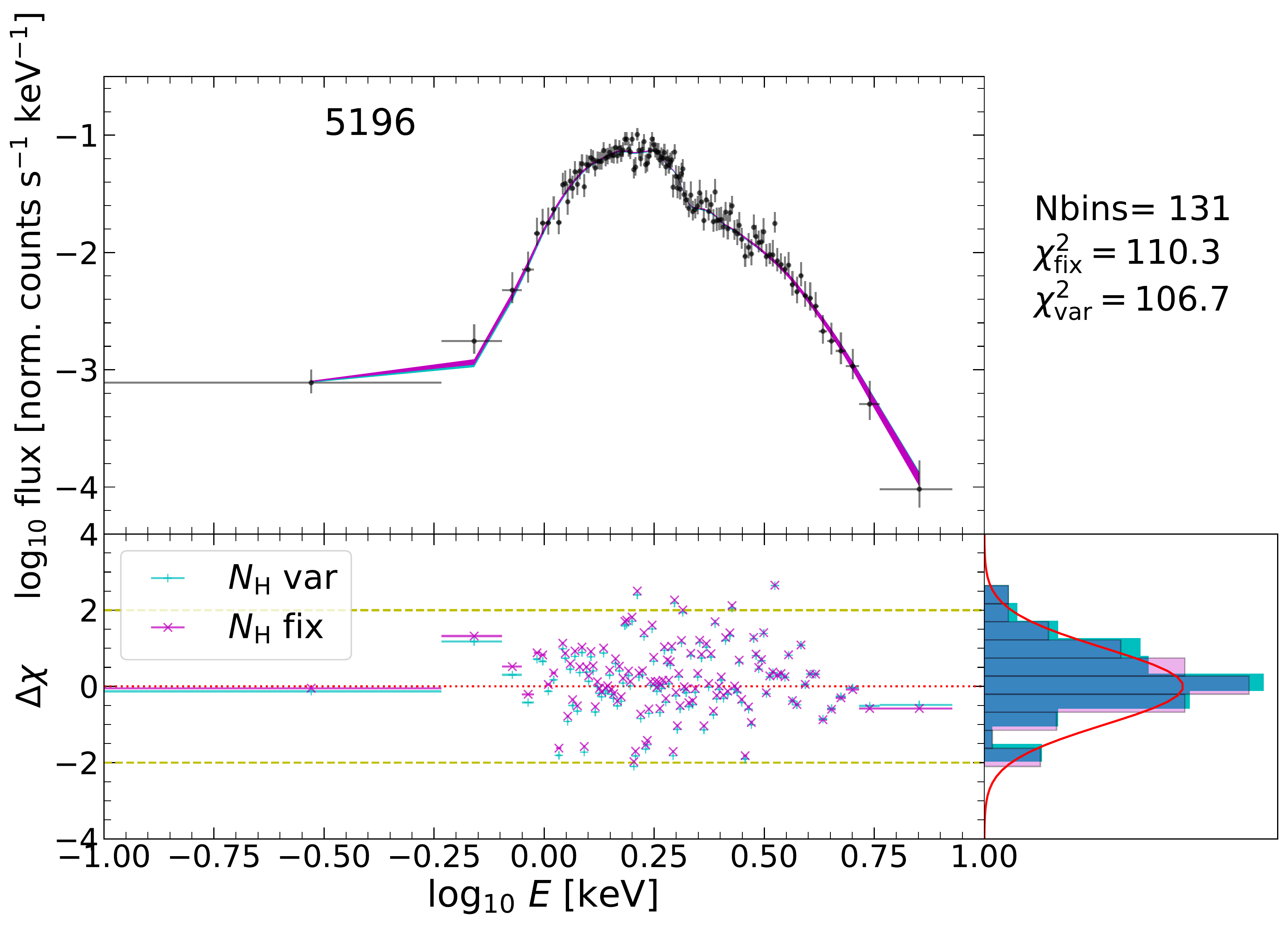}
\includegraphics[width=0.45\textwidth]{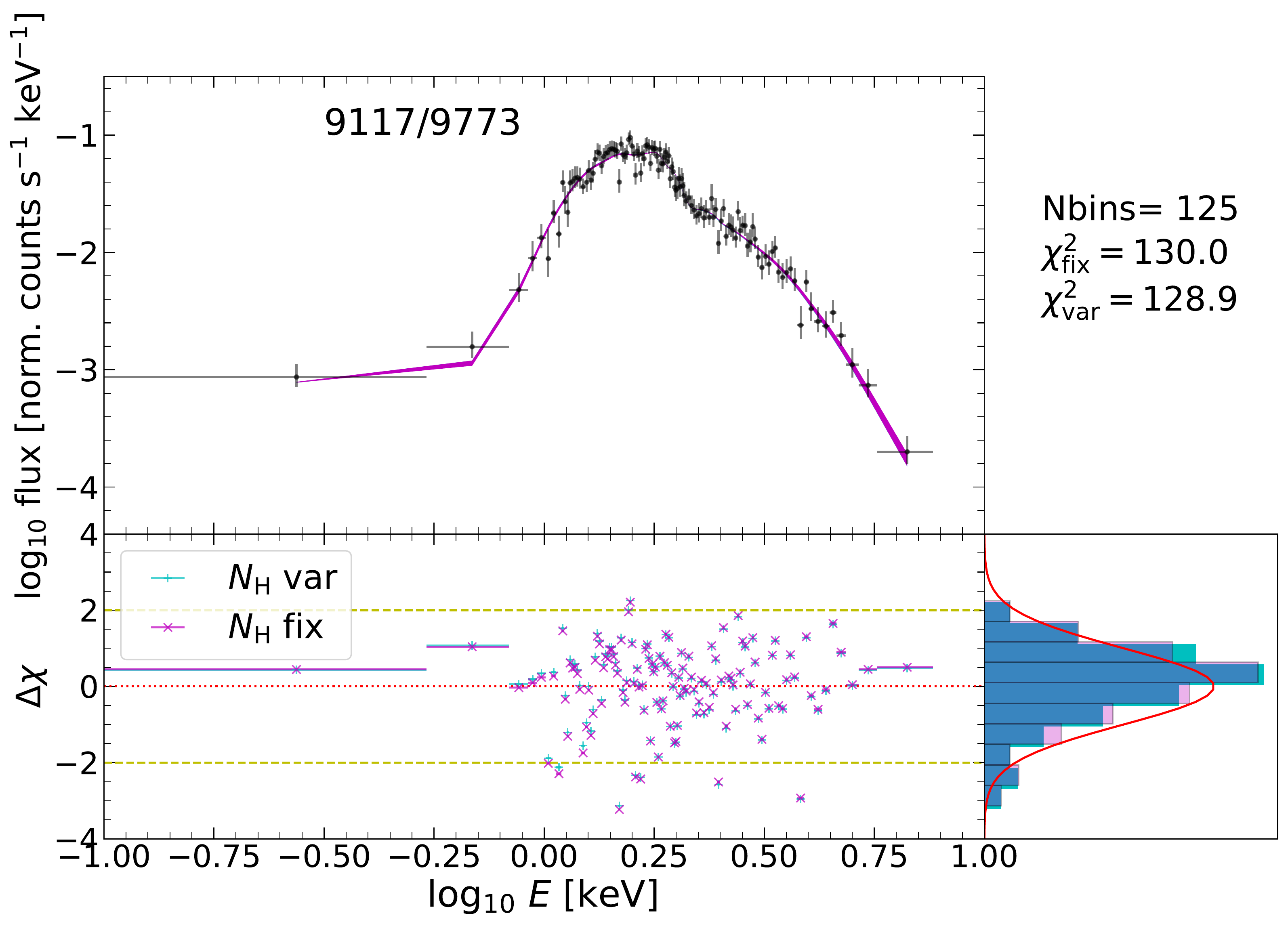}
\caption{Comparison of the spectral data taken in the GRADED mode with the joint fit spectral model. Each panel corresponds to an individual spectrum, which ObsID is indicated in the plot. For each spectrum, the upper panels show the normalised count flux in energy bins compared to the 68 per cent credible intervals of the model predictions. The models with variable and fixed $N_{\mathrm{H}}$ are shown with cyan and magenta colours, but they are practically indistinguishable in the plot. Lower panels show the standardised residuals as described in the text, and their distribution is compared to the standard normal distribution in the lower right panels. For each spectra the number of energy bins $N_{\rm bins}$  and the $\chi^2$ values for fixed and variable $N_{\mathrm{H}}$ models are also indicated.}\label{fig:spec_graded_1} 
\end{figure*}

\begin{figure*}
\includegraphics[width=0.45\textwidth]{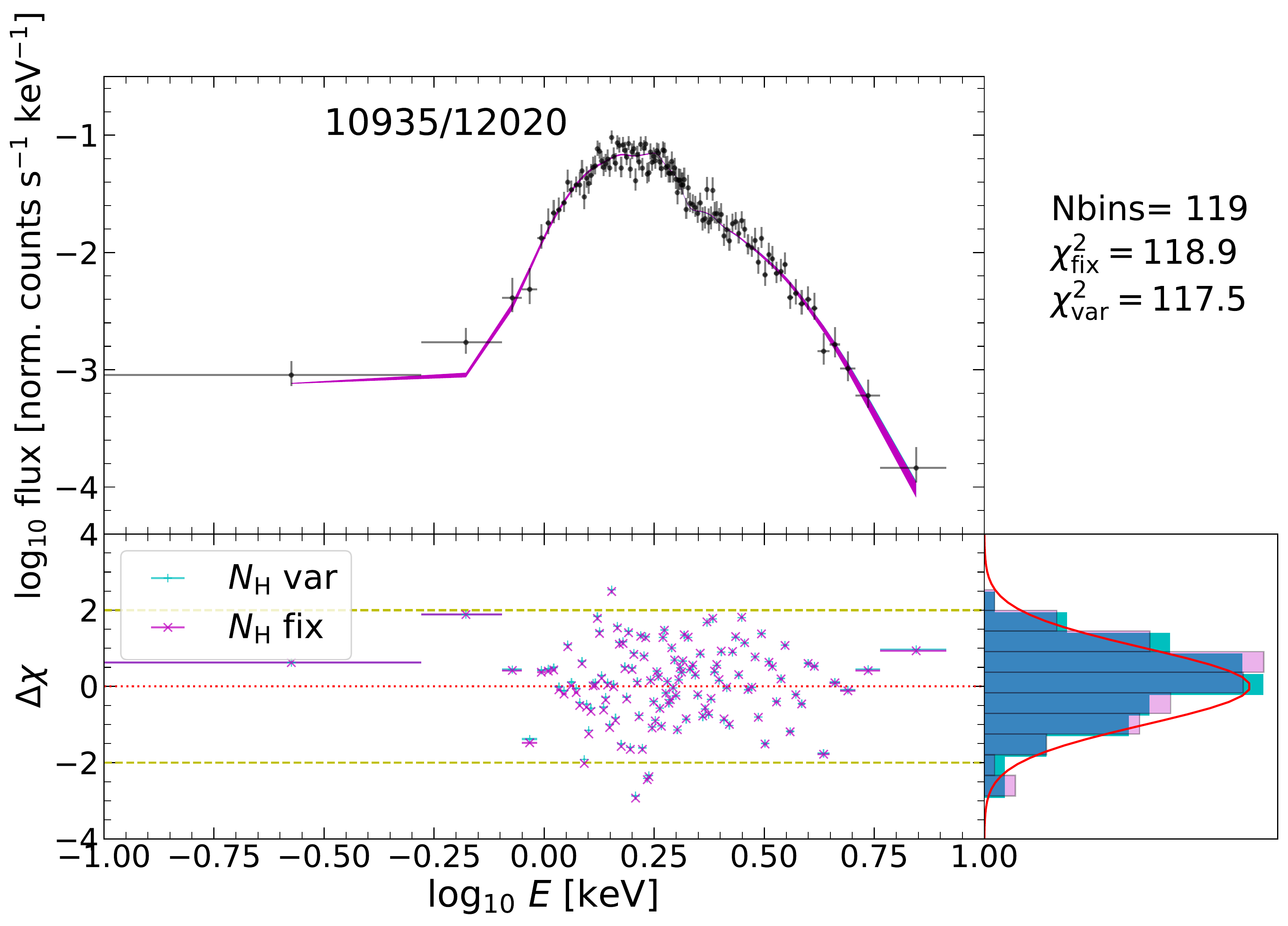}
\includegraphics[width=0.45\textwidth]{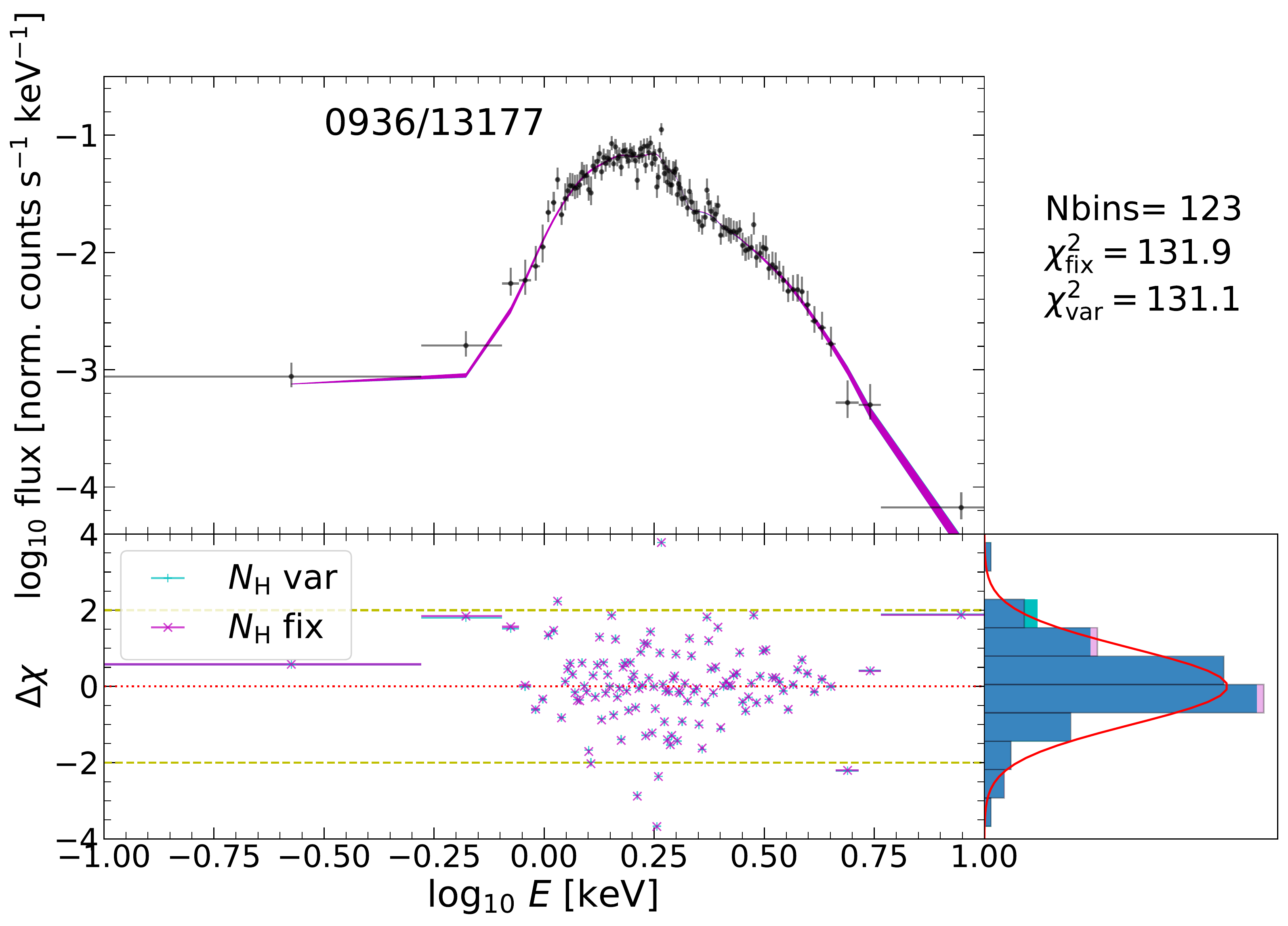}
\includegraphics[width=0.45\textwidth]{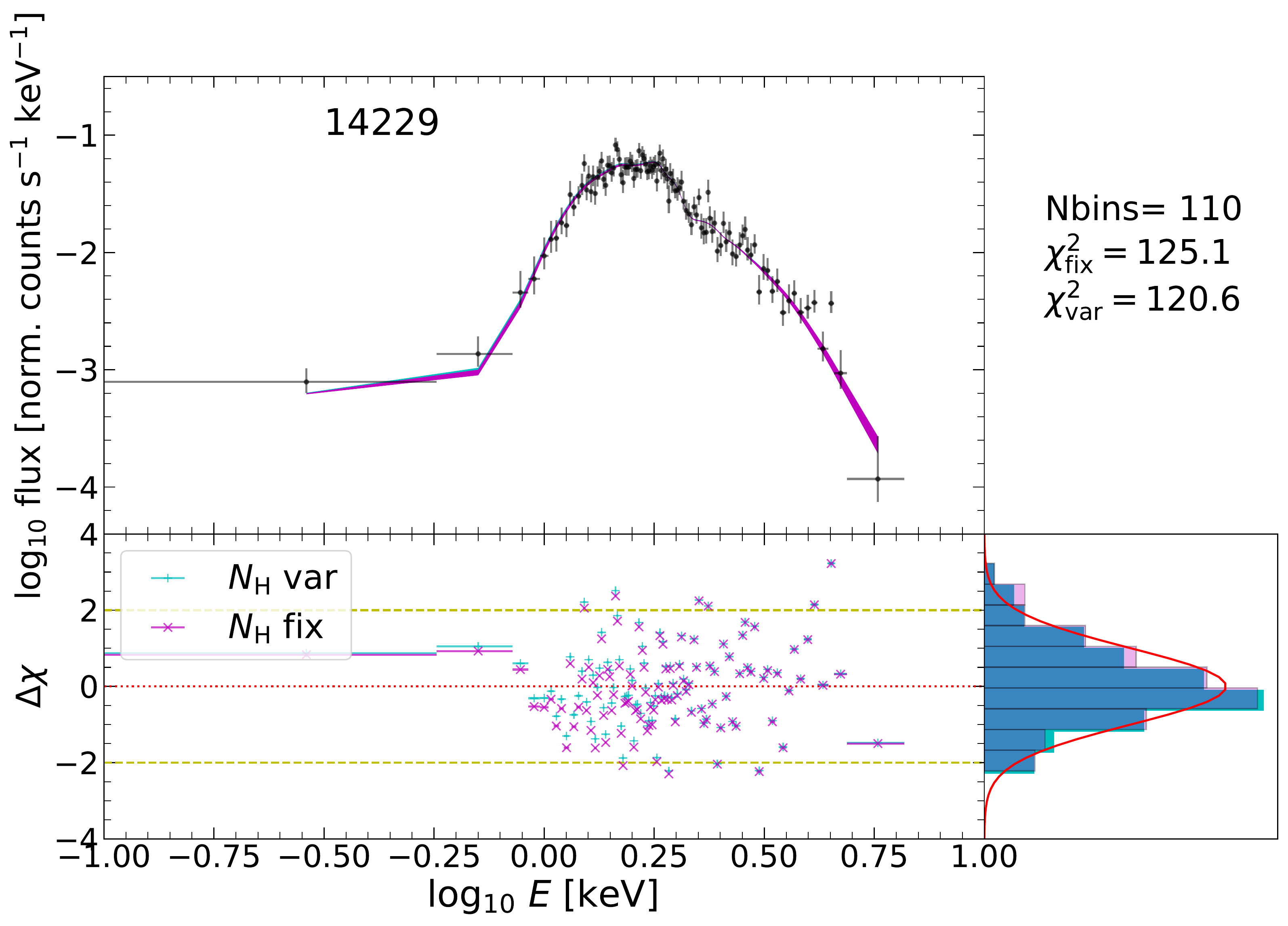}
\includegraphics[width=0.45\textwidth]{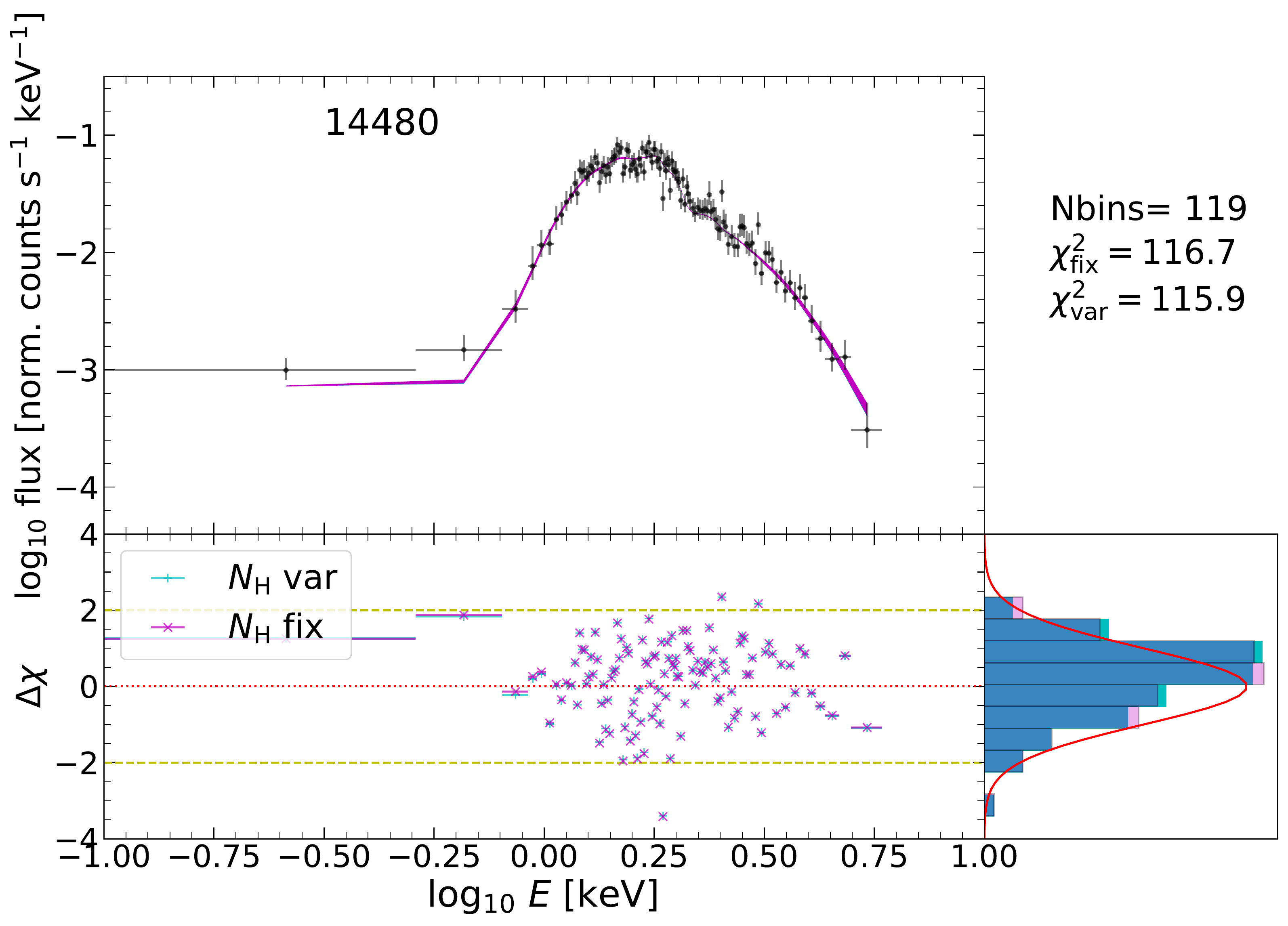}
\includegraphics[width=0.45\textwidth]{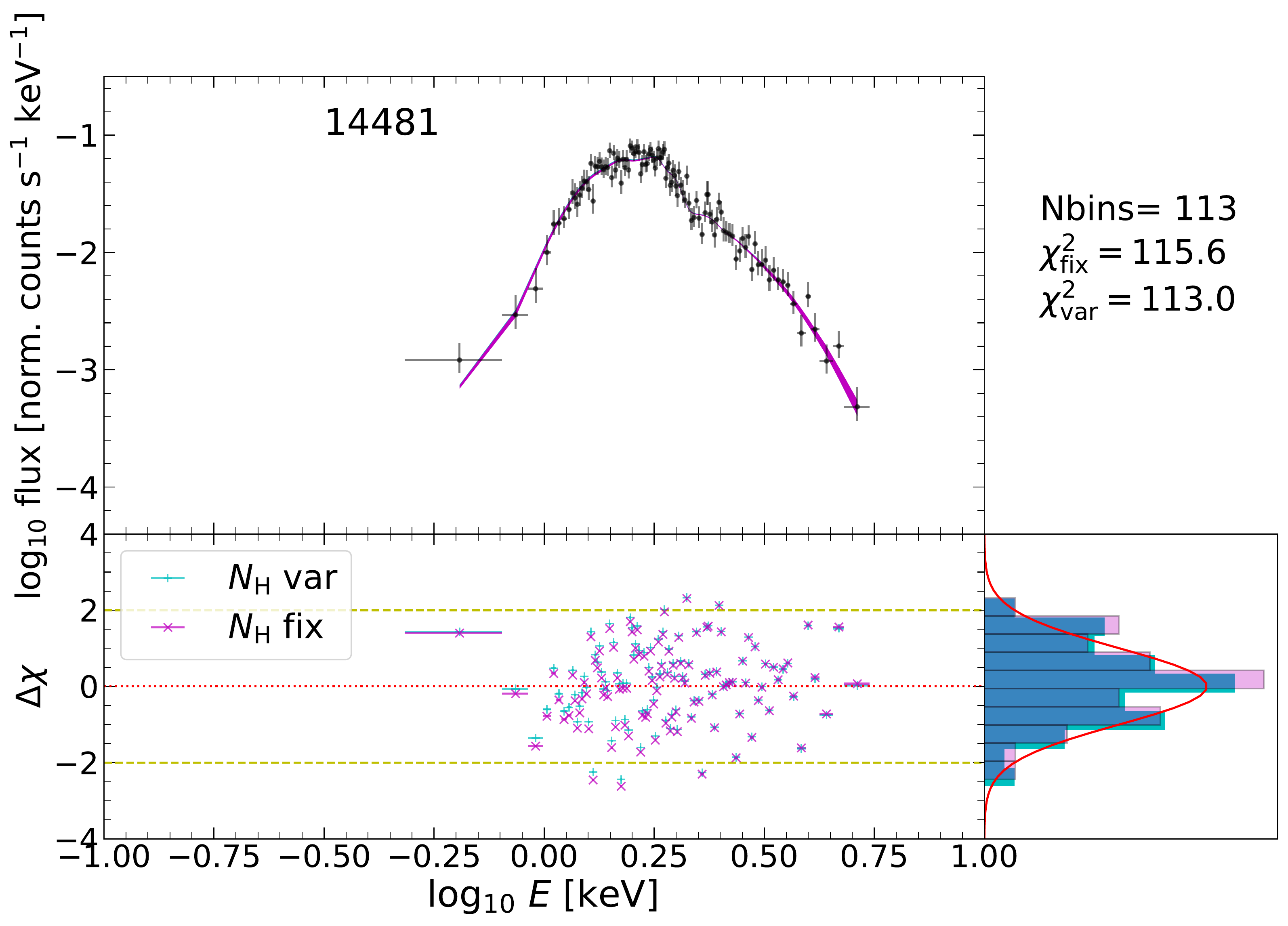}
\includegraphics[width=0.45\textwidth]{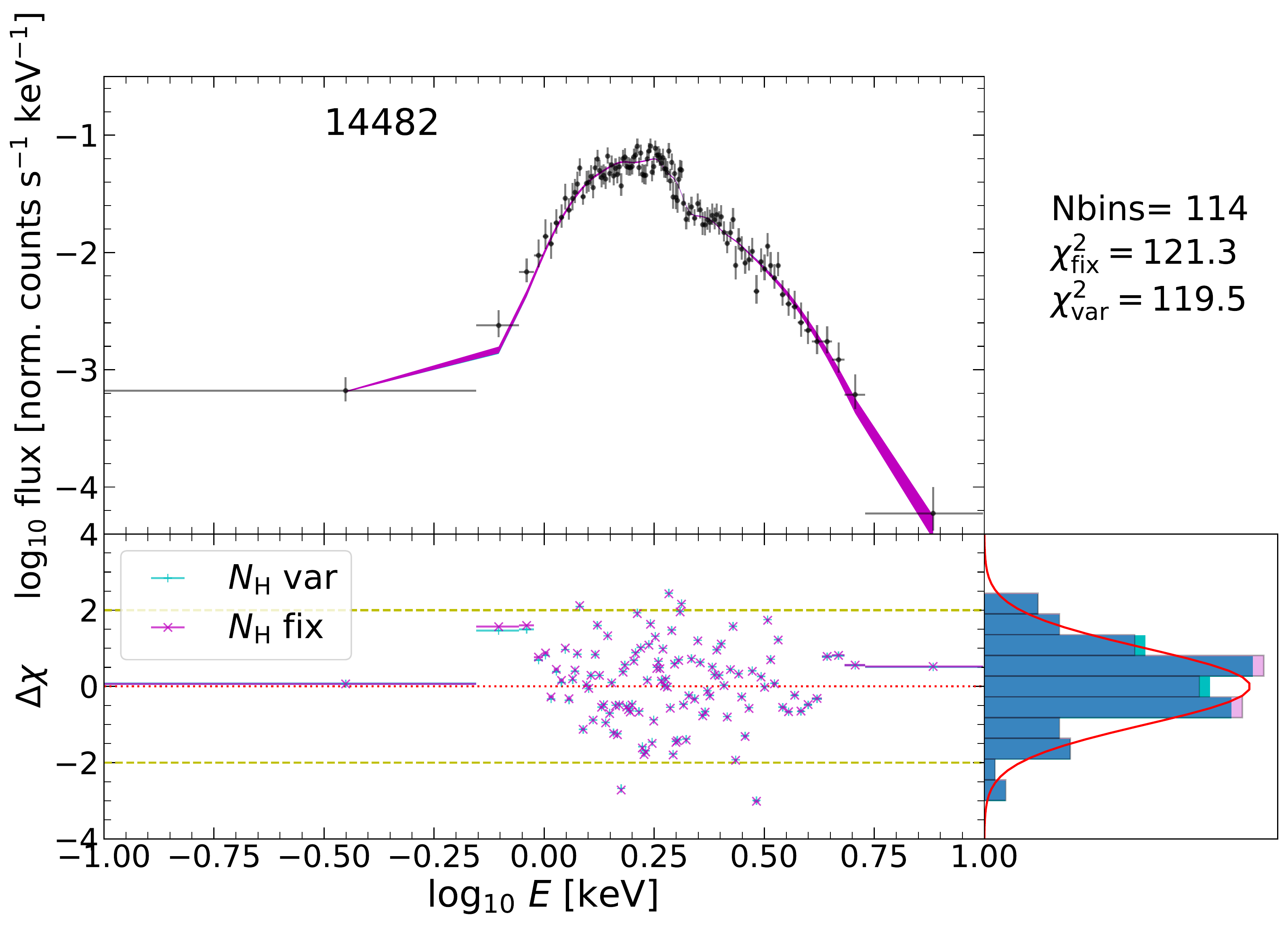}
\includegraphics[width=0.45\textwidth]{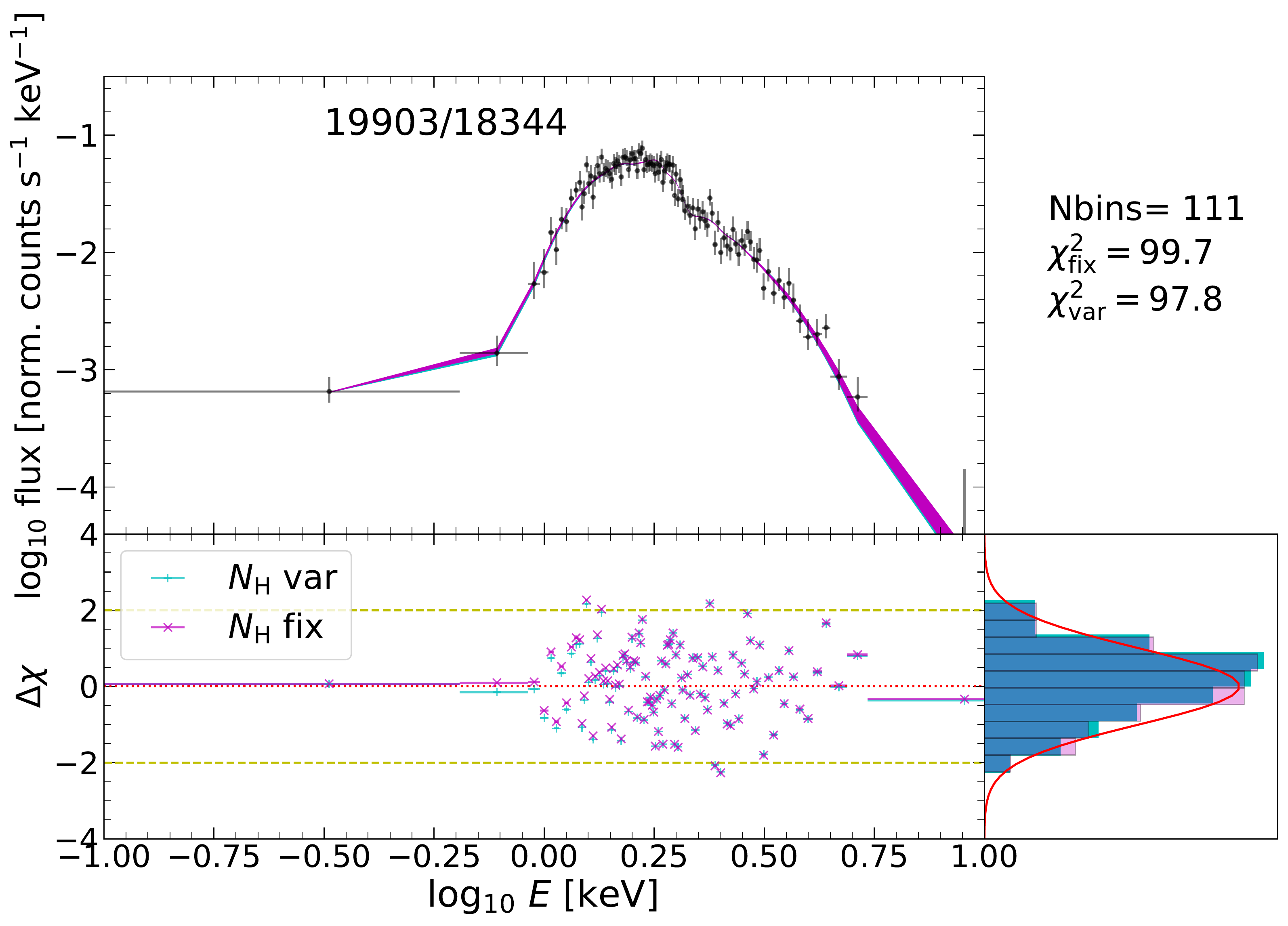}
\includegraphics[width=0.45\textwidth]{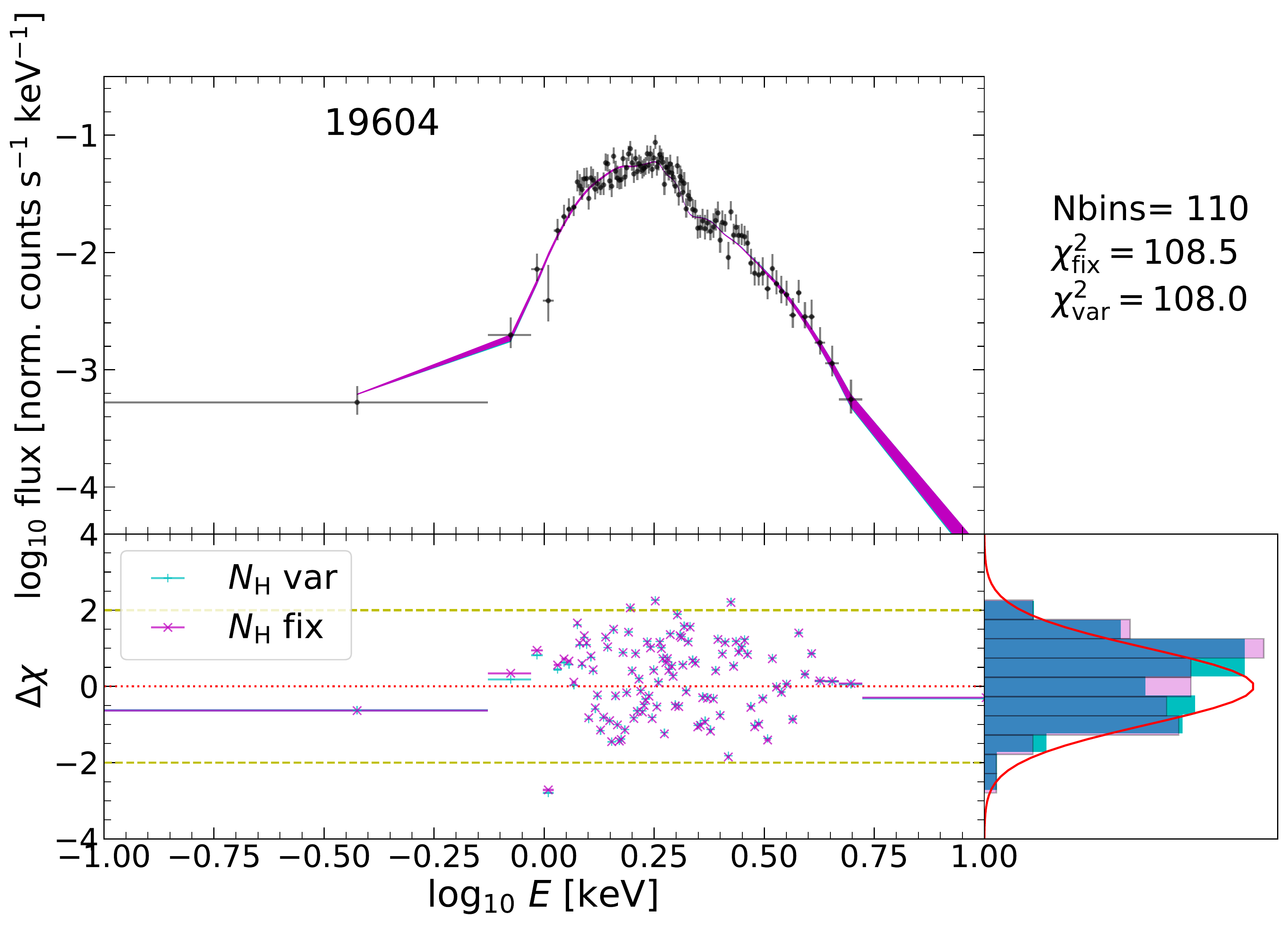}
\caption{Comparison of the spectral data taken in the GRADED mode with the joint fit spectral model. Continuation of Fig.~\ref{fig:spec_graded_1}. 
}\label{fig:spec_graded_1a} 
\end{figure*}

\begin{figure*}
\includegraphics[width=0.45\textwidth]{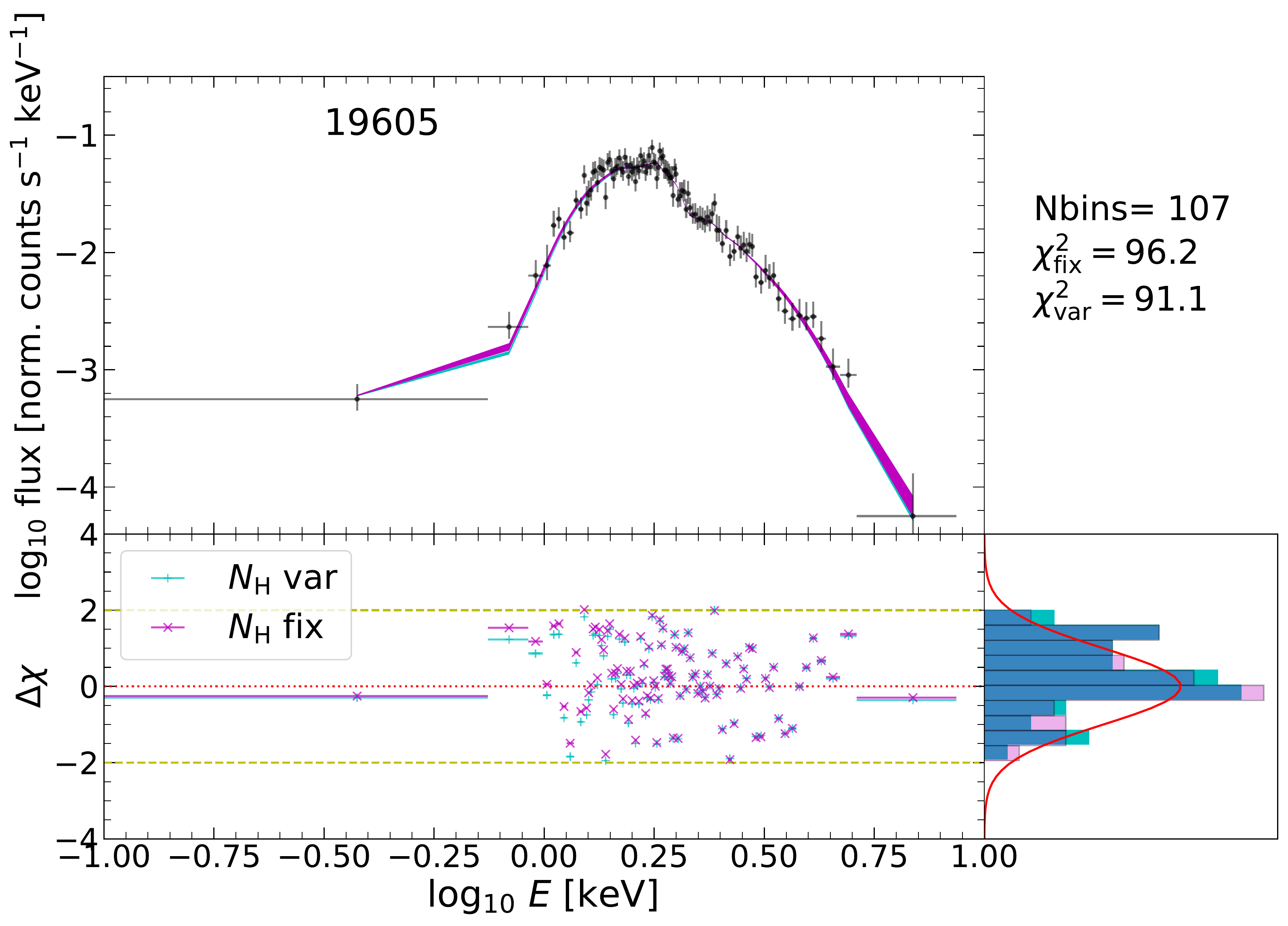}
\includegraphics[width=0.45\textwidth]{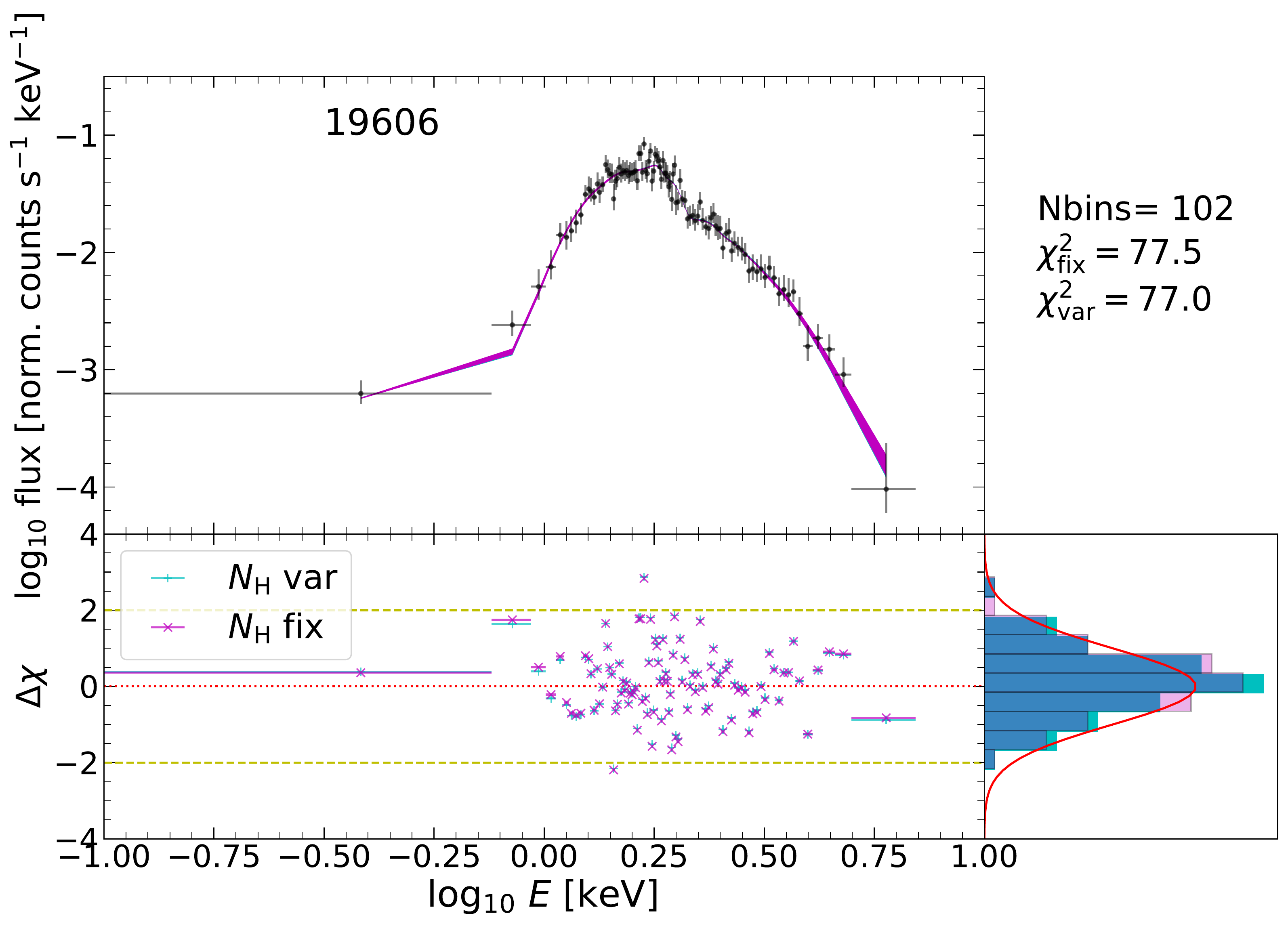}
\caption{Comparison of the spectral data taken in the GRADED mode with the joint fit spectral model. Continuation of Fig.~\ref{fig:spec_graded_1a}}\label{fig:spec_graded_2}
\end{figure*}

\begin{figure*}
    \centering
\includegraphics[width=0.45\textwidth]{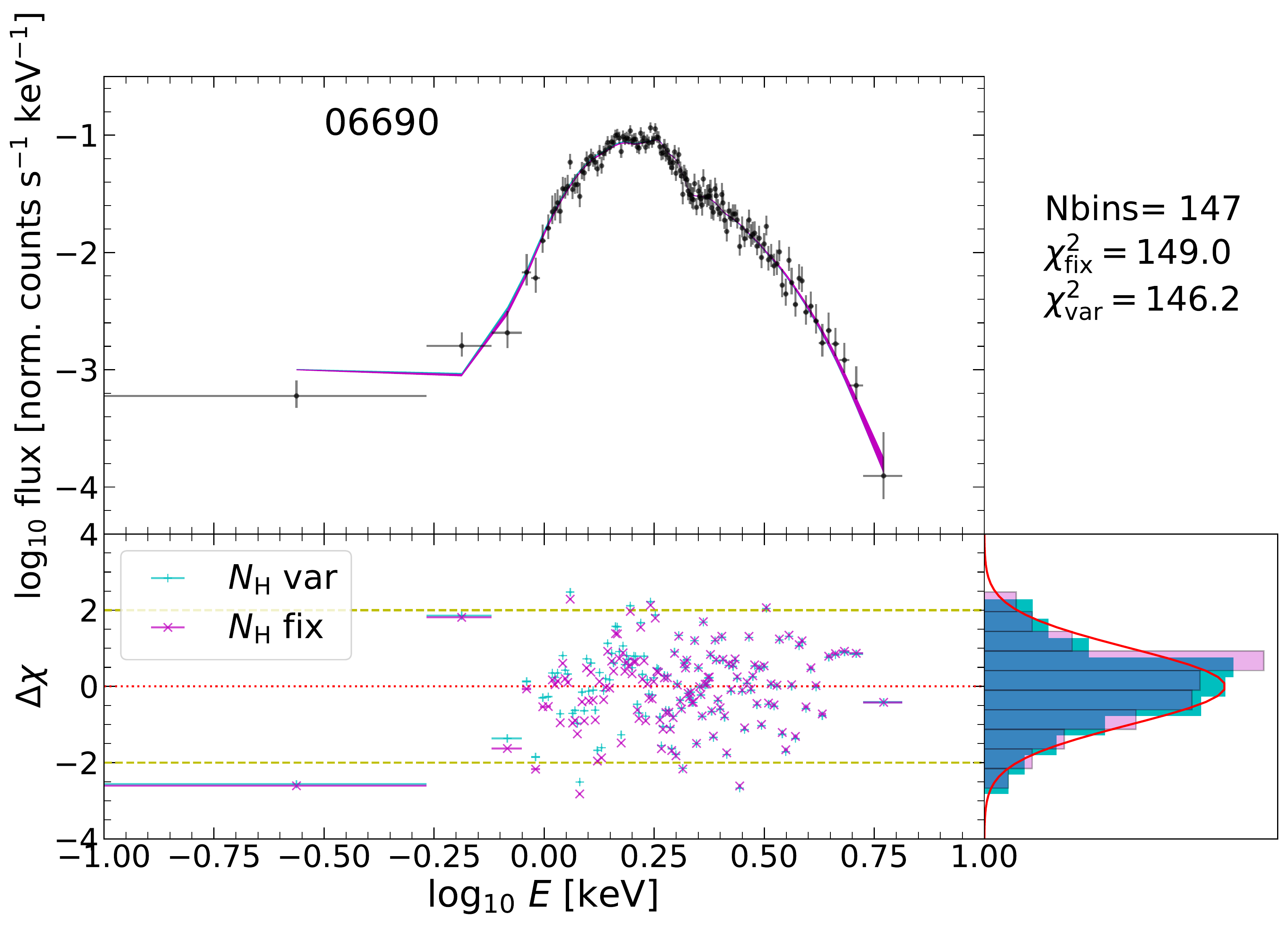}
\includegraphics[width=0.45\textwidth]{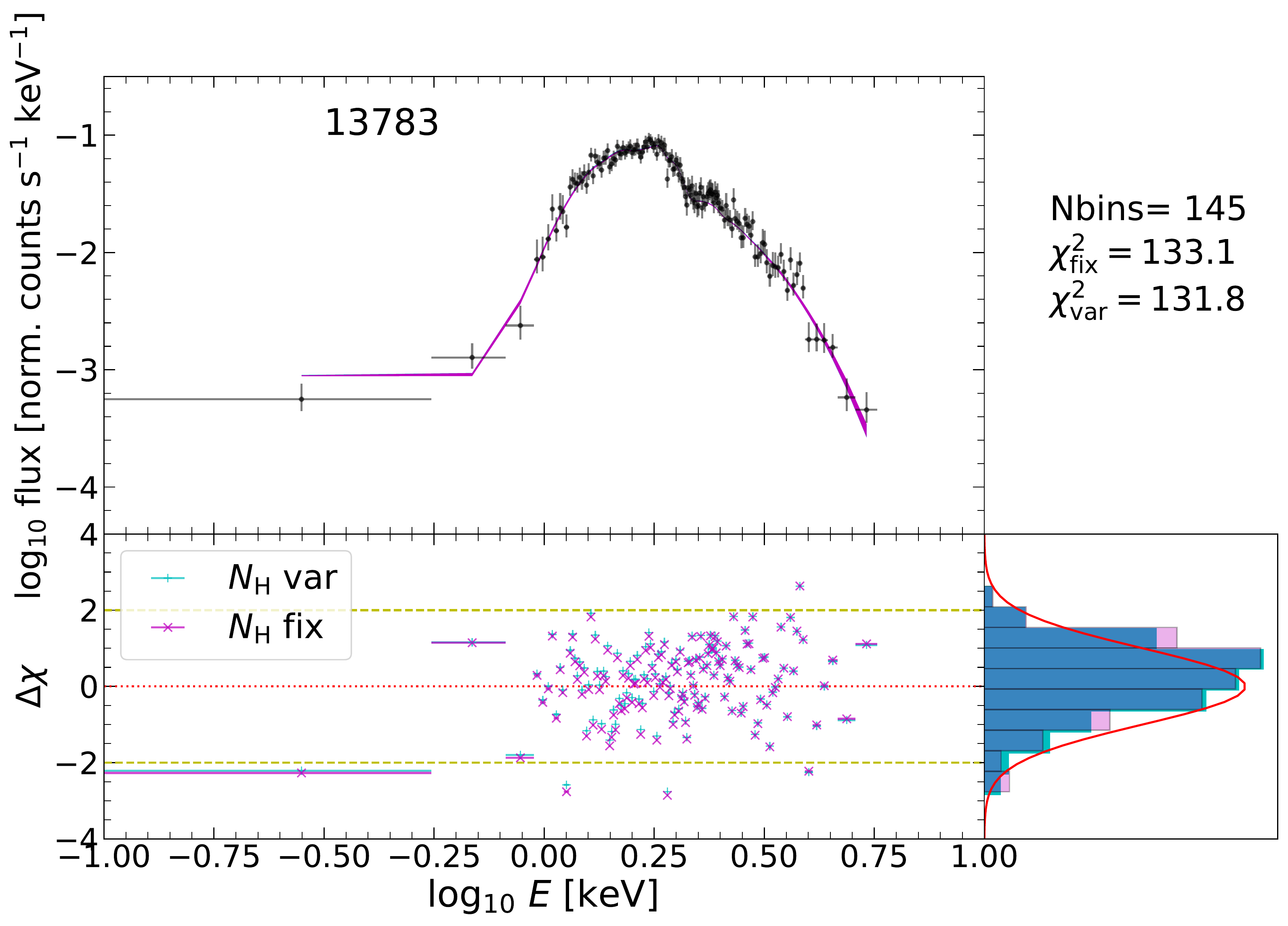}
\includegraphics[width=0.45\textwidth]{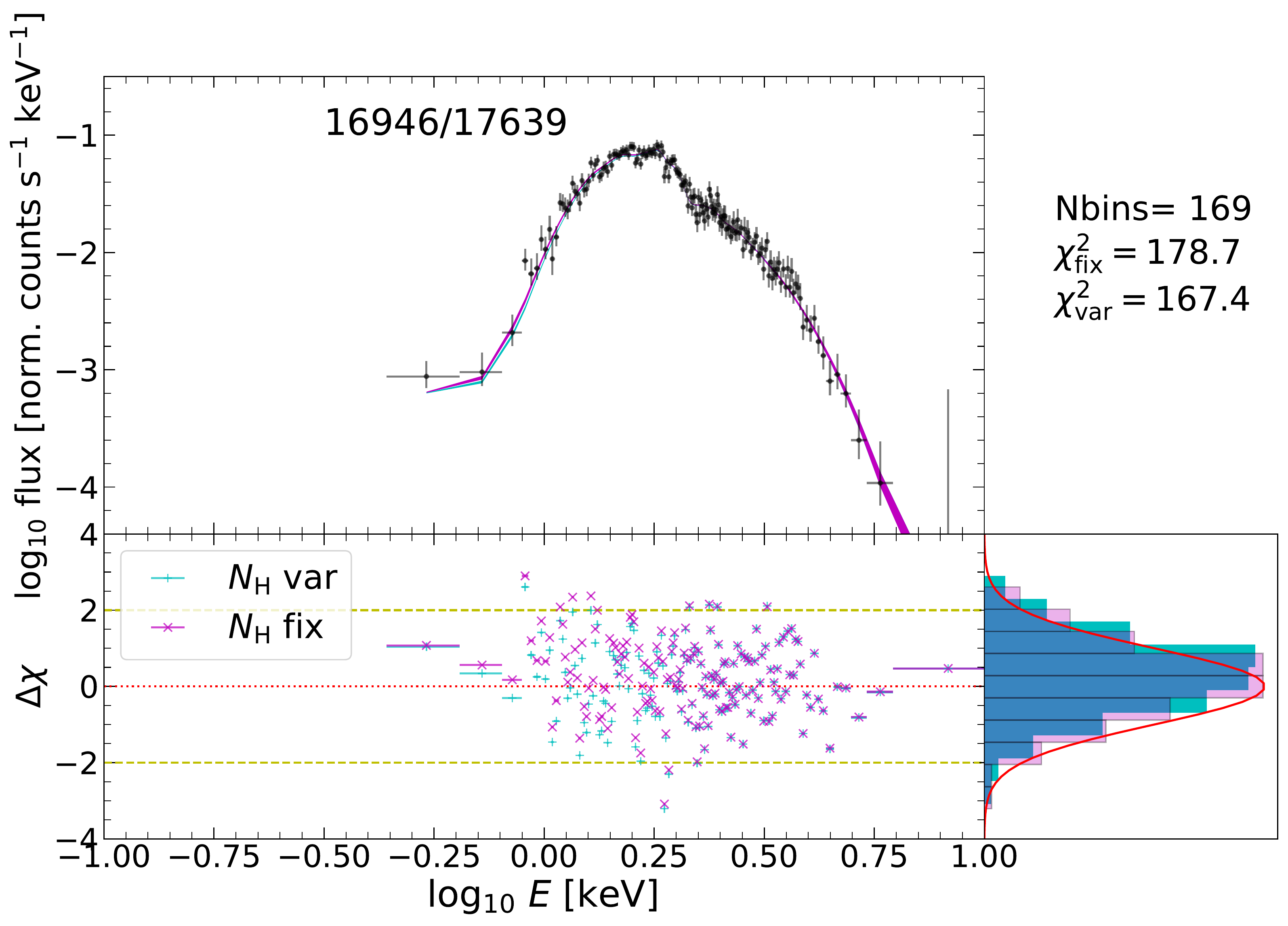}
\includegraphics[width=0.45\textwidth]{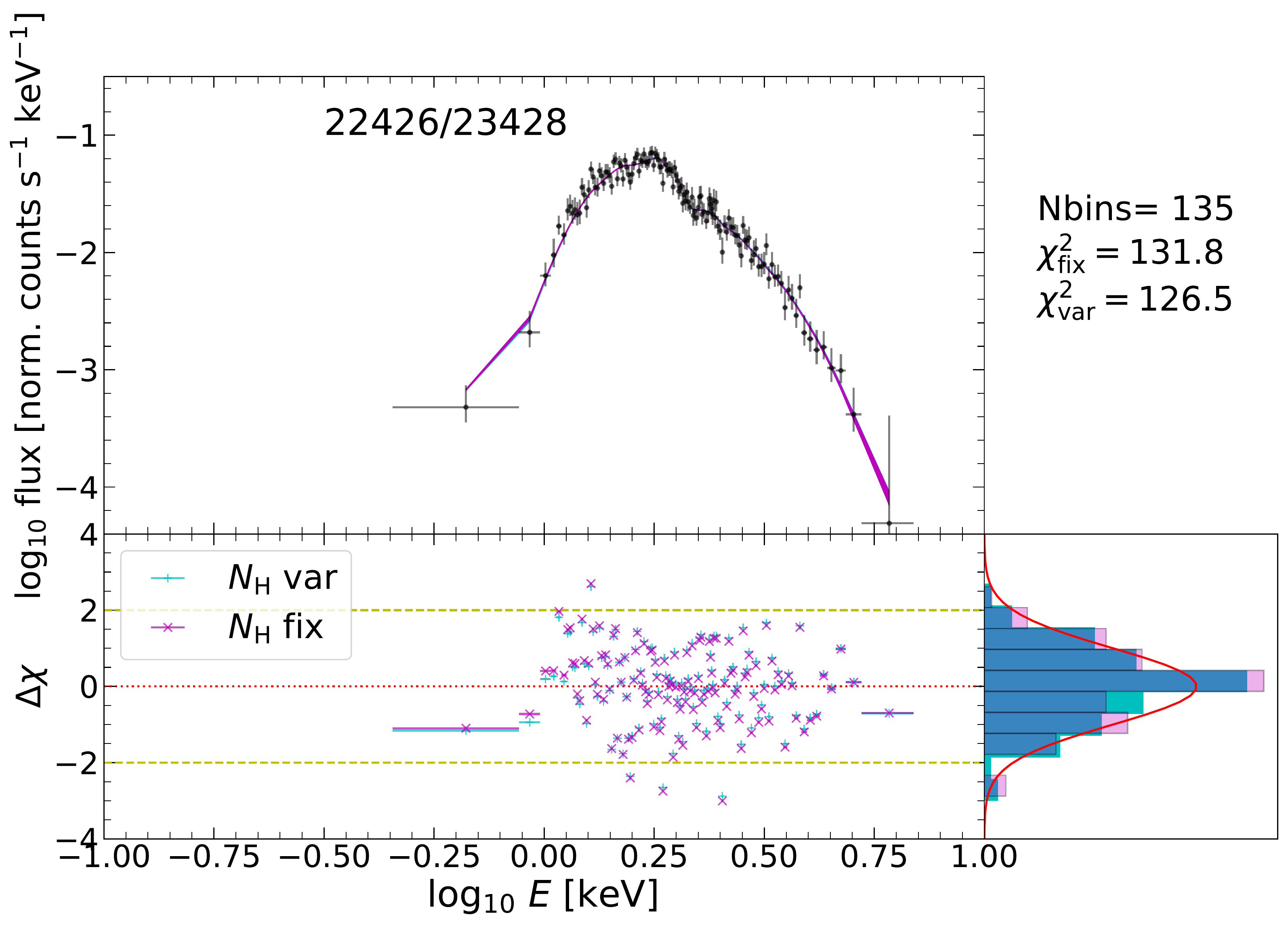}
    \caption{Comparison of the spectral data taken in the FAINT mode with the joint fit spectral model. Notations are the same as in Fig.~\ref{fig:spec_graded_1}.}
    \label{fig:spec_faint}
\end{figure*}

\begin{figure*}
    \centering
    \includegraphics[width=0.35\textwidth]{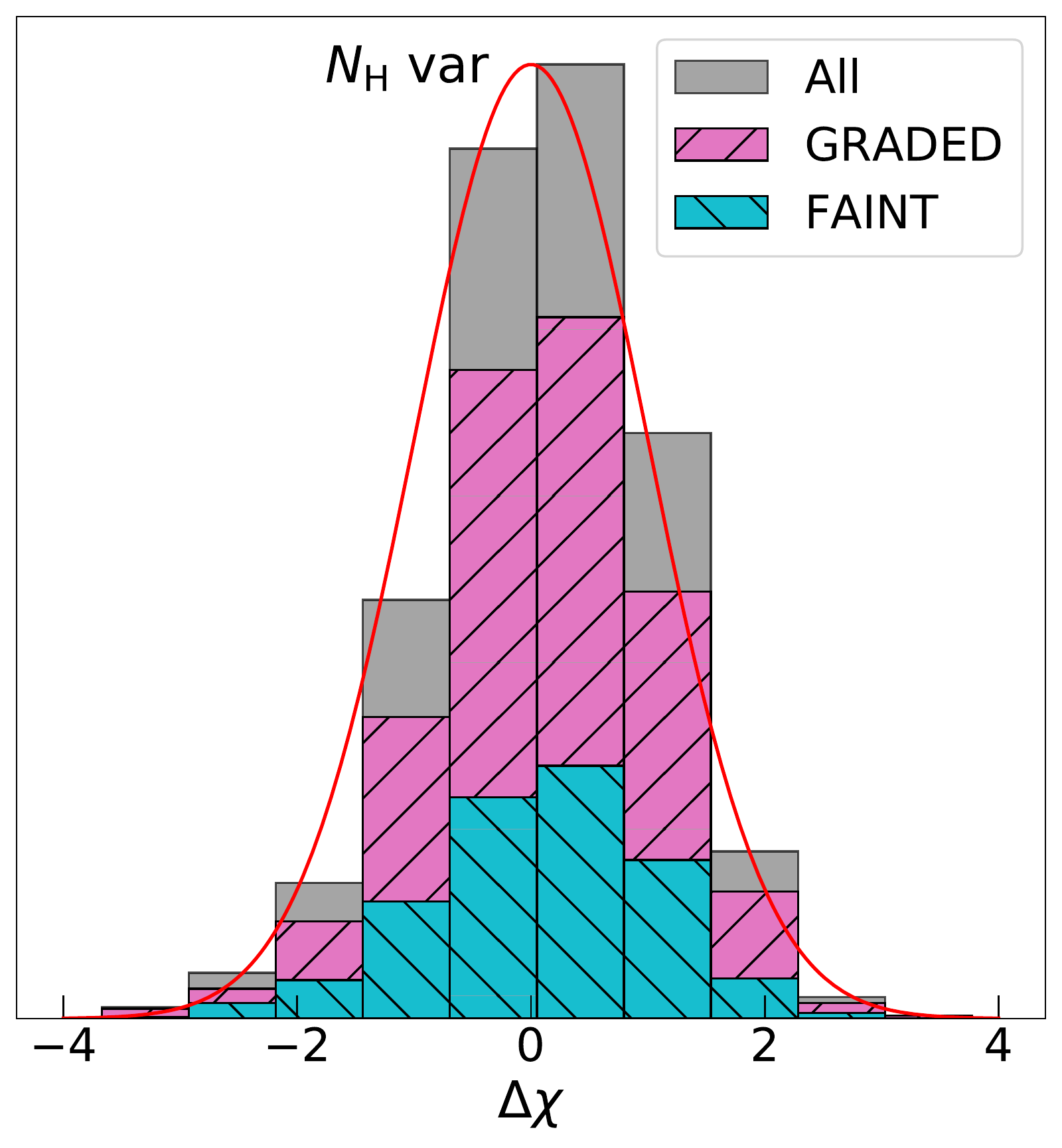}
    \hspace{2cm}
    \includegraphics[width=0.35\textwidth]{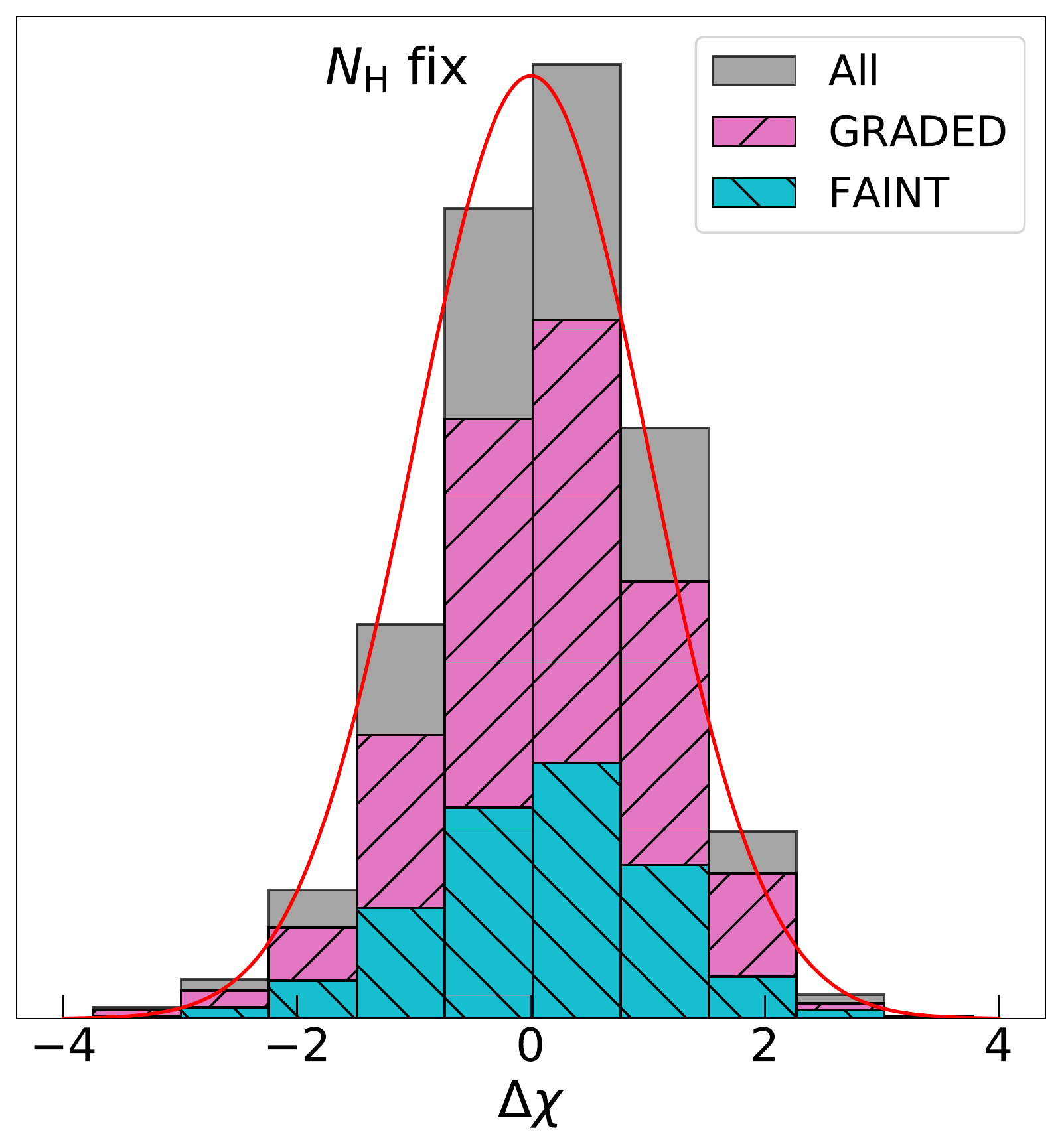}
    \caption{Distribution of the standardised residuals for the joint  fit spectral model with variable $N_{\mathrm{H}}$ (left), and fixed $N_{\mathrm{H}}$ (right). Different hatch styles correspond to contributions from data taken in different modes, as indicated in the legend. Red solid lines show the standard normal distribution for comparison. }
    \label{fig:spec_resid_hists}
\end{figure*}

Finally, Figs.~\ref{fig:spec_graded_1}--\ref{fig:spec_faint} show the comparison between the spectra and joint fit spectral models. Each panel corresponds to an individual spectrum, as indicated with an ObsID number in the plot. The upper panels show the spectral data along with the 68 per cent credible interval for the model predictions for individual data points based on the posterior samples. The filled magenta strips correspond to the model with fixed $N_{\mathrm{H}}$, while filled cyan strips correspond to the model with variable $N_{\mathrm{H}}$. Actually, the two models are indistinguishable by eye. Moreover the variance of the model prediction due to the variance in the parameter posterior distributions is much smaller than the measurement errors, therefore the thickness  of the mode lines (i.e. that this is a region, and not the line) is barely seen.
The lower panels for each spectra show, as in Fig.~\ref{fig:slopes}, the standardised residuals with plus markers for variable $N_{\mathrm{H}}$ models and x markers for fixed $N_{\mathrm{H}}$ models. Here the variance must include both measurement error and the variance due to the variance in model parameters. Indeed, assuming that the $i$th data point $y_i$ due to Gaussian measurement error is distributed as $y_i\sin {\cal N}(\widetilde{y}_i(\theta), \sigma_i)$, where $\widetilde{y}_i(\theta)$ is the model prediction for the parameter set $\theta$ and $\sigma_i$ is the measurement error, one obtains $\mathrm{E}(y_i|\theta)=\mathrm{E}(\widetilde{y}_i|\theta)$ and $\mathrm{Var}(y_i|\theta)=\mathrm{Var}(\widetilde{y}_i|\theta)+\sigma_i^2$. In our case, the measurement error contribution to variance is clearly a dominant one, see Fig.~\ref{fig:slopes}. Therefore, here the standardised residuals for each data point are defined as \citep{GelmanBook}
\begin{equation}
    \Delta\chi_i=\frac{y_i-\mathrm{E}(\widetilde{y}_i|\theta)}{\sigma_i^2+\mathrm{Var}(\widetilde{y}_i|\theta)}.
\end{equation}
Accordingly, $\chi^2$ values given for each spectrum in Figs.~\ref{fig:spec_graded_1}--\ref{fig:spec_faint} and the total $\chi^2$ given in Table~\ref{tab:spectral_params} are calculated as $\chi^2=\sum_i (\Delta \chi_i)^2$. Notice also that the number of degrees of freedom given in Table~\ref{tab:spectral_params} neglects the presence of the hierachial priors for $N_{\mathrm{H}}$ (in which case, the effective number of model parameters is reduced, see, e.g, chapter 6 in the book by \citet{GelmanBook} for details) and all complications due to complex non-linear character of the spectral model \citep*{Andrae2010arXiv}. We also show in Fig.~\ref{fig:spec_resid_hists} the total distribution of the standardised residuals compared with the standard normal distribution.

\renewcommand\arraystretch{1.2}
\begin{table*}
    \centering
    \caption{Results of the joint spectral fit with Cstat statistics. Uncertainties correspond to the 68 per cent highest posterior density credible intervals.}
    \label{tab:spectral_params_cstat}
    \begin{tabular}{lccccccccc}
        \hline
         $N_{\mathrm{H}}$ &$\log_{10} T_{s0}$ & $s$ & $M$ & $R$ & $d$ & $N_{\mathrm{H}0}$ & $\sigma_{N_{\mathrm{H}}}$& $A$ \\
               &(K)&&$(M_\odot)$& (km) &(kpc)&($10^{22}$~cm$^{-2}$)& ($10^{20}$~cm$^{-2}$)&\\
        \hline
         Var & $6.22^{+0.03}_{-0.01}$ & $0.73^{+0.08}_{-0.09}$ & $1.37^{+0.13}_{-0.14}$ & $13.8^{+1.2}_{-1.3}$ & $3.33^{+0.11}_{-0.09}$& $1.609^{+0.027}_{-0.037}$ &  $4.3^{+1.5}_{-1.1}$ &$1.10^{+0.02}_{-0.02}$\\
         Fix & $6.22^{+0.03}_{-0.01}$ & $0.54^{+0.06}_{-0.05}$ & $1.45^{+0.14}_{-0.13}$ & $14.1^{+1.0}_{-1.6}$ & $3.34^{+0.10}_{-0.10}$& $1.618^{+0.034}_{-0.026}$  & -- & $1.09^{+0.01}_{-0.01}
         $\\
                \hline
    \end{tabular}
\end{table*}

\section{Auxiliary figures for superfluidity analysis}\label{app:SF}
\begin{figure*}
    \centering
        \includegraphics[width=0.47\textwidth]{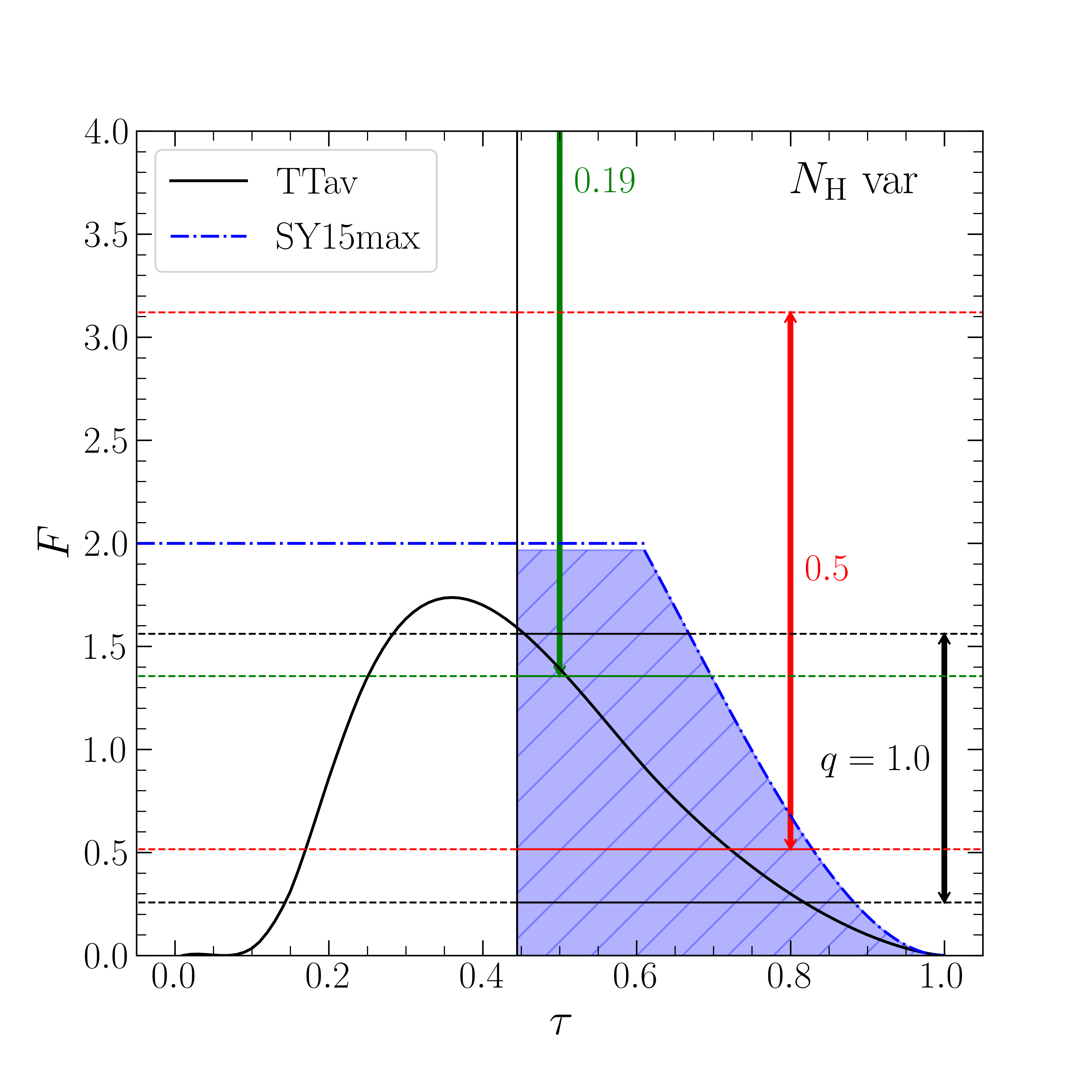}
    \includegraphics[width=0.47\textwidth]{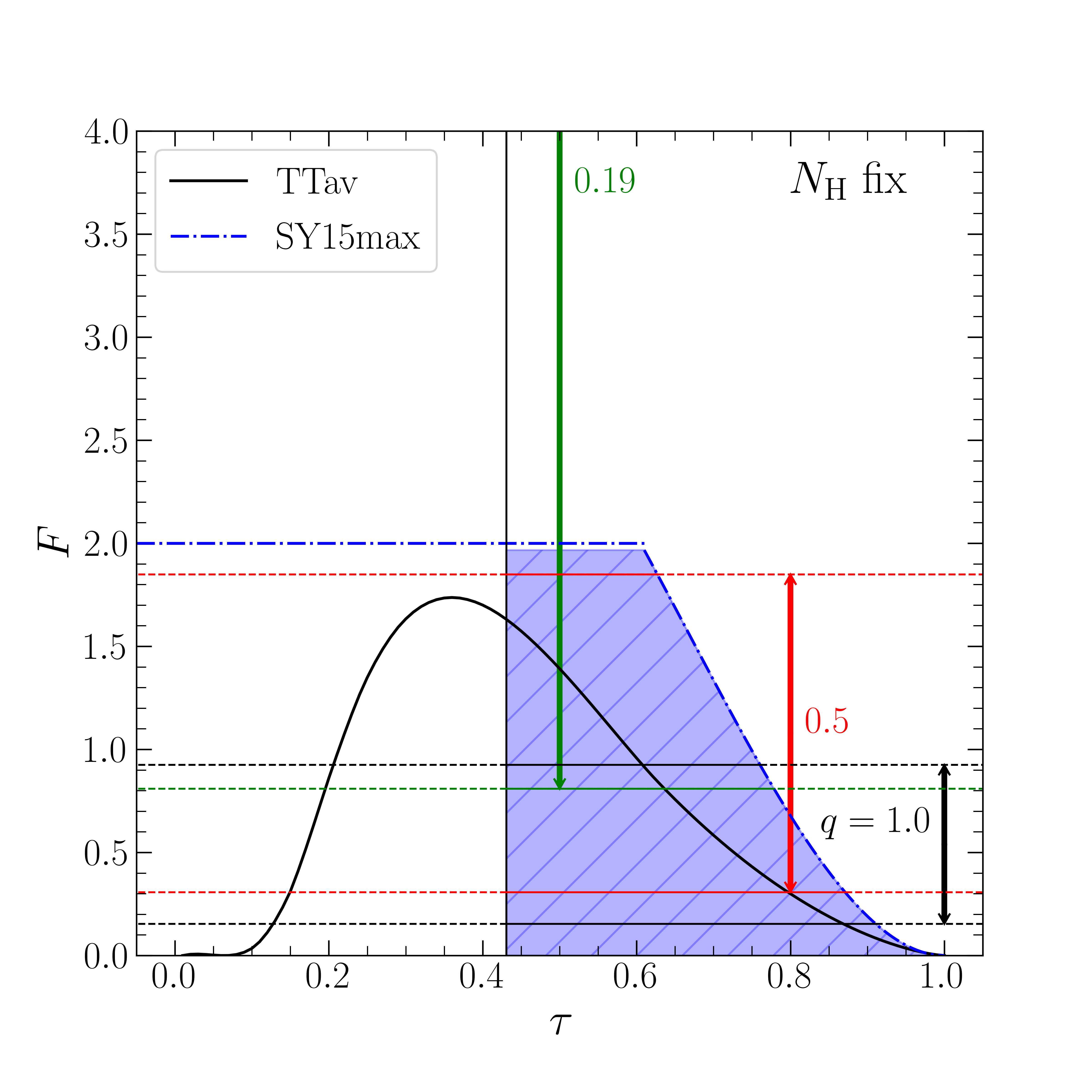}
        \caption{Same as Figure~\ref{fig:boxes}, but for the FAINT mode data alone.}\label{fig:Faintboxes}
\end{figure*}

\begin{figure*}
    \centering
        \includegraphics[width=0.47\textwidth]{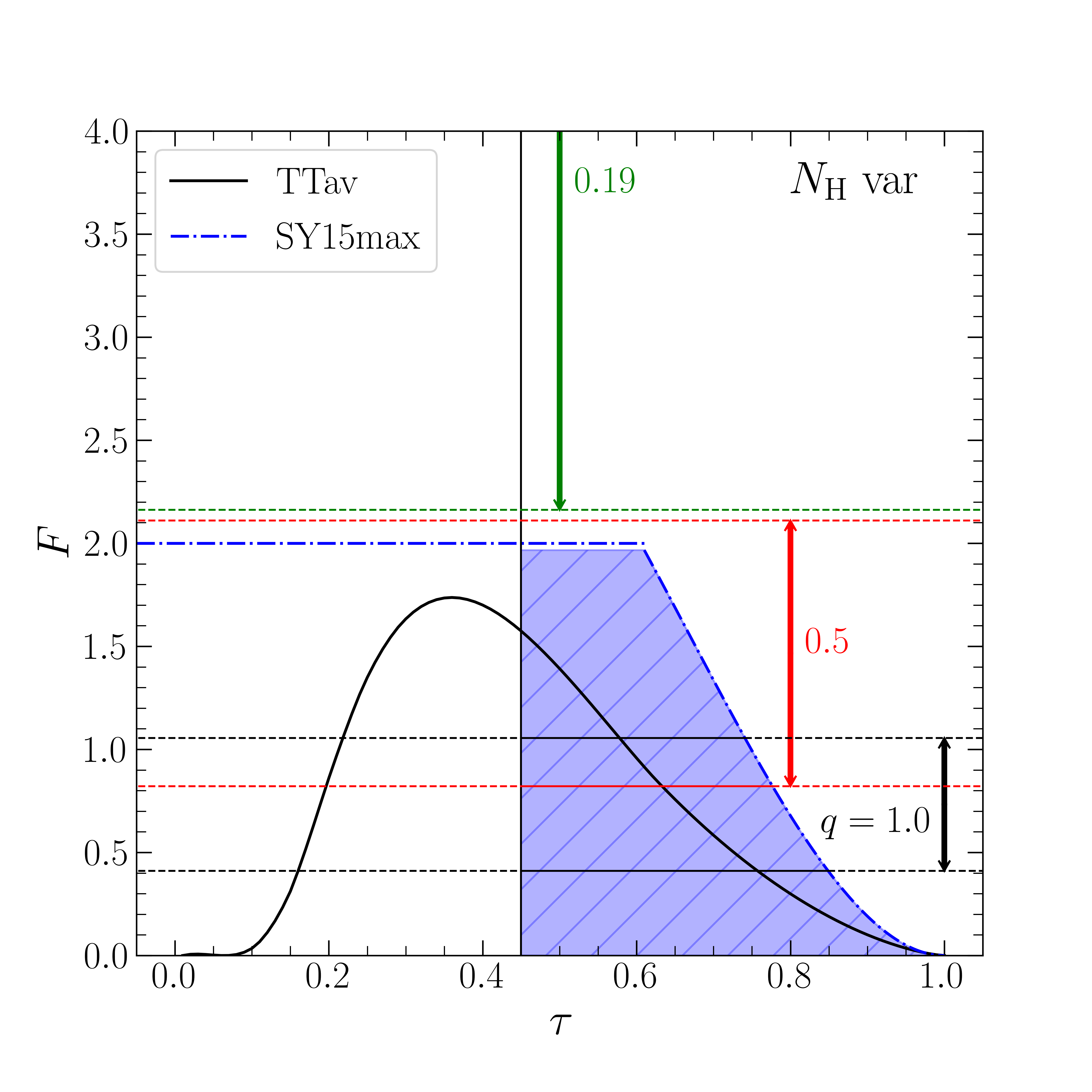}
    \includegraphics[width=0.47\textwidth]{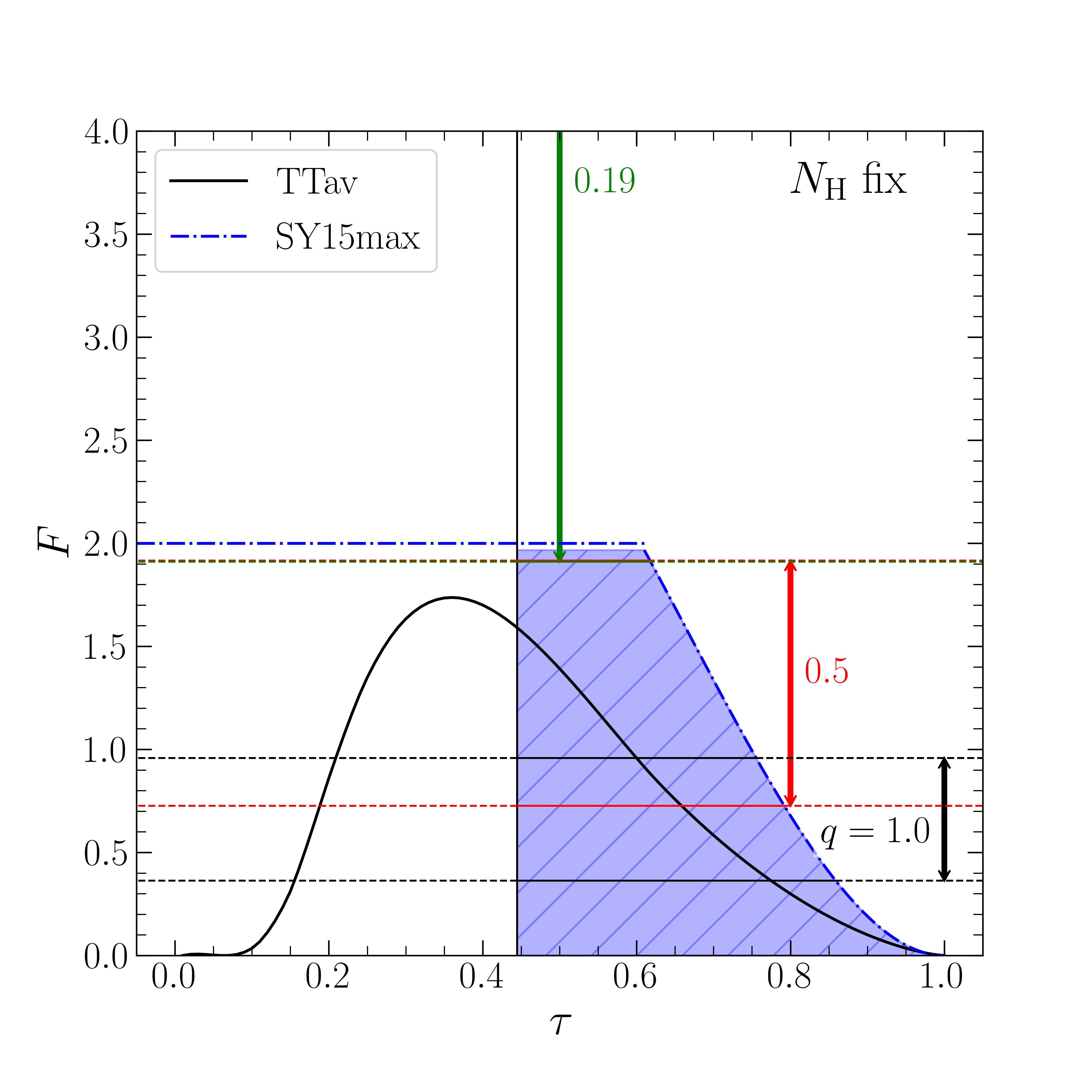}
        \caption{Same as Figure~\ref{fig:boxes}, but for the GRADED mode data alone.}\label{fig:GRADEDboxes}
\end{figure*}

In this appendix we show additional figures one would obtain performing an analysis of Sec.~\ref{Sec:SFanalysis} using only FAINT or GRADED mode data. These figures illustrate the corresponding rows in Tables~\ref{tab:TaudFdBox} and \ref{tab:sf_res}. However, based on the discussion in Appendix~\ref{app:spectral}, at the present stage, preference of the one mode over another probably is not justified, and one should rely on all data. 
Figs.~\ref{fig:Faintboxes} and \ref{fig:GRADEDboxes} repeat Fig.~\ref{fig:boxes} but for FAINT and GRADED modes, respectively. Less constrained data result here in wider limits on $G_d$ and weaker limits on $q$ than in the case of the joint fit. The apparent lowering the limit on $q$ for the FAINT data, see Fig.~\ref{fig:Faintboxes}, is due to the low-$R$, high-$T_s$ tail of the posterior spectral parameter distribution, as discussed in the main text. 
%%%%%%%%%%%%%%%%%%%%%%%%%%%%%%%%%%%%%%%%%%%%%%%%%%%%%%%%%%%%%%%%%%%%%%%%%%%%%%%%%%%%%%%%%%%%%%%%%%%%
%%%%%%%%%%%%%%%%%%%%%%%%%%%%%%%%%%%%%%%%%%%%%%%%%%%%%%%%%%%%%%%%%%%%%%%%%%%%%%%%%%%%%%%%%%%%%%%%%%%%
%%%%%%%%%%%%%%%%%%%%%%%%%%%%%%%%%%%%%%%%%%%%%%%%%%%%%%%%%%%%%%%%%%%%%%%%%%%%%%%%%%%%%%%%%%%%%%%%%%%%
%%%%%%%%%%%%%%%%%%%%%%%%%%%%%%%%%%%%%%%%%%%%%%%%%%%%%%%%%%%%%%%%%%%%%%%%%%%%%%%%%%%%%%%%%%%%%%%%%%%%

\begin{figure*}
    \centering
    \includegraphics[width=0.45\textwidth]{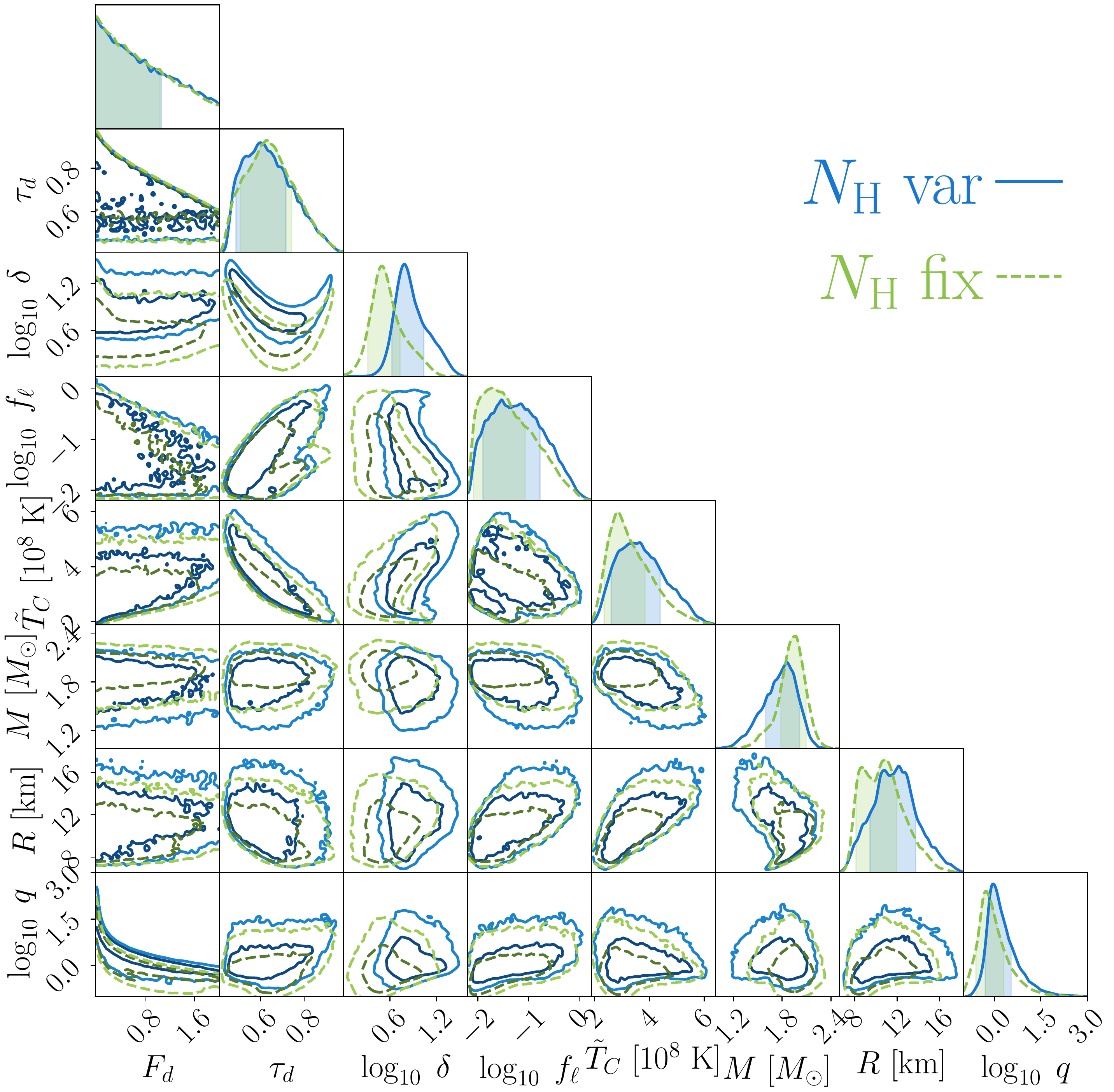}
     \includegraphics[width=0.45\textwidth]{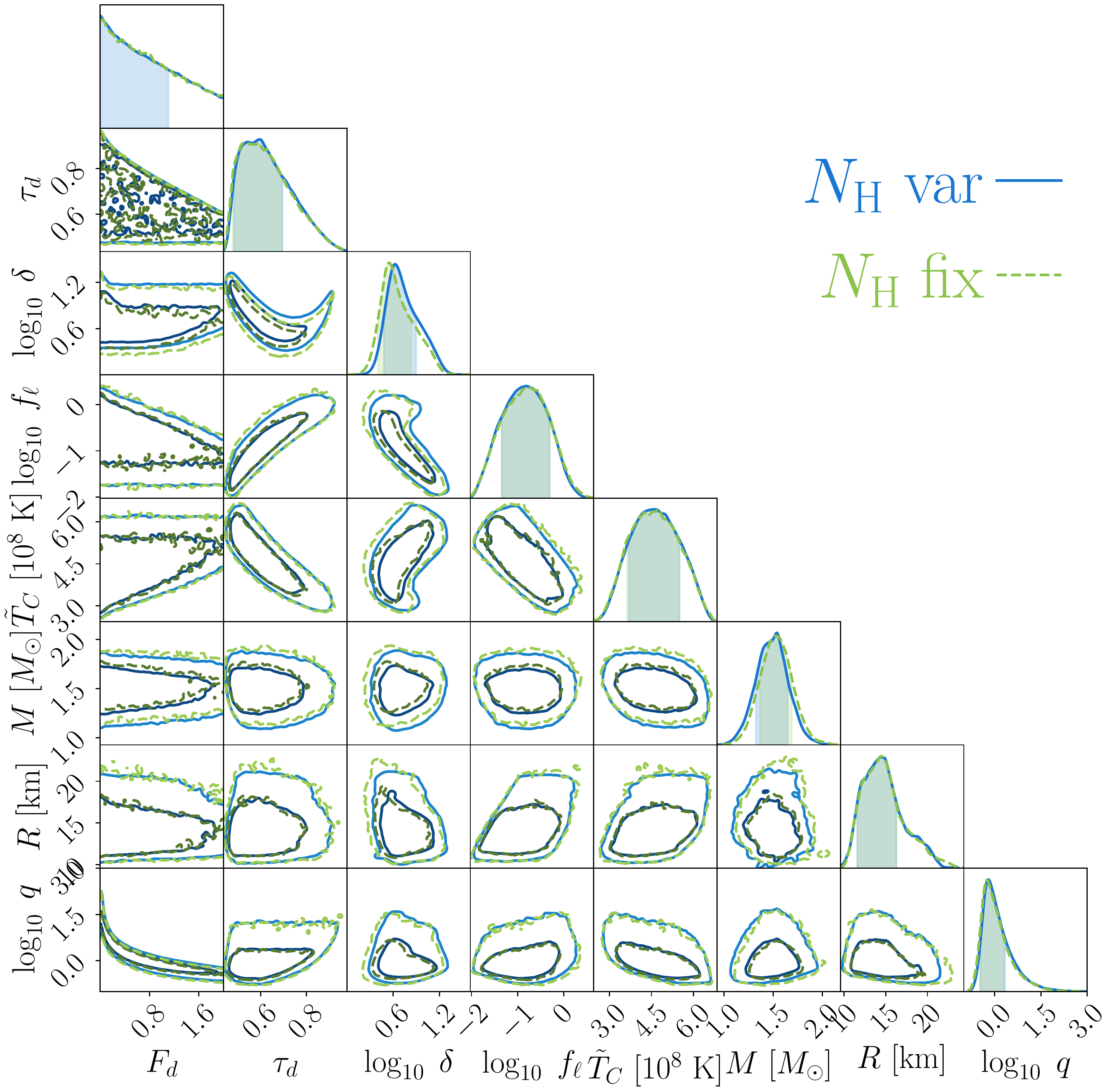}
    \caption{Same as Figure~\ref{fig:tri_SF}, but for FAINT mode data alone (left), and GRADED mode data alone (right). }\label{fig:tri_SF_app}
\end{figure*}

Finally, in Fig.~\ref{fig:tri_SF_app} we show the posterior distributions for the superfluidity parameters obtained from the analysis of the FAINT and GRADED data alone. This figure corresponds to Fig.~\ref{fig:tri_SF} in the main text. Since the derived superfluidity parameters are consistent between the analysed modes, the plots in Fig.~\ref{fig:tri_SF_app} and in  Fig.~\ref{fig:tri_SF} are similar and differ only in minor details. 

%%%%%%%%%%%%%%%%%%%%%%%%%%%%%%%%%%%%%%%%%%%%%%%%%%%%%%%%%%%%%%%%%%%%%%%%%%%%%%%%%%%%%%%%%%%%%%%%%%%%
%%%%%%%%%%%%%%%%%%%%%%%%%%%%%%%%%%%%%%%%%%%%%%%%%%%%%%%%%%%%%%%%%%%%%%%%%%%%%%%%%%%%%%%%%%%%%%%%%%%%

% Don't change these lines
\bsp	% typesetting comment
\label{lastpage}
\end{document}